\documentclass[prb,showpacs,floatfix,superscriptaddress,twocolumn,aps,10]{revtex4-2}
\usepackage[english]{babel} 
\usepackage{bm}
\usepackage{graphicx}
\usepackage{amsmath}
\usepackage{amssymb}
\usepackage{wasysym}
\usepackage{comment}
\usepackage[dvipsnames]{xcolor}
\usepackage{tikz}
\usepackage{adjustbox}
\usepackage{booktabs}
\usepackage{array}
\usepackage{multirow}

\usepackage[hidelinks]{hyperref}

\begin{document}

\title{Pinch-line spin liquids as layered Coulomb phases and applications to cubic models}
    \author{N. Davier}
	\email[]{naimo.davier@u-bordeaux.fr}
    \affiliation{CNRS, Universit\'e de Bordeaux, LOMA, UMR 5798, 33400 Talence, France}
		 
    \author{F. A.  G\'omez Albarrac\'in}
    \affiliation{Instituto de F\'isica de L\'iquidos y Sistemas Biol\'ogicos, CONICET, Facultad de Ciencias Exactas, Universidad Nacional de La Plata, 1900 La Plata, Argentina}
    \affiliation{Departamento de Ciencias B\'asicas, Facultad de Ingenier\'ia, UNLP, La Plata, Argentina}

    \author{H. Diego  Rosales}
	\affiliation{Instituto de F\'isica de L\'iquidos y Sistemas Biol\'ogicos, CONICET, Facultad de Ciencias Exactas, Universidad Nacional de La Plata, 1900 La Plata, Argentina}
    \affiliation{Departamento de Ciencias B\'asicas, Facultad de Ingenier\'ia, UNLP, La Plata, Argentina}
		
	\author{P. Pujol}
	\affiliation{Laboratoire de Physique Th\'eorique, Universit\'e de Toulouse, CNRS, UPS, France}

    \author{Ludovic D.C. Jaubert}
    \affiliation{CNRS, Universit\'e de Bordeaux, LOMA, UMR 5798, 33400 Talence, France}
		
	\date{\today}
		
\begin{abstract} 
Spin liquids form fluctuating magnetic textures which have to obey certain rules imposed by frustration. These rules can often be written in the form of a Gauss law, indicating the local conservation of an emergent electric field. In reciprocal space, these emergent Gauss laws appear as singularities known as pinch points, that are accessible to neutron-scattering measurements. But more exotic forms of electromagnetism have been stabilized in spin liquids, and in a few rare instances, these zero-dimensional singularities have been extended into one-dimensional pinch lines. Here we propose a simple framework for the design of pinch-line spin liquids in a layered structure of two-dimensional algebraic spin liquids. A plethora of models can be build within this framework, as exemplified by several concrete examples where our theory is confirmed by simulations, and where the rank of the tensorial gauge field is continuously varied along the pinch line, opening new avenues in fractonic matter. Then we use our framework to understand how the evolution of the singularity pinch point along the pinch line can be understood as the interference pattern of two emergent electric fields. Finally, we apply our intuition on these emergent electric fields in real space to generic pinch line models beyond our layered framework, and revisit the recently proposed pinch line model on the octochlore lattice.
\end{abstract}
\maketitle

\section{Introduction}
Spin liquids are the hallmark of frustrated magnetism. Being magnetically disordered, they fail to be characterized by order parameters. This is why alternative descriptions have proven to be necessary, such as topological orders \cite{Wen_2002} and fractionalized excitations \cite{castelnovo08a,Kitaev06a}, often related to an underlying emergent gauge field theory \cite{Knolle2019}. Electromagnetism is probably the most natural form of emergent gauge field for spin liquids; frustration induces a microscopic constraint between spins, which can be rewritten as the local conservation rule of an emergent field or, in other words, the zero divergence of {an effective} Maxwell equation \cite{henley2010coulomb}. Such peculiar spin-spin correlations decay algebraically in real space, and take the form of a characteristic singularity in reciprocal space, the so-called pinch point, readily available to neutron-scattering measurements.

The elegance of emergent phenomena is that they are not confined to our natural intuition. Exotic gauge fields can be explored where charges become magnetic \cite{castelnovo08a}, quasiparticles are their own antiparticles \cite{Kitaev06a}, and electromagnetic fields evolve into a tensor \cite{Pretko_2_2017}. Higher-rank (tensorial) gauge fields have recently been actively sought after, as sources for fractons, with potential applications in quantum information \cite{pretko20a,Nandkishore19a}. Although the first models were somewhat complex \cite{Chamon05a,Xu06a,Haah11a}, more realistic Hamiltonians have subsequently been derived \cite{Benton2016,Slagle17a,Yan_2020,Benton_Moessner_2021}, followed by the theoretical proposals of a plethora of higher-rank spin liquids \cite{Niggemann_2023,Desrochers23a,Davier_2023,Yan_2024_short,Yan_2024_long,Fang_2024,Lozano24a,chung_2024_preprint,lozanogomez_2024_preprint}, whose tensorial nature transforms the traditional pinch points into multifold ones \cite{Prem_2018}.

\begin{figure}
    \centering
    \includegraphics[width=\linewidth]{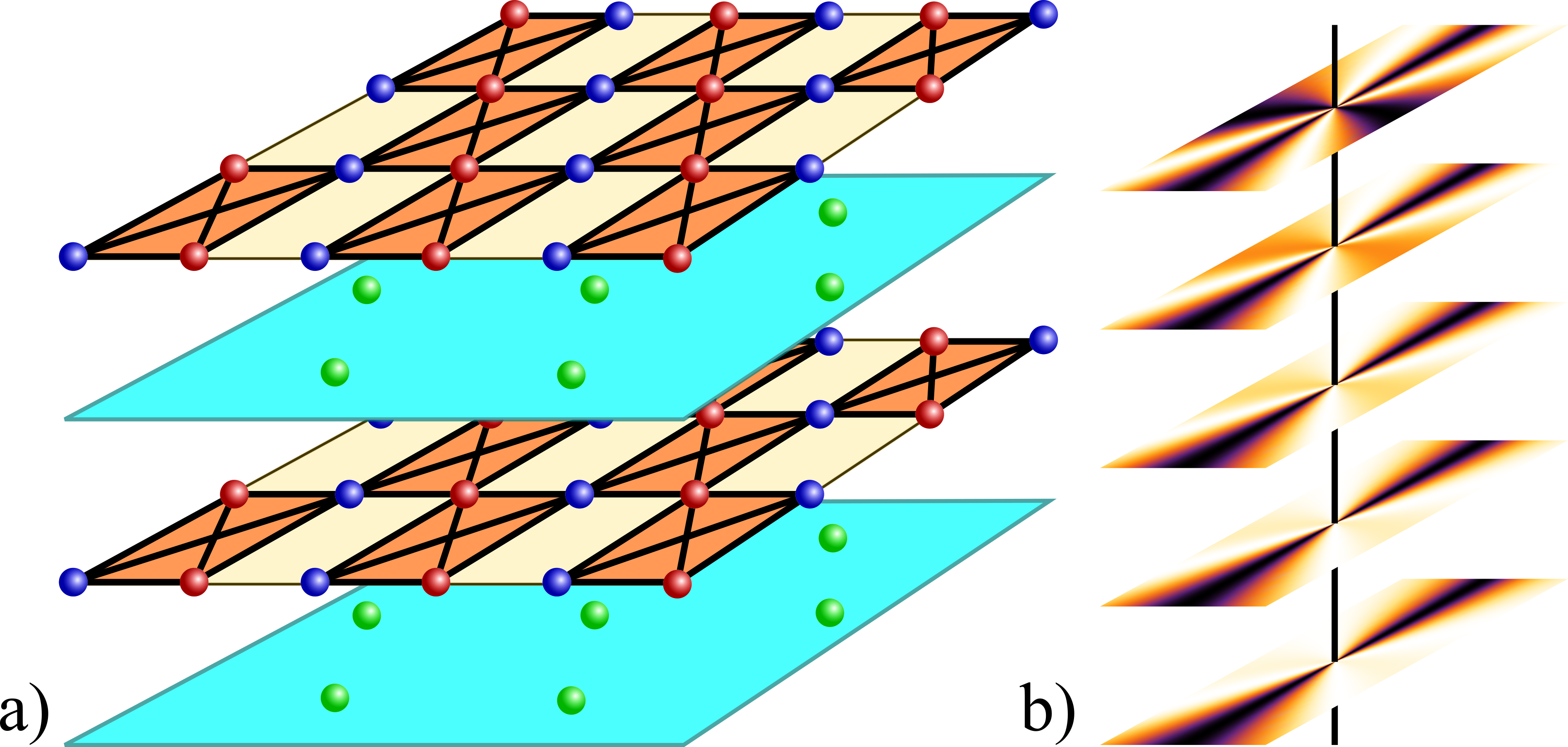}
    \caption{a) Illustration of the pinch-line framework. Planes containing a parent 2D algebraic spin liquid hosting pinch points (depicted as a checkerboard) are placed regularly along a transverse axis, with intermediate layers (in blue) designed to make the pinch points one-dimensional. b) Representation of a pinch line in reciprocal space (black line), with pinch points centered on the line on every transverse plane.}
    \label{fig: stakked system}
\end{figure}

There is, however, a family of spin liquids that has remained rare in the literature: the pinch-line spin liquids \cite{Benton2016}, where the singularity due to the emergent zero divergence forms a line in reciprocal space. As a topological analogy, it is similar to a zero-dimensional monopole becoming a one-dimensional Dirac string. In certain instances, pinch lines can be explained from topological quantum chemistry as a one-dimensional mismatch between two specific band representations of the lattice and Hamiltonian symmetries \cite{Fang_2024}. First derived in a pyrochlore model in the context of Tb$_2$Ti$_2$O$_7$ [\onlinecite{Benton2016}], only a handful of models have been found to support pinch lines so far \cite{Benton2016,Niggemann_2023, Yan_2024_long,Fang_2024}.

In this work we provide a simple and sufficient, but not necessary, framework to design pinch-line spin liquids. Starting from a two-dimensional (2D) algebraic spin liquid, we explain how to build a 3D model able to continuously propagate pinch-point singularities into parallel lines in reciprocal space. The idea is to stack layers of the 2D algebraic spin liquid with intermediate layers whose geometry is designed to propagate each 2D pinch points into a 3D pinch line, as illustrated schematically in Fig.~\ref{fig: stakked system}. We also show how the pinch-line construction is accompanied by an enhancement of zero modes. Our method is fairly generic and only relies on standard Heisenberg exchange couplings. After introducing the necessary literature background, we present the generic aspect of our theory, before applying it to a variety of models of increasing complexity. As a demonstration of the high degree of model design now achievable in frustrated magnetism, we show how pinch lines can become multifold. Our theory is confirmed by Monte Carlo simulations.

The construction as a multi-layer of algebraic spin liquids also offers an intuitive description of the origin of pinch lines in real space. The variation of the pinch points along a pinch line can be understood as the interference between two distinct electric fields; one coming from the parent algebraic spin liquid and another one including the intermediate layers. To conclude we apply our understanding of pinch lines derived from this layered construction to generic pinch lines with cubic symmetry such as on the octochlore lattice.\\

\section{Background} 
While there is no unique way to stabilize a spin liquid, they are known to naturally occur as the ground state of certain cluster-type Hamiltonians describing systems where the spin-spin interactions can all be expressed within small group of spins called cluster. These Hamiltonian can therefore generally be expressed as
\begin{equation}
H = \frac{1}{2}\sum_{n, \alpha} \bm{\mathcal{C}}_{n,\alpha}^2, \qquad \bm{\mathcal{C}}_{n,\alpha} = \sum_{i \in \alpha \cap n} \gamma_i^\alpha\mathbf{S}_i,
\label{eq:hamcl} 
\end{equation}
with classical Heisenberg spins $\mathbf{S}_i$. $n$ labels the unit cells while $\alpha$ labels the different types of clusters among the considered Hamiltonian. The coefficients $\gamma_i^\alpha$ define the different Heisenberg interactions within cluster of type $\alpha$, and the second sum is made over all sites $i$ belonging to the cluster of type $\alpha$ located in the unit cell $n$. The pyrochlore and kagome antiferromagnets are famous members of this class of Hamiltonians with for both systems two types of clusters that are up triangles(tetrahedra) and down triangles(tetrahedra).

The energy of Eq.~(\ref{eq:hamcl}) is minimized by setting the constrainers $\bm{\mathcal{C}}_{n, \alpha}=0, \forall n, \alpha$ (when possible). Following the Benton-Moessner approach \cite{Benton_Moessner_2021}, the Fourier transform of $\bm{\mathcal{C}}_{n, \alpha}=0$ gives
\begin{equation} 
\sum_{u=1}^{n_s} L_u^\alpha(\mathbf{q}) \, \tilde{\mathbf{S}}_u(\mathbf{q}) = 0, \qquad L_u^\alpha(\mathbf{q}) = \sum_{j \in D_u^\alpha} \gamma_j \, e^{i \mathbf{q} \cdot \mathbf{r}_j},
\label{eq:Lu}
\end{equation} 
where $u$ runs over the $n_s$ inequivalent sublattices, $j$ runs over all sites $D_u^\alpha$ of sublattice $u$ within a cluster of type $\alpha$, and $\mathbf{r}_j$ is the position of site $j$ with respect to the cluster center. $\tilde{\mathbf{S}}_u(\mathbf{q})$ is the Fourier transform of the spin configuration on sublattice $u$. The constraint vectors $\mathbf{L}^\alpha(\mathbf{q})\equiv \{L_u^\alpha(\mathbf{q})\}_{u=1,..,n_s}$ encode most of the spin-liquid properties \cite{Benton_Moessner_2021}, allowing for a classification of classical spin liquids \cite{Davier_2023, Yan_2024_short,Yan_2024_long, Fang_2024}. Note that Eq.~(\ref{eq:Lu}) encapsulates in fact three equations, corresponding to the three spin components, which are simply all three identical because of the isotropy of the considered systems. This is why hereafter we never specify the spin index and speak about a single constraint vector for the three spin components. When the dimension of the vector space $V = \text{Vect}(\{\mathbf{L}^\alpha\})$ generated by these constraint vectors is reduced at some point $\mathbf{q}^\star$ of reciprocal space, this indicates the closing of a gap in the energy spectrum, which results in the presence of a pinch point in the structure factor at wave vector $\mathbf{q^\star}$. This is the signature of an algebraic spin liquid\cite{Benton_Moessner_2021,Yan_2024_short, Yan_2024_long, Davier_2023}. This can occur if one of the constraint vectors vanishes at $\mathbf{q}^\star$, or if one of the constraint vectors becomes linearly dependent of the other ones at $\mathbf{q}^\star$. In both cases a critical vector $\mathbf{L}^c(\mathbf{q})$ can be defined as a linear combination of the constraint vectors $\mathbf{L}^c(\mathbf{q})= c_\alpha \mathbf{L^\alpha}(\mathbf{q})$ such that $\mathbf{L}^c(\mathbf{q}^*)=0$.
In 3D, if such a critical point $\mathbf{q}^\star$ is extended as a line, such that $\mathbf{L}^c(\mathbf{q}^*)=0$ for any point $\mathbf{q^*}$ belonging to this line, then the dimension of the vector space $V$ is reduced along the entire line, and this is the signature of a pinch-line spin liquid \cite{Yan_2024_long,Fang_2024}.

The elegance of this formalism is that Eq.~(\ref{eq:Lu}) can be naturally reformulated as a Gauss law in the vicinity of $\mathbf{q}^\star$ when expressed using the critical vector $\mathbf{L}^c$ \cite{Benton_Moessner_2021, Davier_2023, Yan_2024_short}. At no loss in generality, let us consider a rank$-2$ U(1) gauge theory where the Gauss law of the tensor electric field reads $\partial_i\partial_j E_{ij} = 0 \Leftrightarrow q_i q_j \tilde{E}_{ij}=0$ in Fourier space [\onlinecite{Pretko_2_2017}]. The emergent electric field describes the spin degrees of freedom of Eq.~(\ref{eq:Lu}) \footnote{Note that there exists a copy of this electric field for each one of the three spin components.}, while the derivatives of the Gauss law come from the lowest-order expansion of the critical vector $\mathbf{L}^c(\mathbf{q}^\star+\delta \mathbf{q})$; if the first term of Taylor expansion is second order, then the prefactor in front of $\tilde{E}_{ij}$ is proportional to $q_i q_j$. We then have a rank$-2$ Gauss law which imposes that some of the correlators of the tensor electric field (e.g.~$\langle \tilde{E}_{xx}(\mathbf{q}_\perp) \tilde{E}_{yy}(-\mathbf{q}_\perp) \rangle$) support a pinch-point singularity with four-fold symmetry \cite{Prem_2018}.\\

\section{How to design pinch-line models: Generic construction} 
\label{sec:generic}

\subsection{Simple approach}
\label{sec:simple}

\begin{figure}
    \centering
    \includegraphics[width=0.8\linewidth]{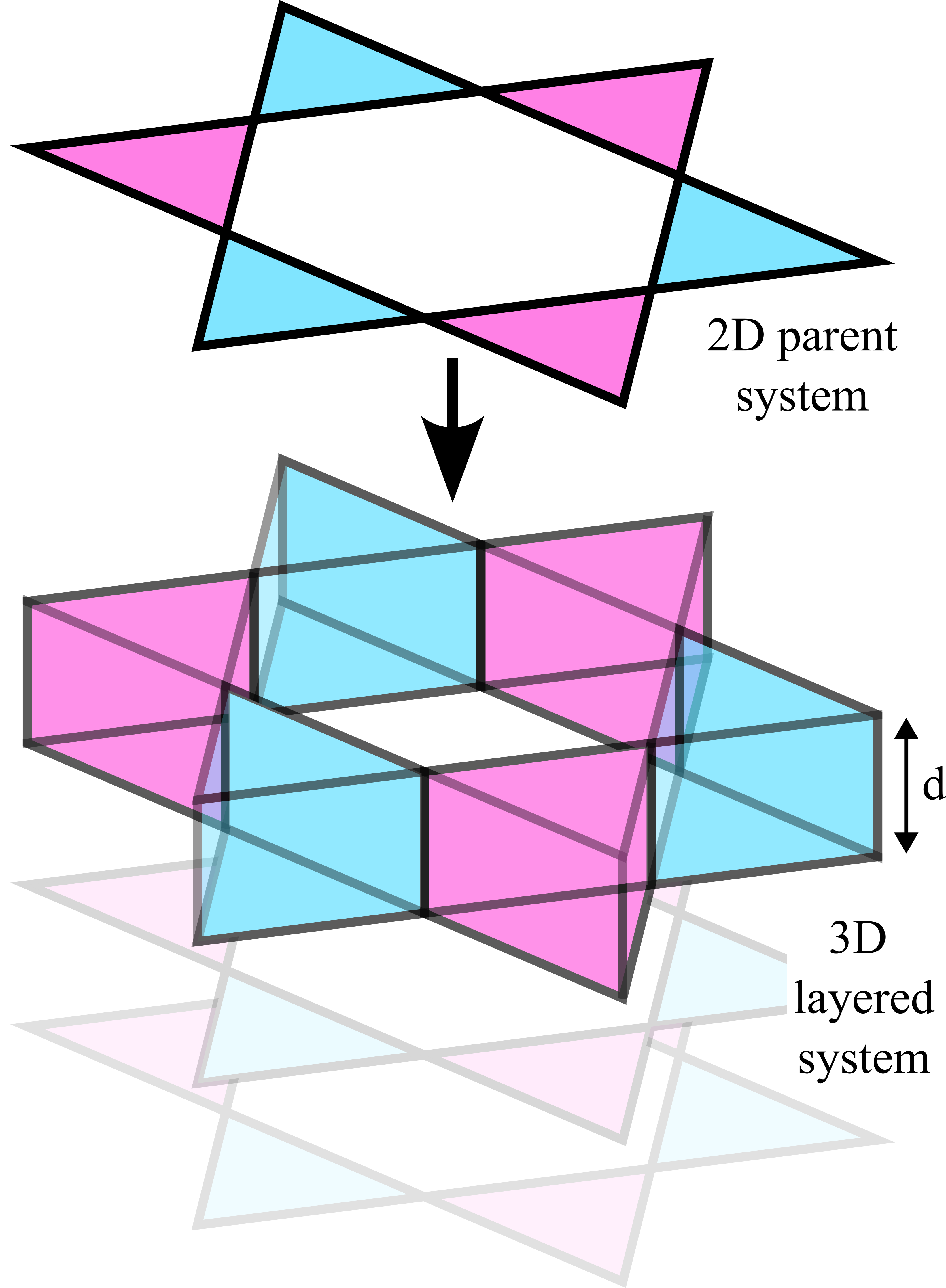}
    \caption{First recipe: Schematic representation of the simple approach to build a spin system hosting a pinch line, presented in section \ref{sec:simple}. First, consider a 2D parent model known to host pinch points; here the kagome antiferromagnet with two types of triangular clusters (pink and blue). Next, build a 3D stacking of this 2D parent model with consecutive layers separated by distance $d$. Finally extend the 2D clusters to 3D by linking together two clusters from two consecutive layers.}
    \label{fig: simple recipe}
\end{figure}

Let us start with a relatively straightforward example, and consider a 2D algebraic spin liquid with $n_s$ sublattices. This is our parent model with Hamiltonian (\ref{eq:hamcl}). In this simple approach, we stack \textit{identical} successive layers of this 2D parent model; e.g. all layers are a kagome lattice [Fig.~\ref{fig: simple recipe}]. Then we couple each cluster of spins with its neighboring cluster in the layer just above. This type of inter-layers interactions preserves the structure of the cluster Hamiltonian (\ref{eq:hamcl}), with a new constrainer defined on two successive layers as 
\begin{equation}
    \bm{\mathcal{C}}_{n, l,\alpha}^{3D} = \bm{\mathcal{C}}_{n, l,\alpha}^{2D} + \delta \, \bm{\mathcal{C}}_{n, l+1,\alpha}^{2D}
\end{equation}
where $\bm{\mathcal{C}}_{n, l,\alpha}^{2D}$ represents the constrainer containing the spins from the 2D cluster $n,\alpha$ among layer $l$, while 
$\delta$ is a real parameter governing the inter-layer interaction strength.  In this situation, since the number of sublattice is the same, the constraint vectors $\mathbf{L}^\alpha$ of the 3D stacked systems can be built directly from the ones $\mathbf{l}^\alpha(q_x, q_y)$ of the parent 2D system, as 
\begin{equation}
    \begin{split}
        \mathbf{L}^\alpha(\mathbf{q}) &= e^{i d q_z /2}\mathbf{l}^\alpha(\mathbf{q}_\perp) + \delta e^{-i d q_z /2} \mathbf{l}^\alpha(\mathbf{q}_\perp)   \\
        &= \left(e^{i d q_z /2} + \delta e^{-i d q_z /2} \right) \mathbf{l}^\alpha(\mathbf{q}_\perp)
    \end{split}
    \label{Eq: L trivial recipe}
\end{equation}
where we consider that planes are stacked along the $z$ direction, and $\mathbf{q}_\perp$ is a shorthand notation for $(q_x, q_y)$. This expression shows that the constraint vector associated with a type of cluster $\alpha$ will simply be renormalized by a function $e^{i d q_z /2} + \delta e^{-i d q_z /2}$ when considering the 3D stacked version of the 2D system. This implies that the structure of the vector space $V$ is left intact when going along the third direction of reciprocal space $q_z$. In such situation any pinch point of the 2D parent system is expected to be extended as a pinch line. As the constraint vectors are only renormalized by a common prefactor when going through such a construction, the low temperature structure factor that is built from the constraint vectors (see Appendix \ref{Appendix B: Definition of the structure factor}) is expected to be invariant along the third direction of reciprocal space. 

This construction allows to extend a 2D structure factor presenting pinch point into 3 dimensions, thus producing pinch lines as expected. It is, however, fairly simple as it does not allow much freedom within the construction of the system. Furthermore the pinch line appears here as a simple expansion of the original pinch point of the 2D parent model, without any possibility of evolution of the singularity along the pinch line; the structure factor in the transverse ($q_x,q_y$) plane looks identical as $q_z$ varies. This is why we propose now a more evolved recipe allowing to engineer pinch line systems with more freedom.

\subsection{Generic recipe}

Let us again consider a 2D algebraic spin liquid with $n_s$ sublattices. This is our parent model with Hamiltonian (\ref{eq:hamcl}) hosting a pinch point at wave vector $\mathbf{q}_\perp^* = (q_x^\star,q_y^\star)$. Now, let us stack successive layers of this 2D parent model as before, but this time inserting intermediate layers as depicted in Fig.~\ref{fig: stakked system}, with inter-layer distance $d$. We impose that these intermediate layers form a unique $(n_s+1)^{\rm th}$ sublattice; this condition is actually not necessary, but it simplifies the reasoning. In order to form a 3D structure, these intermediate layers need to interact with the parent layers above and below them. We add these interactions in Hamiltonian (\ref{eq:hamcl}) via new coefficients $\delta_i^\alpha$ for all clusters $n,\alpha$,
\begin{equation} 
\bm{\mathcal{C}}_{n,\alpha} = \sum_{i \in n \cap \alpha } \left( \gamma_i^\alpha \mathbf{S}_i^{(p)} + \delta_i^\alpha \mathbf{S}_i^{(l)} \right), 
\label{eq:cl3d} 
\end{equation}
where $(p)$ denotes spins from the 2D parent layer including all sublattices from 1 to $n_s$, and $(l)$ represents spins from sublattice $(n_s+1)$ in the intermediate layers located below and above the parent layer. Based on Eq.~(\ref{eq:Lu}), the constraint vectors $\mathbf{L}^\alpha(\mathbf{q})$ of the resulting 3D model have $(n_s+1)$ dimension,
\begin{equation} 
    \mathbf{L}^\alpha =
    \begin{pmatrix}
        \mathbf{l}^\alpha (\mathbf{q}_\perp) \\ 
        L_{n_s+1}^\alpha (\mathbf{q}_\perp, q_z)
    \end{pmatrix}, 
    \label{L 3D} 
\end{equation} 
with $\mathbf{l}^\alpha (\mathbf{q}_\perp)$ constraint vectors from the parent model, and 
\begin{equation}
    \begin{split}
        L_{n_s+1}^\alpha (\mathbf{q}) =\; &e^{i q_z d} \sum_{j \in D_{n_s+1}^{\alpha, a}} \delta_j e^{i \mathbf{q}_\perp \cdot \mathbf{r}_j} \\ 
        +\; &e^{-i q_z d} \sum_{j \in D_{n_s+1}^{\alpha, b}} \delta_j e^{i \mathbf{q}_\perp \cdot \mathbf{r}_j },
    \end{split}
\label{eq:Ln1}
\end{equation} 
where $D_{n_s+1}^{\alpha, a(b)}$ denotes the sites from sublattice $n_s+1$ belonging to the cluster of type $\alpha$ among intermediate layer located above (below) the cluster center. Now, as the 2D parent system hosts a pinch point at $\mathbf{q}_\perp^*$, this means there exists a critical vector $\mathbf{l}_c$ built as a linear combination 
\begin{equation}
    \mathbf{l}^c(\mathbf{q_\perp}) = \sum_\alpha c_\alpha \mathbf{l}^\alpha (\mathbf{q_\perp})
\end{equation}
of the constraint vectors $\mathbf{l}^\alpha$ such that $\mathbf{l}^c(\mathbf{q}_\perp^*) = 0$. Now one can build the 3D version of this critical vector using the same coefficients $c_\alpha$,
\begin{equation}
    \mathbf{L}_c(\mathbf{q}) = \sum_\alpha c_\alpha \mathbf{L}^\alpha (\mathbf{q}) = \begin{pmatrix}
\mathbf{l}^c (\mathbf{q}_\perp) \\ 
L_{n_s+1}^c (\mathbf{q}_\perp, q_z)
\end{pmatrix}.
\end{equation}
Since 2D critical vector $\mathbf{l}^c(\mathbf{q}_\perp)$ does not depend on $q_z$, the pinch points at $\mathbf{q}_\perp^\star$ and equivalent wave vectors extend into parallel pinch lines along $q_z$ if
\begin{equation}
\{L_{n_s +1}^c(\mathbf{q}_\perp^\star,q_z)=0, \forall q_z\}.
\label{eq:Lc0}
\end{equation}
In other words, the necessary condition to form a pinch line in our framework only depends on the positions of, and interactions ${\delta_i}$ with, the intermediate $(n_s+1)^{\rm th}$ sublattice:
\begin{equation} 
    \begin{split}
        \sum_\alpha c_\alpha \sum_{j \in D_{n_s +1}^{\alpha, a}} \delta_j e^{i \mathbf{q}^\star_\perp \cdot \mathbf{r}_j} = 0, \\ 
        \sum_\alpha c_\alpha \sum_{j \in D_{n_s +1}^{\alpha, b}} \delta_j e^{i \mathbf{q}^\star_\perp \cdot \mathbf{r}_j} = 0.
    \end{split}
\label{condition for pinch lines}
\end{equation} 
In the general case where the interlayers contain $m$ additional sublattices, there would be $m$ such conditions to fulfill in order to form a pinch line. Note that for systems with a single cluster type and then a single constraint vector, $\mathbf{l}^c = \mathbf{l}$ and $\mathbf{L}^c = \mathbf{L}$ and the sum over the coefficient $c_\alpha$ in the above expression can be ignored. Additionally, if the $(n_s+1)^{\rm th}$ sublattice is symmetric with respect to the cluster center, which is natural for most 3D structures, Eq.~(\ref{condition for pinch lines}) reduces to a single condition
\begin{equation} 
\sum_{j \in D_{n_s +1}^{\alpha}} \delta_j e^{i \mathbf{q}^\star_\perp \cdot \mathbf{r}_j} = 0,
\label{eq:simplcondition for pinch lines}
\end{equation} 
which will be the case for all concrete examples provided here.

\subsection{Multifold pinch lines} 

We have so far voluntarily remained generic in order to demonstrate the universal breadth of the method. In particular we imposed no condition on the nature of the singularity, and are free to consider multifold pinch points of higher-rank U(1) gauge theories. This freedom raises a follow-up question though: is the multifold symmetry of the parent pinch point preserved along the line ? Based on the discussion below Eq.~(\ref{eq:Lu}), this question can be straightforwardly recast in a mathematical form: do $L_{n+1}^c(\mathbf{q}_\perp,q_z)$ and $\mathbf{l}_c(\mathbf{q}_\perp)$ scale at the same lowest order in $\delta\mathbf{q}_\perp$ near the singularity, for all $q_z$ ? If yes, the multifold singularity is conserved. If not, one obtains a mixture of higher-rank gauge fields \cite{Davier_2023, Yan_2024_long, Lozano24a, lozanogomez_2024_preprint, chung_2024_preprint}. This condition is actually not as constraining as it might seem. Since the $q_z$ contribution of $L_{n+1}^c(\mathbf{q}_\perp,q_z)$ can been factorized out (see Eq.~(\ref{eq:simplcondition for pinch lines})), its Taylor expansion in $\delta\mathbf{q}_\perp$ is automatically the same for all $q_z$.

Now that the generic framework has been defined, we shall apply our approach to three concrete examples.\\

\section{How to design pinch-line models: Examples}
\label{sec:example}

\subsection{Kagome lattice } 
\label{sec:kagex}

\begin{figure}[b]
\centering
\includegraphics[width=\columnwidth]{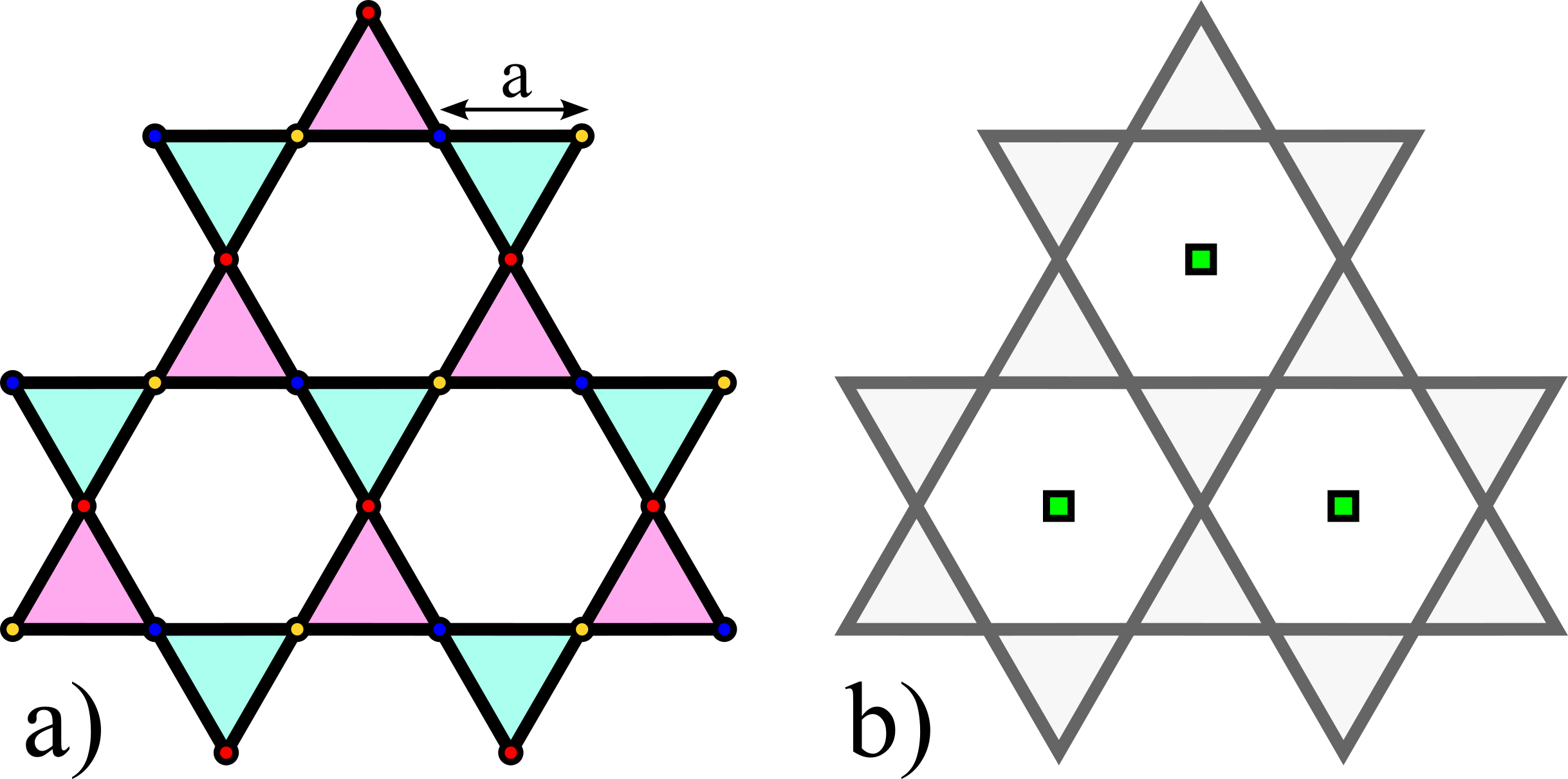}
\caption{Second recipe: (a) Kagome lattice composed of three sublattices, shown in different colors. The lattice can be described as a set of connected clusters, with up and down triangles highlighted in light blue and light magenta, respectively. (b) Projection of the 3D model: the fourth sublattice (green squares) is placed above and below each hexagon of the parent kagome lattice. 
}
\label{fig: 3D kagome lattice}
\end{figure}

Before considering more exotic forms of emergent gauge theories, we shall start with a simple, but nonetheless non-trivial, system that hosts standard twofold-symmetric pinch points. We consider the kagome antiferromagnet, which naturally lends itself to a cluster-based description, where up and down triangles form two distinct types of clusters (see Fig.~\ref{fig: 3D kagome lattice}(a)). In this situation the two types of constrainers can be expressed as
\begin{equation}
    \bm{\mathcal{C}}_{n, \triangledown} = \sum_{i \in \triangledown \cap n} \mathbf{S}_i, \qquad \bm{\mathcal{C}}_{n, \vartriangle} = \sum_{i \in \vartriangle \cap n} \mathbf{S}_i.
\end{equation}
This implies the existence of two distinct constraint vectors, $\mathbf{l}^\triangledown$ and $\mathbf{l}^\vartriangle$, which are complex conjugates $\mathbf{l}^\triangledown = \bar{\mathbf{l}}^\vartriangle$ as the two types of clusters are related by inversion symmetry\cite{Yan_2024_long, Davier_2023}. These two constraint vectors become real -- and therefore equal -- at the center of the Brillouin zone (BZ) and secondary BZs, where the critical vector\footnote{From a general point of view, the correct way to obtain the critical vector $\mathbf{l}^c$ if one wants to compute the associated Gauss law is to look for the eigenvector\cite{Yan_2024_long} of of the dispersive band that admits a contact point in $\mathbf{q}_\perp^*$. In the case of the kagome lattice, this leads to consider $\mathbf{l}^c(\mathbf{q}_\perp) = \mathbf{l}^\triangledown(\mathbf{q}_\perp) - e^{i \phi} \mathbf{l}^\vartriangle(\mathbf{q}_\perp)$ with $\phi(\mathbf{q}_\perp) = \text{Arg} \left( \mathbf{l}^\triangledown(\mathbf{q}_\perp) \cdot \mathbf{l}^\triangledown(\mathbf{q}_\perp) \right)$. }
\begin{equation}
    \mathbf{l}^c(\mathbf{q}_\perp^*) = \mathbf{l}^\triangledown(\mathbf{q}_\perp^*) - \mathbf{l}^\vartriangle(\mathbf{q}_\perp^*).
\end{equation}
thus becomes null. These contact points in the secondary BZs labeled by $\mathbf{q}_\perp^*$ are indeed known to be associated with pinch points in the structure factor. These pinch points can be observed over an intermediary temperature range \cite{Chalker1992, Zhitomirsky2002, Zhitomirsky2008} where the system behaves as an algebraic spin liquid. At lower temperatures, thermal order by disorder selects a submanifold of planar configurations with quartic fluctuations.

Now we shall build intermediate layers between each kagome layers, forming a fourth sublattice, such that condition (\ref{eq:Lc0}) is respected at $\mathbf{q}_\perp^*$. For simplicity, we consider a fourth sublattice whose sites preserve planar symmetry with respect to the kagome plane. This setup generally leads to constraint vectors that can be expressed as
\begin{equation}
    \begin{split}
        L^\vartriangle_4 (\mathbf{q}) = \cos (d q_z) f(\mathbf{q}_\perp), \\ 
        L^\triangledown_4 (\mathbf{q}) = \cos (d q_z) \bar{f}(\mathbf{q}_\perp),
    \end{split}
    \label{eq:kagf}
\end{equation}
with $f(\mathbf{q}_\perp)$ a complex function that encodes the in-plane geometry of the fourth sublattice. Because by definition of the constraint vectors (\ref{eq:Lu}) each component must satisfy $\bar{L}_n(\mathbf{q}) = L_n(-\mathbf{q})$, this function must share the same property $\bar{f}(\mathbf{q}_\perp) = f(-\mathbf{q}_\perp)$. This property imposes that $\mathbf{L}^\vartriangle(\mathbf{q}_\perp^*, q_z) = \mathbf{L}^\triangledown(\mathbf{q}_\perp^*, q_z)$ as $\mathbf{q}_\perp^*$ denotes here the positions of BZs centers. It means that
\begin{equation}
\mathbf{L}^c(\mathbf{q}_\perp^*, q_z) \equiv \mathbf{L}^\vartriangle(\mathbf{q}_\perp^*, q_z) - \mathbf{L}^\triangledown(\mathbf{q}_\perp^*, q_z) = 0,
\end{equation}
imposing the presence of a pinch line along $q_z$.

\begin{figure}[t]
\centering
\includegraphics[width=\linewidth]{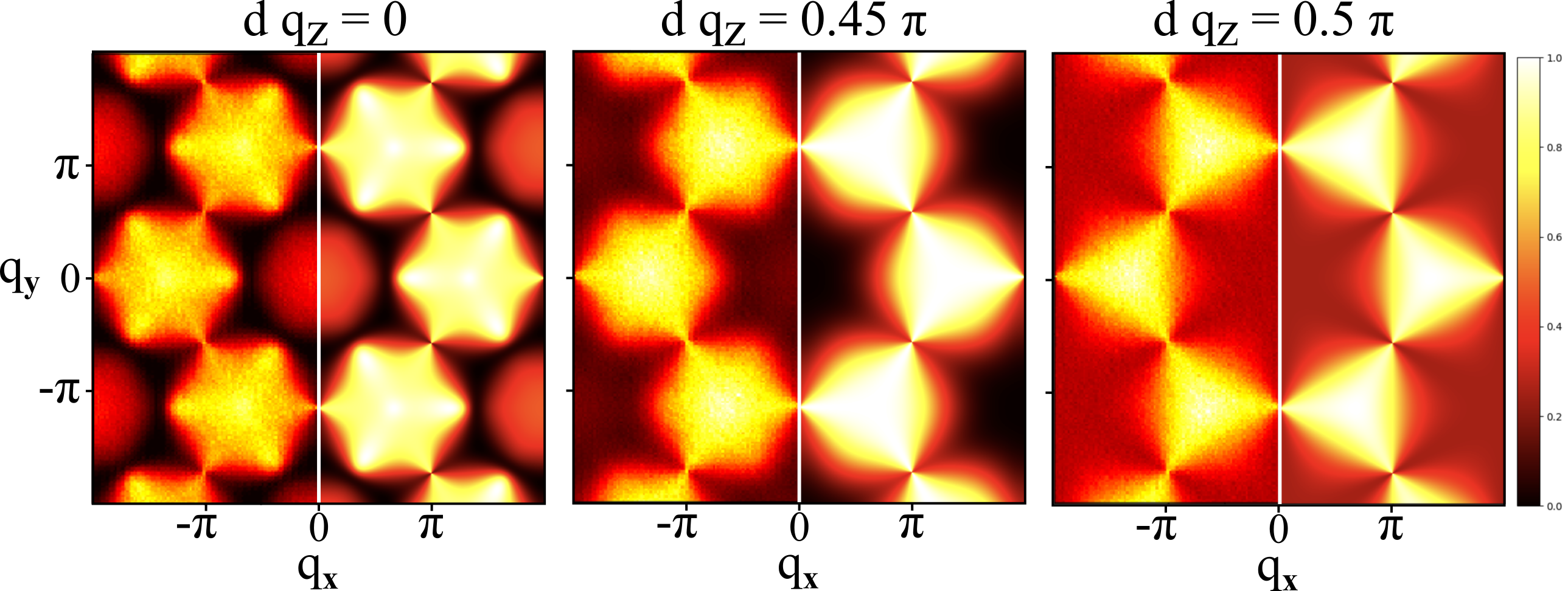}
\caption{Equal-time structure factor for the 3D generalization of kagome lattice presented in Fig.~\ref{fig: 3D kagome lattice}(b). The rows show the evolution in the $(q_x,q_y)$ plane orthogonal to the pinch lines (in $a^{-1}$ units). For each panel, the right side is obtained analytically using Henley's projective method at zero temperature \cite{Henley2005} [Appendix \ref{Appendix B: Definition of the structure factor}], while the left side results from Monte Carlo simulations at very low temperatures $T/J_\text{max} = 2\times 10^{-4}$ [Appendix \ref{Appendix C: Monte Carlo simulations}]. Keeping in mind the difference of temperature between the two methods, the agreement between simulations and theory is excellent; in particular the position and persistence of the pinch-point singularities as a function of $q_z$.}
\label{fig: 3D kagome latticeSQ}
\end{figure}

It is important to understand here that the above geometric conditions are fairly generic; finding the correct ratio between further neighbor exchanges might require some fine tuning, but many 3D geometries of stacked layers would satisfy the above conditions. For example one can build the intermediate layer by placing the fourth-sublattice sites above and below each hexagon of the kagome lattice, as illustrated in Fig.~\ref{fig: 3D kagome lattice}(b). Details on the site positions and interactions are given in Appendix \ref{Appendix A: Lattice definitions}. Each triangle of the kagome lattice is surrounded by three sites in the intermediate layers just above and three sites just below. Coupling these six sites with the kagome triangle in between forms a 3D cluster made of nine spins. Fig.~\ref{fig: 3D kagome latticeSQ} shows the evolution of the structure factor along $q_z$ for this model. Pinch points at the center of the secondary BZs persist for all $q_z$ even though the structure factor in the $(q_x,q_y)$ plane, and thus the spin-spin correlations, does vary. Our 3D model thus support pinch lines with a non-trivial evolution of the structure factor in the transverse plane.



\subsection{Counter-example}
\label{sec:counter}

A natural approach to generalizing the kagome lattice while preserving minimal cluster size—and hence minimizing the number of interaction links—is to place new sites in intermediate planes directly above the centers of the existing triangular clusters. Since the kagome lattice features two types of clusters (up and down triangles), this extension results in two additional sublattices. These new sites are positioned at the vertices of a virtual hexagonal lattice.

We now consider 3D clusters formed by combining each parent triangle with the two additional sites located above and below its center. This leads to constraint vectors of the form
\begin{equation}
    \mathbf{L}^\vartriangle = \begin{pmatrix}
        \mathbf{l}^\vartriangle(\mathbf{q_\perp}) \\ 2 \delta \cos (d q_z) \\ 0
    \end{pmatrix}, 
    \qquad 
    \mathbf{L}^\triangledown = \begin{pmatrix}
        \mathbf{l}^\triangledown(\mathbf{q_\perp}) \\ 0 \\  2\delta \cos (d q_z) 
    \end{pmatrix},
\end{equation}
which are to be associated with the critical vector 
\begin{equation}
    \mathbf{L}^c = \mathbf{L}^\vartriangle - \mathbf{L}^\triangledown = 
    \begin{pmatrix}
        \mathbf{l}^c(\mathbf{q_\perp}) \\ 2 \delta \cos (d q_z) \\ -2\delta \cos (d q_z)
    \end{pmatrix}.
\end{equation}
This critical vector is clearly non zero along the line $\mathbf{q}_\perp = 0$ except for $d q_z = \frac{\pi}{2}$, implying that this 3D construction does not support any pinch line.

This example highlights a counter-example of the pinch-line recipe: adding a pair of sites directly above and below the center of a 2D cluster tends to generate extra components in the constraint vectors that are independent of $\mathbf{q}_\perp$. As a result, there is no tunable in-plane dependence of $L_{n_s+1}^c(\mathbf{q})$ to exploit in order to satisfy the pinch-line condition~(\ref{condition for pinch lines}).\\

We now turn to the study of two other models, looking to build systems hosting exotic pinch lines along which high rank pinch points could be found.

\begin{figure*}[ht]
\centering \includegraphics[width=1.0\textwidth]{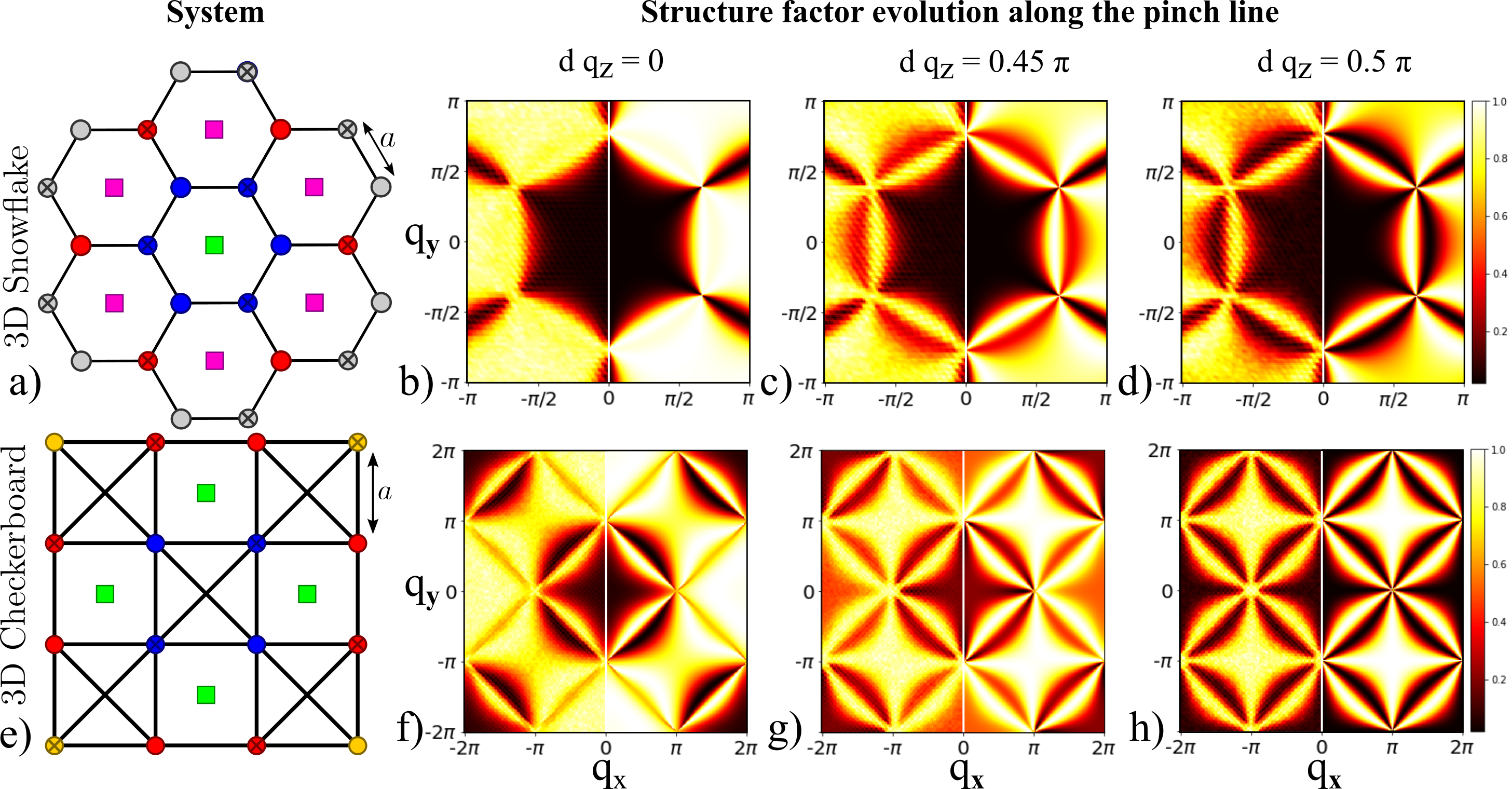}
\caption{The left panels depict the local cluster structure of the two systems studied here. The plain and crossed circles depict the two sublattices of the 2D parent systems. Blue, red and yellow circles respectively appear with coefficient 1, $\gamma_1$ ($\gamma$ for the snowflake) and $\gamma_2$ in the constrainer \bm{\mathcal{C}}. The squares represent the third sublattice located in intermediate planes; locally, they form square (top) and hexagonal (bottom) bipyramids with the 2D parent lattice. The green and pink squares appear with coefficient $\delta_1$ ($\delta$ for the checkerboard) and $\delta_2$ in the constrainer $\bm{\mathcal{C}}$.
The three right columns show the evolution of the equal-time structure factor in $(q_x,q_y)$ planes orthogonal to the pinch lines (in $a^{-1}$ units), with $\gamma = 1/2, \delta_1 = 1$ for the 3D snowflake and $\gamma_1 = 1, \gamma_2 = -1/3$ and $\delta = 1$ for the 3D checkerboard model. For each panel, the right side is obtained analytically using Henley's projective method {at zero temperature} \cite{Henley2005} [Appendix B], while the left side results from Monte Carlo simulations {at very low temperatures} [Appendix C]. {Keeping in mind the difference of temperature between the two methods,} the agreement between simulations and theory is excellent, {in particular the position and persistence of the pinch-point singularities as a function of $q_z$}.
}
\label{fig:Sq}
\end{figure*}

\subsection{3D snowflake honeycomb model}

The 2D snowflake honeycomb model \cite{Benton_Moessner_2021, Yan_2024_long} consists of Hamiltonian (\ref{eq:hamcl}) with a unique type of hexagonal spin clusters defined in Appendix A and illustrated in Fig.~\ref{fig:Sq}(a), and associated to the constrainer
\begin{equation} 
\bm{\mathcal{C}}_{\hexagon} = \sum_{i\in \textcolor{blue}{\hexagon}} \mathbf{S}_i + \gamma \sum_{i\in \textcolor{red}{\langle\hexagon\rangle}} \mathbf{S}_i.
\label{eq:Csnow}
\end{equation} 
Note that for systems with a single cluster type the sum over the unit cells amounts to a sum over clusters, and therefore the unit cell index $n$ can be omitted in Eq.~(\ref{eq:hamcl}). 
The ground state of this 2D parent model is known to support an algebraic spin liquid for any $\gamma>0$  \cite{Benton_Moessner_2021}. For $\gamma = 1/2$, this spin liquid evolves into a rank$-2$ gauge theory with fourfold pinch points at $K$ points in the BZ.

There are multiple ways to extend this singularity into a line. We construct a 3D lattice by inserting intermediate triangular layers (see the square sites of Fig.~\ref{fig:Sq}(a)). Within each cluster, inter-layer interactions take the form of two additional terms in $\bm{\mathcal{C}}_{\hexagon}$ of Eq.~(\ref{eq:Csnow}): sites above and below the respectively central (green) and surrounding (pink) hexagons appear with coefficient $\delta_1$ and $\delta_2$. The third component of the unique constraint vector then becomes
\begin{equation} 
L_3(\mathbf{q}) = 2 \delta_1 \cos(d q_z) \left[ 1 + \frac{\delta_2}{\delta_1} \sum_{i=1}^6 \cos(\mathbf{r}_i \cdot \mathbf{q}_\perp) \right],
\label{eq:Lsnow}    
\end{equation} 
where the sum runs over the six pink sites of the cluster [Fig.~\ref{fig:Sq}(a)]. Since the fourfold pinch points of the parent model appear at $\mathbf{q}^K_\perp=(0,\frac{4\pi}{3\sqrt{3}a},0)$ (and equivalent K points), we get $\{L_3^c(\mathbf{q}^K_\perp,q_z) = L_3(\mathbf{q}^K_\perp,q_z) = 0,\forall q_z\}$ under the condition
\begin{equation}
1 + \frac{\delta_2}{\delta_1} \sum_{i=1}^6 \cos(\mathbf{r}_i \cdot \mathbf{q}^K_\perp) = 0
\quad \Leftrightarrow \quad \delta_1=3 \delta_2.
\label{cond snowflake} 
\end{equation} 
Furthermore, Taylor expansion of Eq.~(\ref{eq:Lsnow}) gives
\begin{equation} 
L_3\left(\mathbf{q}_\perp^K + \delta\mathbf{q}_\perp, q_z\right) = 2\delta_2 \cos(d q_z)\frac{3}{4}( \delta q_x^2+ \delta q_y^2) 
\label{eq:L3snow}
\end{equation} 
up to $\mathcal{O}(\delta q_i^3)$ terms, which is of second order in $\delta\mathbf{q}_\perp$ as, by definition, for the rank$-2$ parent model. Our model thus supports fourfold pinch lines at $(\mathbf{q}^K_\perp,\forall q_z)$ for $\gamma = 1/2$ and $\delta_1 = 3 \delta_2$, with a degree of freedom on the ratio $\gamma/\delta_1$.

The tensor electric field $\tilde{E}_{2D}$ of our parent model is traceless, as derived in Ref.~[\onlinecite{Yan_2024_long}]. Injecting Eq.~(\ref{eq:L3snow}) into Eq.~(\ref{eq:Lu}), it can be generalized to our pinch-line spin liquid as
\begin{equation}
\tilde{E}_{3D} = \tilde{E}_{2D} \,+\, \eta \tilde{S}_3\,\mathrm{Id}_2,
\label{eq:Esnow}
\end{equation}
where $\eta \propto \delta_2 \cos(d q_z)$ and $\tilde{S}_3$ is the Fourier transform of the spin configuration on sublattice $u=3$. This tensor undergoes the same type of second order Gauss laws 
\begin{equation}
    \partial_i\partial_j E_{3D}^{ij} = 0, \qquad i,j = x,y,
\end{equation}
than for the parent model. The major difference is that the electric field now depends continuously on $\eta \propto \cos(dq_z)$, meaning there are in fact a infinite number of such Gauss laws. This imposes that there are in fact two independent electric fields emerging form the spins degrees of freedom, one relying on parent layers spins and the other on interlayers ones. This point is discussed in details in Sec.~\ref{sec : What do pinch lines look like in real space ?}.

As these two electric fields are rank-2, the pinch points observed along the pinch line are expected to be a mixing of fourfold pinch points. However, the actual shape of the fourfold singularity in the structure factor $\mathcal{S}(\mathbf{q})$ is subtle. Four-fold pinch points are properly defined in the correlators of any rank-2 tensor electric field $\tilde{E}_{ij}$ [\onlinecite{Prem_2018}] and appear in some of them (e.g.~$\langle \tilde{E}_{xx}(\mathbf{q}_\perp) \tilde{E}_{yy}(-\mathbf{q}_\perp) \rangle$) but not all of them (e.g.~not $\langle \tilde{E}_{xx}(\mathbf{q}_\perp) \tilde{E}_{xx}(-\mathbf{q}_\perp) \rangle$). The subtlety is that $\mathcal{S}(\mathbf{q})$ is the Fourier transform of \textit{all} spin-spin correlations, and thus receives contributions from all correlators a priori \cite{Henley2005}. This is why multifold pinch points can be somewhat hidden in $\mathcal{S}(\mathbf{q})$. Performing a Taylor expansion in the $\mathbf{q}_\perp$ plane around the pinch line, $\mathcal{S}(\mathbf{q})$ adopts the form
\begin{equation} 
\mathcal{S}(\mathbf{q}) \propto \alpha + \beta \frac{q_x^2q_y^2}{(q_x^2+q_y^2)^2} + \chi \cos(dq_z) \frac{q_x^4-q_y^4}{(q_x^2+q_y^2)^2},
\label{eq:Sq24fold}
\end{equation} 
where $\alpha$, $\beta$, and $\chi$ are real parameters. This expression is the sum of two terms with distinct symmetries: the term proportional to $\beta$ exhibits fourfold symmetry, while the term proportional to $\chi$ only has twofold symmetry {because it can be rewritten as
\begin{equation}
    \frac{q_x^4-q_y^4}{(q_x^2+q_y^2)^2}\equiv 2 \frac{q_1q_2(q_1^2+q_2^2)}{(q_1^2+q_2^2)^2} = 2 \frac{q_1q_2}{q_1^2+q_2^2},
\end{equation}
after a local $\pi/4$ rotation in the Fourier plane, $q_x = q_1+q_2$, $q_y = q_1-q_2$}. Both terms are allowed in phases governed by rank-2 divergenceless tensors $E_{ij}$ [\onlinecite{Prem_2018}]. As a consequence, and even though we have a rank$-2$ gauge theory with a fourfold singularity propagating along the pinch line in the appropriate correlator sectors, the pinch points in the structure factor $\mathcal{S}(\mathbf{q})$ continuously evolve from twofold to fourfold symmetry {because the former vanishes as a function of $\cos(dq_z)$ [Eq.~(\ref{eq:Sq24fold})]}, as confirmed in Figs.~\ref{fig:Sq}(b-d) both analytically and numerically.\\

\subsection{3D generalized checkerboard model} 
\label{subsec: 3D ckb}

The 2D generalized checkerboard model is known to support a diversity of algebraic spin liquids \cite{Davier_2023}. This diversity offers several pinch points at distinct wave vectors that can be considered as potential pinch lines within our framework. The Hamiltonian of this 2D parent model follows Eq.~(\ref{eq:hamcl}) with unique type of square spin clusters defined in Appendix A and illustrated in Fig.~\ref{fig:Sq}(e), associated with constrainer
\begin{equation}
\bm{\mathcal{C}}_{\square} = \sum_{i\in\textcolor{blue}{\square}} \mathbf{S}_i + \gamma_1\sum_{i\in\textcolor{red}{\langle\square\rangle}} \mathbf{S}_i + \gamma_2 \sum_{i\in\textcolor{Goldenrod}{\langle\langle\square\rangle\rangle}} \mathbf{S}_i.
\label{Cckb}
\end{equation}
We have again several options to build the 3D extension. We choose to place the third sublattice above (and below) each empty plaquette of the checkerboard lattice (see the squares in Fig.~\ref{fig:Sq}(e)), whose sites appear in $\bm{\mathcal{C}}_{\square}$ with a unique coefficient $\delta$. In that case, there is a unique constraint vector and Eq.~(\ref{eq:Ln1}) becomes
\begin{equation}
L_3(\mathbf{q}) = 4\delta \cos(d q_z) \left[ \cos(a q_x)+\cos(a q_y) \right].
\label{eq:L3check}
\end{equation}
The necessary condition to observe a pinch line, $L_3^c(\mathbf{q}^\star_\perp, q_z) = L_3(\mathbf{q}^\star_\perp, q_z)=0$, is satisfied for all $q_y = \pm q_x \pm \pi/a$, i.e. for any point on the BZ boundary. Hence, any pinch point of the 2D parent model located on the BZ edge extends into a pinch line in this 3D layered structure. According to Ref.~[\onlinecite{Davier_2023}], this offers several pinch points from which to choose from. For the sake of originality, we analyzed the Taylor expansion of $L_3(\mathbf{q})$ along the BZ edge; we found that the first-order term vanishes at $M$ points (BZ corners), giving rise to a diagonal traceless rank$-2$ electric field 
\begin{equation}
    \tilde{E}^{3D}_{xx}=-\tilde{E}^{3D}_{yy} = 6 \delta \cos(q_z d) \tilde{S}_3,
    \label{Eq: E il checkerboard}
\end{equation}
with $\tilde{S}_3$ the Fourier transform of the coarse grained spin field associated with third sublattice.
Remarkably\cite{Davier_2023}, for $\gamma_2 = (1-2\gamma_1)/3$, the structure factor of the 2D parent model presents sixfold singularities at $M$ points, attached to a rank$-3$ field $E^{2D}$ satisfying 
\begin{equation}
    \partial_i \partial_j \partial_k E^{2D}_{ijk} = 0, \qquad i,j,k = x,y,
\end{equation}
and relying on the two first sublattices coarse grained spin variables. Hence, we expect in our 3D model an exotic pinch line mixing rank$-2$ and rank$-3$ Gauss laws 
\begin{equation}
    \partial_i\partial_jE_{ij}^{3D} + \partial_i \partial_j \partial_k E^{2D}_{ijk} = 0, \qquad i,j,k = x,y,
\end{equation}
in any plane orthogonal to the $z$ direction. This is confirmed in the structure factor of Figs.~\ref{fig:Sq}(f-h) both analytically and numerically. As for the previous example, as these Gauss laws are valid for any value of $q_z$ along the line, and because $\tilde{E}^{3D}$ depends explicitly on $q_z$, this impose that the two fields $E^{3D}$ and $E^{2D}$ each fulfill a Gauss law. As these fields possess different structure, this means the system will possess distinct types of fracton excitations\cite{Pretko_2_2017,Yan_2024_long}. This point is discussed in more details in Sec.~\ref{sec : What do pinch lines look like in real space ?}.

Since the rank$-2$ term disappears in the structure factor when $q_z=\pi/d$ (mod. $\pi$), it explains the perfect sixfold symmetric observed in Fig.~\ref{fig:Sq}(h), as it is the signature of the rank-3 field $E^{2D}$ alone. Note that according to Refs.~\cite{Pretko_2_2017,Yan_2024_long}, the rank$-3$ gauge theory implies the conservation of the charge-, dipole- and quadrupole moments, further restricting the movements of {potential} fracton excitations.\\

\section{Enhanced zero modes}

At the exception of the first trivial example of Fig.~\ref{fig: simple recipe}, all models presented so far have introduced an additional sublattice (in the intermediate layer) while keeping the effective number of constraints unchanged ($\bm{\mathcal{C}}_\alpha = 0$). We will now demonstrate that it leads to an enhancement of zero modes in the ground state, by computing the effective number $F$ of such zero modes per cluster \cite{Moessner_1998_counting, Rehn_Moessner_2016, AlbaPujol2018, AlbaRosales2021}.\\

In order to compute $F$, we first need to define the effective number $m$ of spins belonging to each cluster. If all spins are equivalent, or at least if they all belong to the same number $b$ of clusters, this is straightforward: $m=q/b$ where $q$ is the number of sites per cluster \cite{Moessner_1998_counting}. But this formula doesn't work if some spins belong to more clusters than others. This is why the general formula is $m=N/N_c$, i.e. the total number of spins $N$ divided by the total number of clusters $N_c$ in the system.

\begin{figure}
    \centering
    \includegraphics[width=0.9\linewidth]{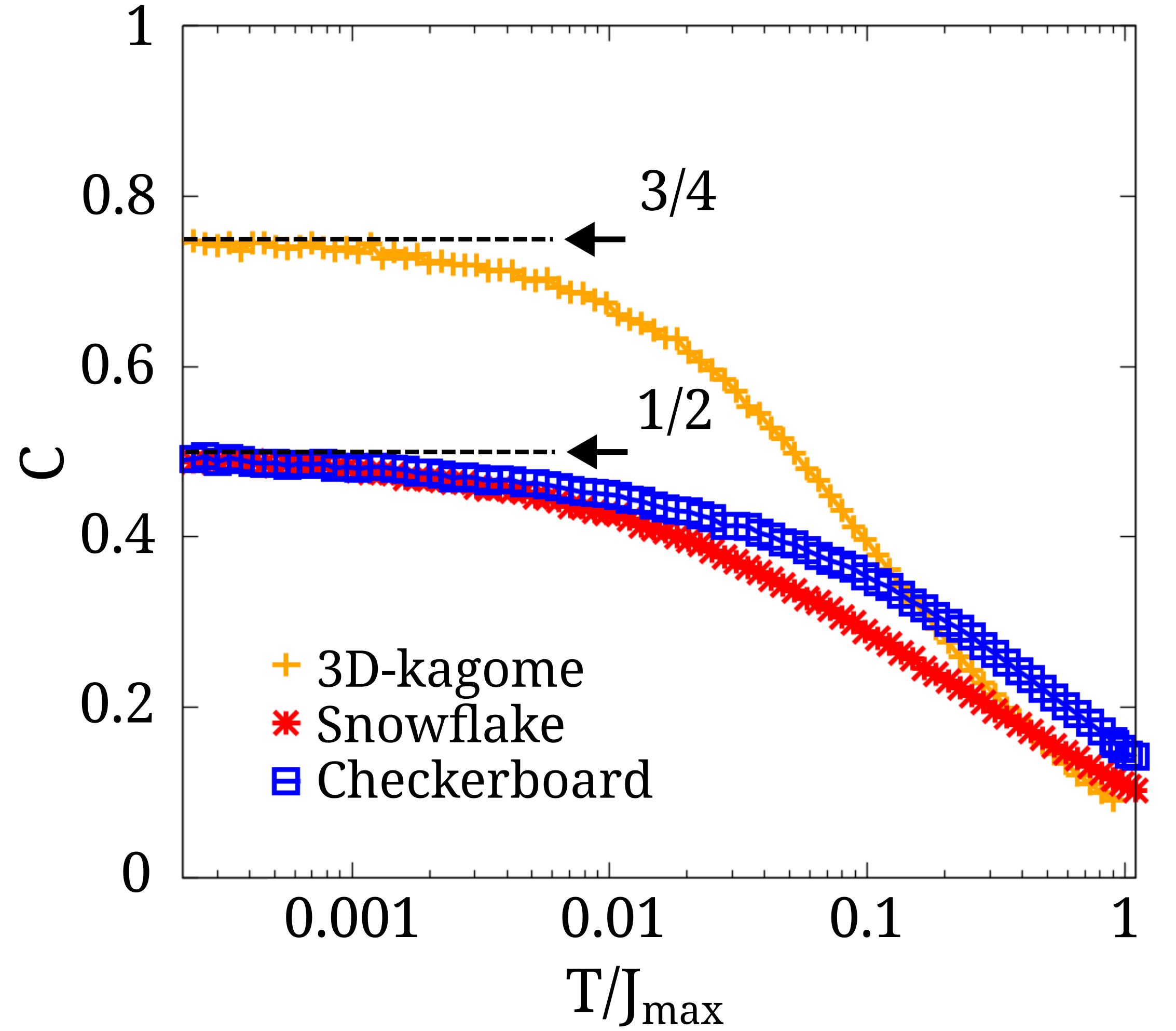}
    \caption{Specific heat per spin $C_h$ as a function of temperature, computed through classical Monte Carlo simulations for all 3D lattices. The black arrow indicates that in all cases, the specific heat reaches the value expected from the zero-mode counting argument of Eq.~(\ref{Eq: Analytical C expectation}) at the lowest temperature ($T/J_{max}=2\times10^{-4}$). These results have been obtained for a system composed of $N=3\times 18\times 18\times 6$ sites for kagome model, and $N=3\times 15\times 15\times 10$ sites for honeycomb and checkerboard models.}
    \label{fig:specific_heat}
\end{figure}

Each spin possesses $n-1$ degrees of freedom, where $n$ is the spin dimension (here $n=3$ for classical Heisenberg spins) and the $-1$ subtraction comes from the spin length constraint, $|\mathbf{S}_i|=1$. Hence, having an average of $m$ spins per clusters, with $n-1$ degrees of freedom per spin, and $n$ constraints per cluster, coming from the $n-$dimensional constrainer $\bm{\mathcal{C}}_\alpha = 0$, the number of zero modes per cluster is
\begin{equation}
    F = m(n-1) - n.
\end{equation}
Since (i) $n=3$ is fixed in our models and (ii) our pinch-line construction amounts to increasing the effective number $m$ of spins per cluster, our construction automatically increases the number of zero modes.\\

But how to measure this increase of zero modes ? As temperature $T\rightarrow 0^+$, it is possible to relate these zero modes to the value of the specific heat \cite{Chalker1992}. For pedagogical reasons, let us start the reasoning with a standard ferromagnet. Each spin fluctuates in the plane transverse to the magnetization. As $T\rightarrow 0^+$, these are small quadratic fluctuations. Hence, there are $2N$ quadratic modes of excitations, and the equipartition theorem gives an energy of $2N(\frac{1}{2}k_B T)$, thus a specific heat per spin $C_h=1$ in units of the Boltzmann constant $k_B$. This is true for all systems whose ground states only possess quadratic modes.

However, being \textit{infinitely} soft modes, zero modes contribute neither to the energy nor the specific heat. In other words, only the $n$ constrained modes per cluster contribute to the specific heat. Assuming these modes to be quadratic, this leads to a specific heat per spin
\begin{equation}
    C_h = \frac{N_c}{N}\frac{n}{2} = \frac{n}{2 m}.
    \label{Eq: Analytical C expectation}
\end{equation}
Our pinch-line construction thus lowers the value of the specific heat as $T\rightarrow 0^+$. The explicit values for the number of zero modes and specific heat per spin of the models considered here are given in the following table.
\begin{center}
    \begin{tabular}{c | c  c  c | c  c  c }
        & & \, $2D$ & & & \, $3D$ & \\
    \hline
        Model \, & \, $m$  & \, $F$ & \, $C_h$ & \, $m$ & \, $F$ & \, $C_h$ \\
    \hline
        Kagome & 3/2 & 0 & 1 \footnote{This calculation does not account for weather-wane modes on kagome responsible for quartic modes; as $T\rightarrow 0^+$, $C_h\rightarrow 11/12$ \cite{Chalker1992}.} & 2 & 1 & $3/4$ \\
        Snowflake & 2 & 1 & 3/4 & 3 & 2 & 1/2 \\
        Checkerboard & 2 & 1 & 3/4 & 3 & 2 & 1/2 \\

    \end{tabular}
\end{center}
These theoretical values are in excellent agreement with the low temperature limit of the specific heat obtained with Monte Carlo simulations, comparing the last column of the table with Fig.~\ref{fig:specific_heat}. This thus confirms the existence of an extensive quantity of zero modes, consistent with an extensively vast ground state manifold \cite{Chalker1992, Zhitomirsky2008}.

 Within this context, the spin liquid recipe can be seen as a way to extend the two dimensional lattice in a fashion that leaves the number of constraints, and in fact also the spacial structure of the constraints, unchanged while adding more degrees of freedom to fulfill these constraints.

\section{What do pinch lines look like in real space ?}
\label{sec : What do pinch lines look like in real space ?}

A pinch line in reciprocal space implies the existence of pinch points in any plane orthogonal to the pinch line, and no pinch points in planes containing the pinch line. In other words, in real space, we have algebraic correlations within parent and intermediate layers, associated with two dimensional Gauss laws, and short range correlations between spins belonging to different planes.

This raises a subtle question: when the structure factor changes along $q_z$, what kind of Coulomb phase lives inside the layers ? To be more specific, a unique Coulomb phase with a unique effective electric field gives rise to a unique shape of pinch points, which should be invariant along $q_z$ as found in our trivial model Fig.~\ref{fig: simple recipe}. However, since the pinch points shape of Figs.~\ref{fig: 3D kagome latticeSQ} and \ref{fig:Sq} change along $q_z$, it means we must have the co-existence of at least two distinct electric fields with different weights as one moves along $q_z$. Is it possible to build a generic, real-space, description of these electric fields within our layered framework ? This qualitative view can be supported for any pinch line system by looking at the expression of the effective electric field that can be built from the critical vector $\mathbf{L}^c$. Let us start by analyzing the case of a pinch line hosting only regular twofold pinch points, as the 3D kagome model presented above, that is easier to illustrate, before to present the general case of rank-$n$ and multirank pinch lines.

\subsection{Two fold pinch lines underlying physics}
\label{subsec: Two fold pinch lines underlying physics}

\subsubsection{Generic description for two fold pinch lines}
\label{subsubsec: Generic description for two fold pinch lines}

For a twofold pinch point, the coarse grained constraint $\bm{\mathcal{C}}(\mathbf{r}_c) = 0$ can be expanded around each cluster center $\mathbf{r}_c$ to obtain an explicit expression for the effective electric field\cite{Davier_2023}. For a pinch line, such expansion can be made at any point $\mathbf{q}^\star$ along the pinch line, leading to
\begin{equation}
        \mathbf{E}_{\mathbf{q}^\star}^j(\mathbf{r}) = -i \sum_u \bm{\nabla}_\mathbf{q} \left. L_u^c \right|_\mathbf{q^\star}  \chi_u^j (\mathbf{r})
        \label{Eq : Electric vector q^*}
\end{equation}
where $\chi_u^j(\mathbf{r}) \equiv S_u^j (\mathbf{r})e^{-i \mathbf{q}^\star \cdot \mathbf{r}}$ denotes a vector field encoding the fluctuations of the coarse grained spins $S_u^j (\mathbf{r})$ around the contact-point $\mathbf{q}^\star$ configurations. The coarse grained spin $S_u^j (\mathbf{r})$ is obtained from the $j^\text{th}$ spin component and is such that $S_u^j (\mathbf{r}_{u,i}) = S_{u,i}^j$ for a spin of sublattice $u$ located at position $\mathbf{r}_{u,i}$ in unit cell $i$. The existence of a pinch line along a direction given by the unit vector $\mathbf{e}_t$ imposes by definition $\left( \bm{\nabla}_\mathbf{q} \left. L_{u}^c \right|_\mathbf{q^\star}\right) \cdot \mathbf{e}_t = 0$ for any $u$, which imply that $\mathbf{E}^j\cdot \mathbf{e}_t = 0$. This means that for any pinch line the associated effective electric field cannot have any component along the pinch line direction.\\

The case of layered pinch line systems obtained with the two recipes presented in this work give good examples of systems where these considerations can be made explicit. 
For the first trivial recipe of section \ref{sec:simple}, the 3D electric field can be simply expressed directly from the critical vector (\ref{Eq: L trivial recipe}) as 
\begin{equation}
    \mathbf{E} = \left(e^{i d q_z /2} + \delta e^{-i d q_z /2} \right) \mathbf{E}_p
    \label{Eq: E trivial recipe}
\end{equation}
where $\mathbf{E}_p$ is the electric field associated with the 2D parent layer. The fields obtained when varying $q_z$ along the pinch line are thus simply renormalization of the parent electric field, satisfying the exact same unique Gauss law $\bm{\nabla}\cdot \mathbf{E}_p = 0$, and then simply replicating the 2D Coulomb phase of the parent system in successive layers.\\

For the second recipe with intermediate layers, exemplified by section \ref{sec:kagex}, the 3D electric field can be decomposed as 
\begin{equation}
    \begin{split}
        \mathbf{E}^j(\mathbf{r}) &= -i\sum_{u = 1}^{n_s} \bm{\nabla}_{\mathbf{q}_\perp} \left. l_u^c \right|_\mathbf{q^\star}  \chi_u^j (\mathbf{r}) \\
        &\; - i\bm{\nabla}_{\mathbf{q}_\perp} \left. L_{n_s+1}^c \right|_\mathbf{q^\star}  \chi_{n_s+1}^j (\mathbf{r}) \\
        &\; - i\bm{\nabla}_{q_z} \left. L_{n_s+1}^c \right|_\mathbf{q^\star}  \chi_{n_s+1}^j (\mathbf{r}) \\
        &= \mathbf{E}^j_\text{p}(\mathbf{r}) 
        + \mathbf{E}^j_{\text{il}, \perp}(\mathbf{r}) 
        + \mathbf{E}^j_{\text{il}, z}(\mathbf{r})
        \label{Eq : Electric filed components}
    \end{split}
\end{equation}
where $\mathbf{E}^j_\text{p}$ is the 2D electric field associated with the parent system, and $il$ stands for "inter-layer". The last field contribution is trivial $\mathbf{E}^j_{\text{il}, z} = 0$ because the presence of a pinch line imposes $\bm{\nabla}_{q_z} \left. L_{n_s+1}^c \right|_\mathbf{q^\star} = 0$. The interlayer sites contribution is then encapsulated in $\mathbf{E}^j_{\text{il}, \perp}$, that can be expressed as
\begin{equation}
    \begin{split}
        \mathbf{E}^j_{\text{il}, \perp}(\mathbf{r}) &= \sum_\alpha c_\alpha \left( e^{i d q_z} \sum_{j \in D_{n+1}^{\alpha, a}} \delta_j e^{i \mathbf{q}_\perp \cdot \mathbf{r}_\perp^j} \mathbf{r}_\perp^j \right. \\
        & \;+ \left.
         e^{- i d q_z} \sum_{j \in D_{n+1}^{\alpha, b}} \delta_j e^{i \mathbf{q}_\perp \cdot \mathbf{r}_\perp^j} \mathbf{r}_\perp^j \right) \chi_{n_s+1}^j (\mathbf{r}).
    \end{split}
\end{equation}
For systems with interlayers planes symmetric relative to parent planes, and with a single constraint vector, this expression reduces to 
\begin{equation}
    \begin{split}
    \mathbf{E}^j_{\text{il}, \perp} (\mathbf{r})
    &= 2\cos(d q_z) \left(\sum_{j \in D_{n+1}^{\alpha}} \delta_j e^{i \mathbf{q}_\perp \cdot \mathbf{r}_j} \mathbf{r}_\perp^j \right) \chi_{n_s+1}^j (\mathbf{r}).\\
    &= \cos(d q_z)\, \tilde{\mathbf{E}}^j_{\text{il}, \perp} (\mathbf{r}),
    \label{Eq : E interlayer sites}
    \end{split}
\end{equation}
which rewrites the total electric field as 
\begin{equation}
    \mathbf{E}^j =  \mathbf{E}^j_\text{p} + \cos(d q_z) \tilde{\mathbf{E}}^j_{\text{il}, \perp}
    \label{Eq: explicit E for second recipe applied to rank 1 PL}
\end{equation}
where $\tilde{\mathbf{E}}^j_{\text{il}, \perp}$ depends only on interlayers spins (sublattice $n_s +1$), and does not depend on $q_z$. Since the Gauss law
\begin{equation}
    \bm{\nabla}\cdot\mathbf{E}^j = 
    \bm{\nabla}\cdot \mathbf{E}^j_\text{p} + \cos(d q_z)  \bm{\nabla}\cdot \tilde{\mathbf{E}}^j_{\text{il}, \perp} = 0
\end{equation}
must be satisfied for all values of $q_z$, it imposes that each electric field respects its own Gauss law
\begin{equation}
    \bm{\nabla}\cdot \mathbf{E}^j_\text{p} = 0 
    \quad\&\quad
    \bm{\nabla}\cdot \tilde{\mathbf{E}}^j_{\text{il}, \perp} = 0.
\end{equation}
%
Hence, such systems with an intermediate layer between parent algebraic spin liquids, as exemplified in  section \ref{sec:kagex}, host a superposition of two distinct Coulomb phases, which explains the evolution of the pinch point along $q_z$.

This picture is similar to the one associated with a system hosting a multiple band touching point in its band structure. For such systems there is indeed a distinct Gauss law associated with each band touching, that are all located in the same point in reciprocal space, in a very similar way to the case of a pinch line where the 2D Gauss laws are all obtained in the vicinity of a unique point $\mathbf{q}_\perp^\star$. For 3D systems however, the physics is different as the Coulomb phases associated with a multiple contact points are 3D, while the ones associated with a pinch line are necessary 2D.

With these analytical elements in our possession, there are now two key points to elucidate in real space. First, why, despite the 3D nature of the clusters, the emerging electric field is only 2D ? And second, what is the geometrical construction supporting the unique electric field for the former recipe, and the two independent electric fields for the second one ?

\subsubsection{Two dimensional nature of the electric field}

\begin{figure}
    \centering
    \includegraphics[width=\linewidth]{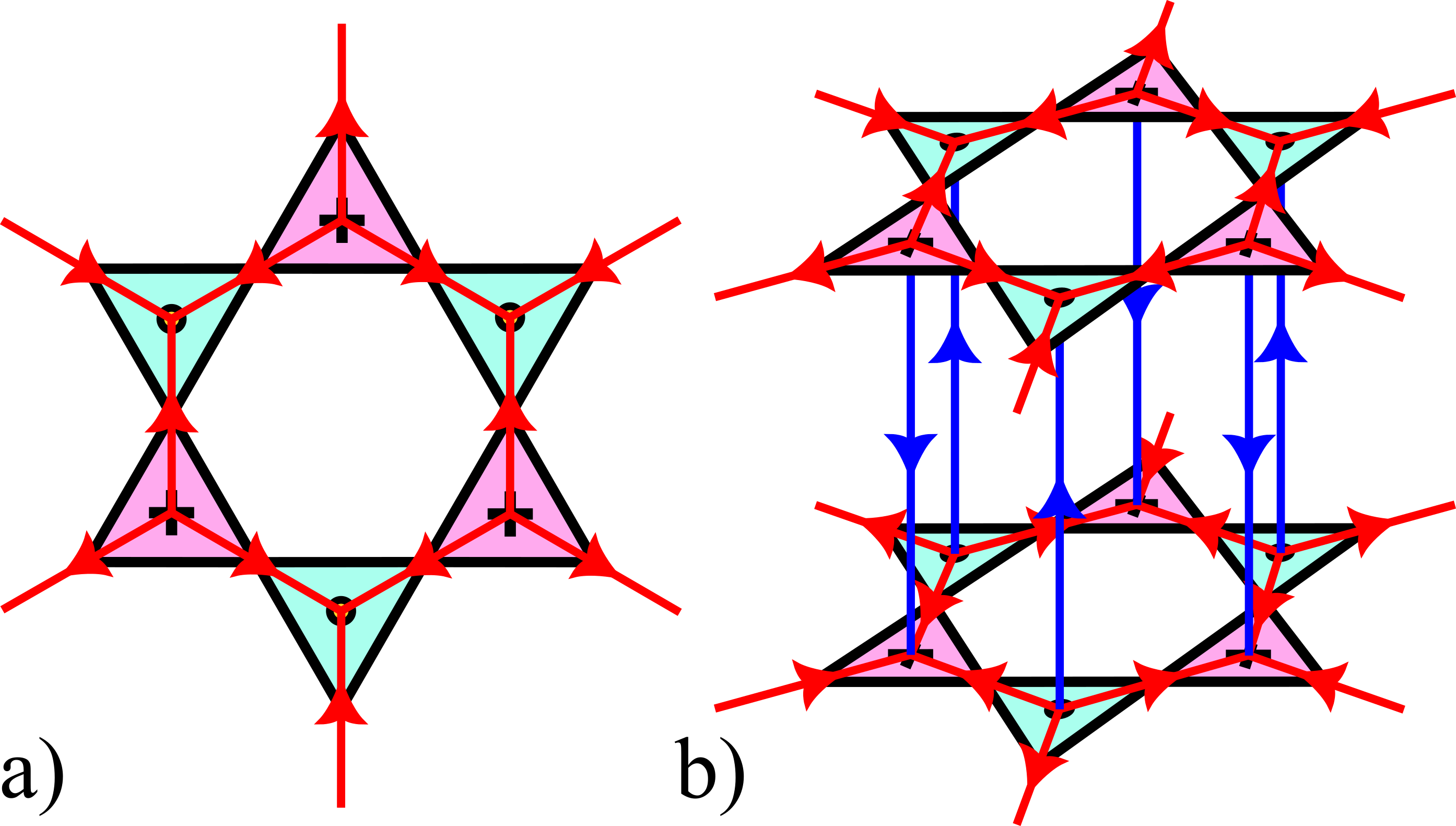}
    \caption{a) Flux construction for the 2D kagome model. The two type of premedial lattice sites are depicted by circles and crosses. Each premedial lattice bond is oriented from circles to crosses. To each of these bonds is attached a flux composed by the spin components of the underlying kagome lattice site. b) This construction can be naturally extended into 3D, simply adding vertical links linking consecutive parent planes clusters. Alternating the bond direction from one layer to the next allows to give an orientation to the vertical bonds, such that for each cluster there are only entering/outgoing bonds. }
    \label{fig: 2D parent model flux construction}
\end{figure}

The intrinsic 2D nature of the electric field can be illustrated geometrically for the two types of recipes of sections \ref{sec:simple} and \ref{sec:kagex}; one first needs to build the relevant fluxes for the 2D parent system \cite{Davier_2023}. Let us consider the premedial lattice of the 2D parent system, that is obtained by placing a site at the center of each cluster, and by linking these sites together. For example the premedial lattice of kagome is the honeycomb. If the premedial lattice is bipartite\footnote{If the premedial lattice is not bipartite, such a construction is still possible yet more complicated, see for example Fig. 26 from \cite{Davier_2023} for the construction associated with $K$ pinch points in snowflake model.}, which is the case for the honeycomb, its links can be oriented from sublattice A to sublattice B, as illustrated in Fig.~\ref{fig: 2D parent model flux construction}. Each of these oriented links can be attached with a vector flux
\begin{equation}
 \bm{\Pi}_i^j \equiv \gamma_i \; S_i^j \; \mathbf{u}_i,
    \label{Eq:Piflux}
\end{equation}
where $S_i^j$ is the $j^{\text{th}}$ component of the spin sitting on kagome site $i$ (or alternatively on the oriented honeycomb link $i$), $\mathbf{u}_i$ is the bond vector and $\gamma_i$ is the corresponding coefficient appearing in the constrainer definition (\ref{eq:hamcl}). The fluxes surrounding a 2D cluster of type $\alpha$ sitting at position $\mathbf{r}_c$ can be summed into an electric field\cite{Henley2005}
\begin{equation}
    \mathbf{E}^j_{2D}(\mathbf{r}_c) = \sum_{i \to c,\, 2D} \bm{\Pi}_i^j
    \label{Eq: Field flux construction}
\end{equation}
such that the coarse grained version of this field satisfies 
\begin{equation}
    \bm{\nabla}\cdot \mathbf{E}^j_{2D}(\mathbf{r}_c) \sim \sum_{i \to c,\, 2D} \bm{\Pi}_i^j\cdot \mathbf{u}_i = \mathcal{C}^j_{\alpha,\, 2D} = 0.
    \label{Eq: Flux field divergence}
\end{equation}

\begin{figure}
    \centering
    \includegraphics[width=\linewidth]{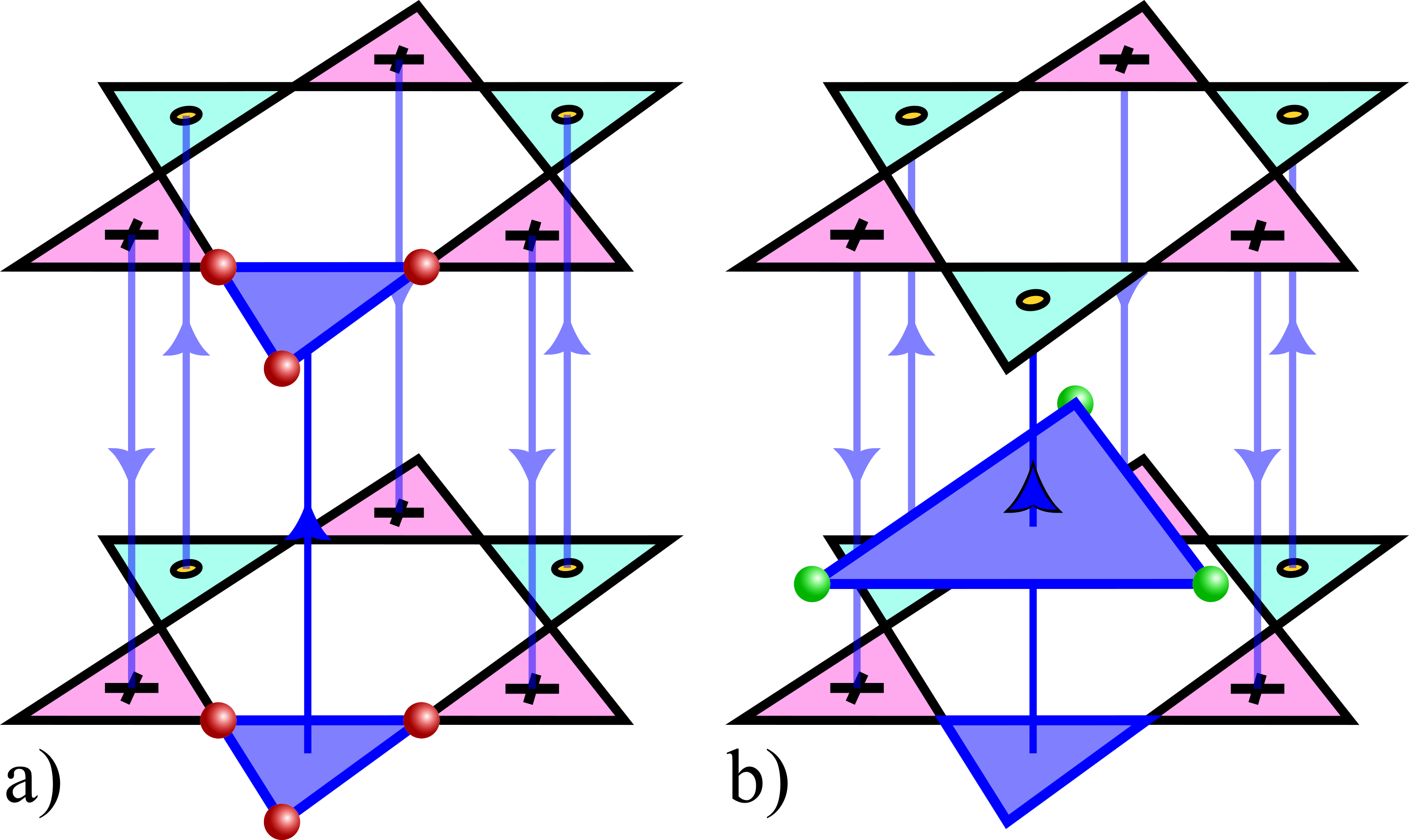}
    \caption{a) 3D flux construction for the 3D kagome model built with the trivial recipe. Each vertical flux carries the spin components of the six spins composing the starting triangle and the arrival triangle, depicted here as red spheres. To be able for the flux sum in one cluster to equate the constrainer, these sites are weighted by coefficients $\delta_i$. b) Similar flux construction but for the 3D kagome model obtained thanks to the second recipe. The time vertical fluxes carry only intermediate layer sites, depicted as green spheres. the three spins carried by a given flux being the three sitting the closest from it. }
    \label{fig: 3D flux construction}
\end{figure}

The next step is to extend this 2D field to 3D, which requires to define a link, and an orientation, between clusters of successive parent layers along the third dimension. Since parent layers are all equivalent, any orientation from one to the next is necessarily arbitrary. In order to conserve a natural definition of the Gauss law, we choose to alternate the orientation of the in-plane links from a parent layer to the next; then we connect the parent layers by vertical links connecting the centers of two clusters siting above one another. These vertical links are oriented the same way as the planar links; for a given 3D cluster, either all links are oriented inwards, or they all are oriented outwards, as illustrated in Fig.~\ref{fig: 2D parent model flux construction}(b). This is what will make the Gauss law rewriting natural. Finally we attach to these vertical links a flux made from neighboring spins. 

For the trivial recipe, these fluxes must encapsulate the spins from the cluster below and above, with weights $\delta_i$, as illustrated on Fig.~\ref{fig: 3D flux construction}(a). On horizontal bonds $i$, the fluxes weights can be rescaled from $\gamma_i$ to $\gamma_i - \delta_i$, this way summing the horizontal and vertical fluxes entering in a given cluster allows to satisfy a 3D version of Eq.~(\ref{Eq: Flux field divergence}) that would write
\begin{equation}
    \bm{\nabla}\cdot \mathbf{E}^j_{3D}(\mathbf{r}_c) \sim \sum_{i \to c,\, 3D} \bm{\Pi}_i^j\cdot \mathbf{u}_i = \mathcal{C}^j_{\alpha,\, 3D} = 0.
    \label{Eq: Flux field divergence 3D}
\end{equation}
With this 3D construction, each spin is encapsulated in two pairs of opposite vertical fluxes that are entering/outgoing above and below the two neighboring triangles, see Fig.~\ref{fig: 3D flux construction}(a). Hence, neighboring pairs of fluxes compensate and, independently of the spin configuration, a coarse-graining of the electric field necessarily gives a zero vertical component. 

For the second recipe including additional interlayer sites, the reasoning is similar. The 3D oriented lattice can be built the same way for the trivial recipe, see Fig.~\ref{fig: 2D parent model flux construction}(b), but this time the vertical bonds will carry fluxes composed of the intermediate spins sitting around it. As an illustration, for the kagome model, the vertical fluxes associated with an upward vertical bond $i$ will be expressed as 
\begin{equation}
    \bm{\Pi}^j_i = \delta \left(\sum_{p \in \langle i \rangle } S_p^j \right) d\mathbf{e}_z
\end{equation}
where the sum runs over the three sites belonging to the 3D cluster and located in the intermediate layer, see Fig.~\ref{fig: 3D flux construction}(b). As a result, a 3D electric field can be defined from these fluxes, and will by construction respect the 3D Gauss law (\ref{Eq: Flux field divergence 3D}). Similarly to the trivial recipe case, each interlayer spin belongs to pairs of clusters with opposite orientation convention; in Fig.~\ref{fig: 3D flux construction}(b) for the kagome model, the green dots belong to 6 clusters, three of whom have a vertical link oriented upwards, the three others being oriented downwards (see Fig.~\ref{fig: 3D flux construction}(b)). Hence, and independently of the spin configuration, the coarse grained electric field necessarily have a zero vertical component; the electric fields produced using the second recipe are thus also planar.

The conclusion to draw from these constructions is that for pinch line systems, \textit{the 3D electric field is constrained to be only 2D because the fluxes pointing along the pinch line direction cancel each other}. \\

We shall now address the existence of an alternative 2D flux construction, allowing for the emergence of distinct electric fields. 

\subsubsection{A 2D construction of the electric field}

Let us start again from the oriented premedial lattice on the parent system. However, the in-plane fluxes will now include spins from different layers, such that the 2D divergence of the produced electric field could equate the entire 3D constrainer. 

For the trivial recipe, this can be achieved by placing on each horizontal link $i$ belonging to the layer $l$ a flux that is composed of the underlying spin and the spin sitting directly above 
\begin{equation}
 \bm{\Pi}_{i,l}^j \equiv \left( \gamma_i \; S_{i,l}^j + \delta_i S_{i,l+1}^j \right) \mathbf{u}_i.
    \label{Eq: Pi flux 2D}
\end{equation}
Since $\mathbf{u}_i$ is the bond vector of link $i$ in the parent layer, it is a two-dimensional planar vector; and thus, so is the $\bm{\Pi}_i^j$ flux and the corresponding emergent electric field. We recover an electric field with no $z$ component as expected, fulfilling the Gauss law 
\begin{equation}
    \bm{\nabla}\cdot \mathbf{E}^j_{2D}(\mathbf{r}_c) \sim \sum_{i \to c,\, 3D} \bm{\Pi}_i^j\cdot \mathbf{u}_i = \mathcal{C}^j_{\alpha,\, 3D} = 0.
    \label{Eq: 2D Flux field divergence}
\end{equation}
This construction amounts to build a 2D electric field $\mathbf{E}_l$ living in the layer $l$, and relying on a lattice that share no bonds with any other layer. To form an electric field spanning the entire 3D lattice then requires to consider a superposition of these layered fields 
\begin{equation}
    \mathbf{E} = \sum_l a_l \mathbf{E}_l.
    \label{Eq: E interference}
\end{equation}
Because any site in the trivial recipe is shared vertically between two clusters, there is an effective overlap between consecutive layered fields, that can thus interfere depending on the weights $a_l$ between vertical structure. For example, for the trivial example of layered kagome, considering the special case $\delta = 1$ together with alternating weights $a_l = -a_{l+1} = 1$ leads to destructive interference. Indeed in this case each spin is carried with equal weights within two opposed fluxes, one sitting in its layer, and the other sitting in the above layer. The resulting electric field is thus trivial, explaining why for $\delta = 1$ and $d q_z = \pi$ the field (\ref{Eq: E trivial recipe}) becomes trivial. Except from these special parameters, the layered fields interfere to produce an effective divergence free field that shall be associated with a parameter $d q_z$ value depending on the spatial periodicity of the weights $a_l$, explaining the general electric field expression (\ref{Eq: E trivial recipe}). As the geometrical structure of the field is identical in all layers, this explains why there is a unique Coulomb phase replicated in all layers, and no structure factor evolution along the pinch line for a system built within the trivial recipe. 

\begin{figure}[h]
    \centering
    \includegraphics[width=0.5\linewidth]{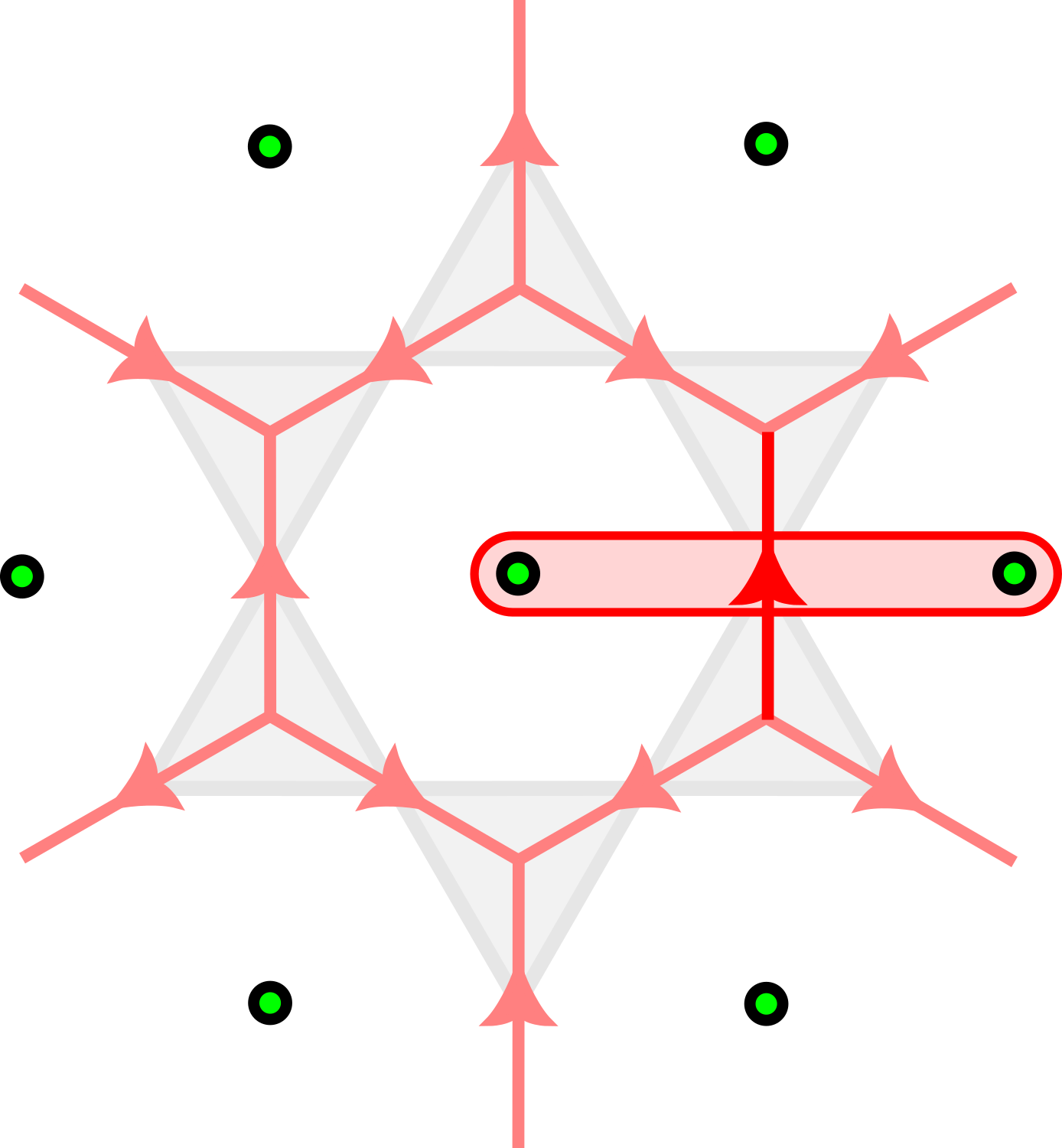}
    \caption{2D construction for the interlayer planes fluxes for 3D kagome model obtained with second recipe. The relevant oriented lattice to build in this case is in fact identical to the one of the parent layer, that is an oriented honeycomb lattice depicted in red. The sites from this intermediate layer, depicted as green dots, can be encapsulated in fluxes carried by these oriented bonds and containing the two nearest sites, as depicted for one flux with a red region around the underlying bond.  }
    \label{fig: Kagome interlayer flux construction}
\end{figure}

For the second recipe the story is expected to change, as analytical calculations have predicted the existence of two distinct electric fields. The proper 2D construction can be seen as an extension of the one presented for the trivial recipe. It is a layered construction, relying on the three planes intersecting a cluster, and with each of these planes carrying a geometrical flux structure. The key is that these flux structures, once projected into the central parent layer, that is the layer containing the cluster centers, must combine in such a way that the fluxes sum to zero in any point of the projected lattice. Let us illustrate it with the case of the 3D kagome lattice built with the second recipe. In this case any cluster is shared between three consecutive planes that are two interlayers and one central parent layer. We will now build fluxes on these three planes. The flux construction for the parent layer has been already done, see Fig.~\ref{fig: 2D parent model flux construction}. For the intermediate layers, the spins are located on a triangular lattice, whose sites are located above the parent layers hexagons. We now search for an oriented lattice, which possesses a site above/below each cluster center, and which is such that for each of its lattice sites, the sum of the incoming/outgoing fluxes is zero. This can happen for two reasons, first if the sum of these fluxes is geometrically trivial, or second, if once the three layers are all superimposed together, in each cluster center the fluxes from the three planes sum to equate the cluster constrainer, exactly as for the previous constructions. In the case of the 3D kagome interlayers, the good lattice to build appears to be identical to the one associated with parent planes, that is the oriented honeycomb lattice, see Fig.~\ref{fig: Kagome interlayer flux construction}. Indeed, attaching to each bond $i$ a flux $\bm{\Pi}_i$ containing the two neighboring sites as 
\begin{equation}
    \bm{\Pi}_i^j = \frac{1}{2}\sum_{k \in \langle i \rangle} \delta_k S_{k}^j \mathbf{u}_i
    \label{eq:pi2ndflux}
\end{equation}
allows, when summing in a cluster center the incoming fluxes coming from the two interlayers and the parent layer, to obtain the usual relation
\begin{equation}
    \sum_{p=1}^3\sum_{i \to c}\bm{\Pi}_{p,i}^j\cdot \mathbf{u}_{p,i} = \mathcal{C}^j = 0
    \label{eq: fluxes sum 2D layered fluxes}
\end{equation}
at each cluster center, with $p$ labeling the planes. The factor $1/2$ in front of the fluxes comes from double counting since each interlayer site is adjacent to two different fluxes. As before, these fluxes can be coarse grained to form an effective electric field. Since the fluxes $\bm{\Pi}_i^j$ are planar vectors because $\mathbf{u}_i$ are, so is the corresponding emergent electric field. With this construction we thus again build a layer electric field $E_l$ for each layer parent $l$, that can interfere to form the complete electric field following Eq.~(\ref{Eq: E interference}). 

This construction appears to be pretty similar to the one presented for the trivial recipe. However, in the present case the sites from a parent layer $l$ enter only in the field $\mathbf{E}_l$, while the intermediate layer sites sitting between the parent layers $l$ and $l+1$ belong both to the fields $\mathbf{E}_l$ and $\mathbf{E}_{l+1}$. This means that when the layer electric fields $\mathbf{E}_l$ will interfere, only their contributions from the interlayers sites will truly interfere, resulting in different effective electric fields for the parents and interlayers sites. Explicitly, looking at the 3D kagome model obtained with the second recipe, if one takes alternated weights $a_l = - a_{l+1} = 1$ from one parent layer to the next, this leads to cancellation of interlayers sites flux contribution, producing an electric field that does not rely on intermediate layer spins. This corresponds to considering Eq.~(\ref{Eq: explicit E for second recipe applied to rank 1 PL}) with $2d q_z = \pi$ as the distance between two interlayers is $2d$, that gives indeed a field $\mathbf{E}_p$ emerging from the parent layers spins only. If now one considers the $q_z = 0$ mode associated with uniform weights $a_l = 1$, the electric field produced will maximally depend on the intermediate layers sites, in good agreement with Eq.~(\ref{Eq: explicit E for second recipe applied to rank 1 PL}) that gives a maximal contribution of $\mathbf{E}_{il, \perp}$ for $q_z = 0$. 

The main message from this discussion is thus that for a pinch line to host distinct Coulomb phases, all 2D fluxes must \textbf{not} all interfere the same way when making interfering layer electric fields, allowing for these intereference to produce distinct electric fields.

Now the remaining question is; is there a way to use these types of geometrical constructions to understand the physics host by a rank-$n$ or even a multirank pinch line ?

\subsection{Exotic pinch line real space interpretation}

For a pinch line hosting not only two fold pinch points, the effective electric field for a point $\mathbf{q}^\star$ can no more be obtained following Eq.~(\ref{Eq : Electric vector q^*}). For such a pinch line, the critical vector $\mathbf{L}^c$ admits different order expansions depending on its component. As these are associated with different sublattices, this means the multi-order Gauss law deriving form the critical vector $\mathbf{L}^c$ acts on electric fields with different rank, and emerging from coarse grained spins associated with distinct group of sublattices. These Gauss laws can be generally expressed as
\begin{equation}
    \sum_i a_i(q_t) \mathcal{D}^{n(i)} E_i\left[\{\mathbf{S}\}_{\Omega_i}\right] = 0
\end{equation}
where $i$ labels the different groups of sublattices $\Omega_i$ that are associated with same order $n(i)$ critical vector components, and where $\mathcal{D}^{n} E$ denotes the order $n$ divergence\footnote{Examples of order $n$ divergences :
\begin{equation*}
    \begin{split}
        &n = 1 : \qquad \bm{\nabla}\cdot \mathbf{E} \\
        &n = 2 : \qquad \partial_i \partial_j E_{ij} \\
        &n = 3 : \qquad \partial_i \partial_j \partial_k E_{ijk} \\
    \end{split}
\end{equation*}
with Einstein's summation rule over repeated indices.}
 of the rank-$n$ tensor $E$. For example in the cases of the 3D checkerboard and 3D snowflake models, there are two terms entering in this sum. The first is the parent electric field $E_p$ that only relies on parent planes spins, and comes with a coefficient $a_p(q_z) = 1$. The second one is the interlayer electric field $E_{il}$ relying on the interlayer spins components, which comes with a weight $a_{il}(q_z) = \cos(dq_z)$, see Eqs~(\ref{eq:Esnow}) and (\ref{Eq: E il checkerboard}). For such layered systems obtained within the second recipe, these two electric field emerge from 2D fluxes living in parent planes, as for the case of a two fold pinch line. If the pinch line only host rank-2 electric fields, as for the 3D snowflake model, the situation is almost identical to the one of a two fold pinch line. Rank-2 fluxes can be placed on consecutive 2D oriented lattices, to build a rank-2 coarse grained field $\mathbf{E}^l$ living in the parent layer $l$, and satisfying
 \begin{equation}
     \sum_{i, j = x,y} \partial_i\partial_j \mathbf{E}_{ij}^l(\mathbf{r}_c) \sim  \bm{\mathcal{C}}_c = 0.
 \end{equation}
A global electric field can finally be obtained by summing these layer fields as for twofold pinch lines, see Eq.~(\ref{Eq: E interference}). Similarly, as spins components from the interlayers are here also shared by fluxes encapsulated in consecutive layer electric fields $\mathbf{E}^l$ and $\mathbf{E}^{l+1}$, any alternate configuration $2dq_z = \pi$ with weights $a_l = -a_{l+1} = 1$ will wash out the interlayer spins components from the global electric field, and produce a field relying only on parent layers spins. $q_z=0$ zero modes with $a_l = 1$ will again lead to a global electric field mixing all spin components. In this case the system would thus host two distinct rank-2 electric fields, both a priori associated with fractonic excitations. Note that for the 3D snowflake model the electric field relying on intermediate layers is traceful, that is not the case of the electric field emerging from parent layer sites only. The fracton excitations of these two distinct fields are thus expected to be different\cite{Pretko_2_2017}, being associated with different conserved quantities. The present discussion is not specific to this model, and should remain valid for any rank-$n$ pinch line. 

For the case of a multirank pinch line mixing different fields with ranks $m>n$, there must exist a layered rank $n$ flux construction which is such that for a given type of layer, the rank-$n$ fluxes become trivial. The electric field associated with this rank-$n$ construction will then only encapsulate spins from the other layers producing non trivial fluxes. To study the field emerging for the remaining spins will next require to build a rank-$m$ flux construction, producing a rank-$m$ electric field encapsulating all spins components, that will be subject to interference as discussed for the case of a rank-$n$ pinch line. This can be illustrated with the example of the 3D generalized checkerboard model built with the second recipe, which rank-2 flux construction is presented in Appendix. \ref{Appendix D : ckb rank 2 flux construction}. 

Note that these geometric flux constructions allow to get a physical understanding of the previous statement about the fact that considering only interlayer sites located above the cluster center were not a good solution to generate pinch lines. This is because the existence of a 2D flux construction relying on intermediate planes, allowing for the emergence of a pinch line, requires that neighboring clusters share intermediate layer spins in order to have fluxes exchanges between these neighboring clusters.
 
These constructions have been illustrated for the case of layered systems, but the key ingredients of these constructions are applicable to any pinch line system, as we will now illustrate with the case of a non layered pinch line system known to host pinch lines.

\section{Application to generic pinch lines}

As explained in the introduction, our motivation in this paper is double. First, we wanted to present a generic recipe to build spin models with pinch lines in the structure factor. This is what has been done in sections \ref{sec:generic} and \ref{sec:example}, showing how to stack 2D lattices in order to propagate the 2D pinch points along the $q_z$ direction. This construction brought to light the property that, depending on the model, the pattern of the pinch points may vary as one moves along the pinch line. This suggested the co-existence of multiple two-dimensional effective electric fields; an intuition that was confirmed and put into a theoretical framework in section \ref{sec : What do pinch lines look like in real space ?}. This theoretical framework naturally emerges from the layered structure of our models, where the two-dimensional electric fields live.

But not all pinch-line models are layered with pinch lines parallel to a given axis. In some cases they form a grid in Fourier space \cite{Yan_2024_long} or propagate along (111) directions for systems with cubic symmetry \cite{Benton2016,Niggemann_2023}. The second message of our paper is that the theoretical framework of the previous section to build the electric field(s) can be directly applied to these other pinch-line models, even if the pinch lines are not parallel and if the system is not layered. This is what we shall explain in the rest of this paper.

\begin{figure}[ht]
    \centering  \includegraphics[width= 0.8 \columnwidth]{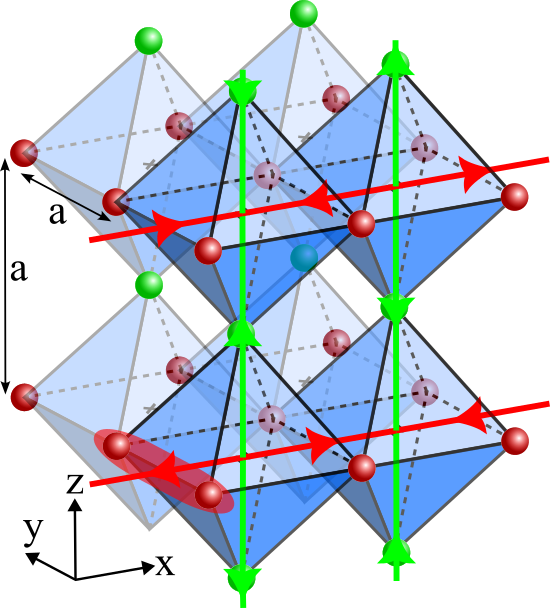}
    \caption{Octahedral lattice, composed of edge sharing octahedra in the $(x,y)$ plane and corner sharing along the vertical $z$ direction. This lattice possesses two sublattices depicted as red and green spheres. This lattice can support 2D flux constructions located in $(x,z)$ planes passing through octahedra centers, leading to formation of pinch line along the orthogonal direction $y$.  }
    \label{Fig: octahedral}
\end{figure}
\subsection{Octahedral lattice}

In Ref.~\cite{Yan_2024_long}, the authors introduced a 3D lattice made of octahedra, corner-sharing along the $z-$ direction and edge-sharing in the $(x,y)$ plane [Fig.~\ref{Fig: octahedral}], composed of two sublattices and with each octahedra associated with the simple constrainer 
\begin{equation}
    \bm{\mathcal{C}}_o = \sum_{i \in o} \mathbf{S}_i.
    \label{Eq: constrainer octahedral alltice}
\end{equation}
This lattice could be seen as a stacking of a $J_1-J_2$ square lattice forming the first sublattice, with an intermediate layer forming the second sublattice. Since the second sublattice sits in the center of the 2D square clusters, it cannot, however, form pinch lines along $q_z$ (see section \ref{sec:counter}) and thus does not belong to our recipe exposed in this paper. Indeed, pinch lines form a grid in the $(q_x,q_y)$ reciprocal planes \cite{Yan_2024_long}. This can be understood from a 2D flux construction similar to the ones presented above that we shall detail now.

Let us consider a plane of octahedra as the one depicted at the forefront of Fig.~\ref{Fig: octahedral}, and project the octahedra in the plane $(x,z)$ passing through the centers of these octahedra. This projection forms a checkerboard lattice, where pair of sites form the first sublattice, depicted in red on Fig.~\ref{Fig: octahedral}, are superimposing. This checkerboard lattice possesses a premedial lattice that is bipartite and can therefore be oriented as depicted with red and green arrows on Fig.~\ref{Fig: octahedral}. These oriented bonds can next be attached with fluxes composed of the the two neighboring spins from first sublattice for red bonds, and from the underlying spin from second sublattice only for green bonds. This way in each octahedron center the sum of the incoming/outgoing fluxes equates the constrainer (\ref{Eq: constrainer octahedral alltice}) of the octahedron. Making this construction 3D by placing additional vector fluxes along the $y$ direction requires to alternate the construction from one $(x,z)$ plane to the next, leading to fluxes cancellation along both $x$ and $y$ directions, explaining why it is not possible to have a field having both $x$ and $y$ components for this system presenting pinch line along both of these directions \cite{Yan_2024_long}. The 2D construction can finally be extended along the entire 3D space by replicating it in any $(x,z)$ plane, and superposing the layer electric fields associated with each of these planes as in Eq.~(\ref{Eq: E interference}). Similarly to the case of layered systems obtained with the second recipe, only sites from the first sublattice are able to interfere as they are the only ones to be encapsulated in consecutive layer fields. This will then produce two independent fields, one relying on first sublattice spins components and the other on second sublattice spins only, in good agreement with a direct application of Eq.~(\ref{Eq : Electric vector q^*}) giving 
\begin{equation}
    \mathbf{E} \simeq \cos\left(\frac{a q_y}{2}\right) \bm{\chi}_1 \mathbf{e}_x + \bm{\chi}_2 \mathbf{e}_z = \cos\left(\frac{a q_y}{2}\right) \mathbf{E}_1 + \mathbf{E}_2
\end{equation}
for a pinch line along $y$ direction and with $a$ the distance between two unit cells along $y$ direction. Because this field expression is valid for any point $q_y$ along the pinch line, it implies the existence of two independent fields $\mathbf{E}_1$ and $\mathbf{E}_2$. Because $x$ and $y$ directions are equivalent for this system, the exact same derivation can be made for $(y,z)$ planes, explaining the presence of pinch lines along $x$ direction also.

As explained in Ref.~\cite{Yan_2024_long}, such pinch lines are expected to be common to 3D classical spin liquids on inversion-symmetric lattices with two sublattices and one constraint per unit cell. We expect the present electric field construction to conserve the same degree of generality.

\begin{figure*}[ht]
    \centering  \includegraphics[width=\linewidth]{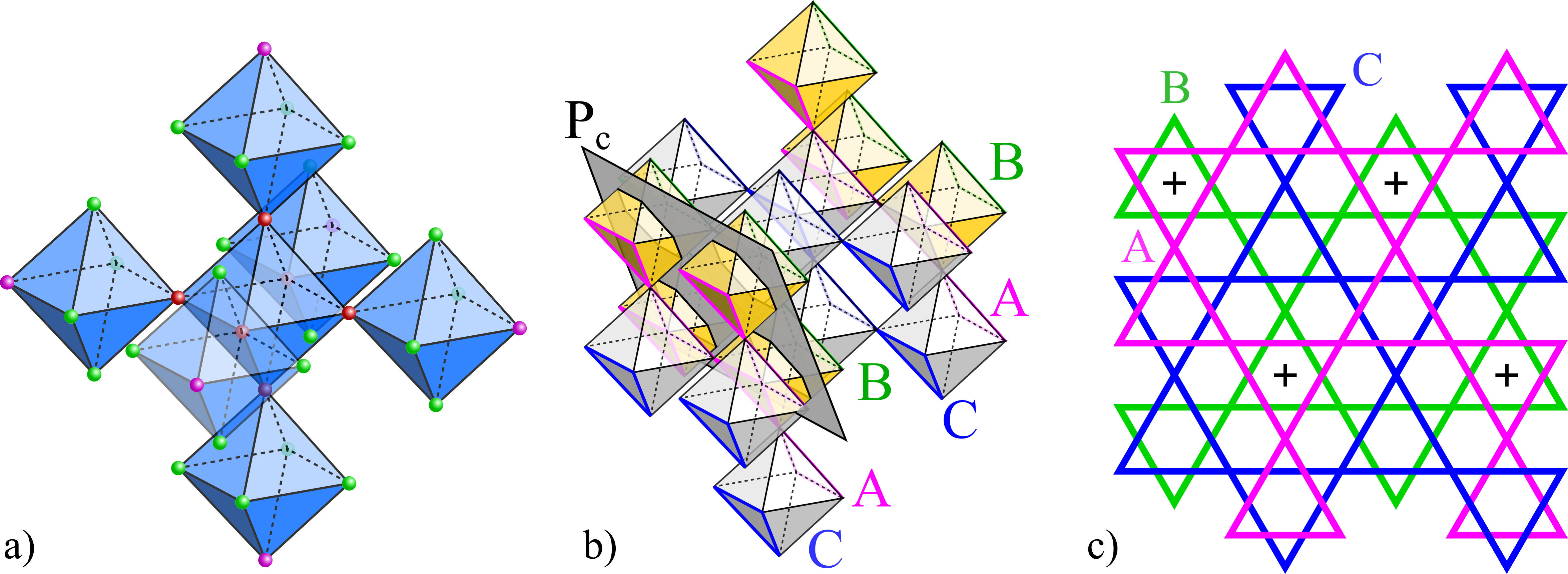}
    \caption{Octochlore model geometry. a) Cluster definition, the sites composing the central octahedron ($o$),  depicted in red, are accounted with coefficient 1 in the constrainer definition Eq.~(\ref{Eq: constrainer octochlore}), while the 24 green sites ($\langle o \rangle$) are considered with coefficient $\alpha$ and the six magenta sites ($\langle \langle o \rangle \rangle$) are weighted by a coefficient $\beta$. b) Illustration of the stacking of the three kagome planes $A,B$ and $C$ along the $111$ direction. The plane $P_c$, depicted in gray, is the plane passing through the center of the yellow octahedra. It is also the plane located at mid distance between $A$ and $B$ layers. c) Projection of four consecutive stacked kagome planes $CABC$ into the plane $P_c$ that is parallel to these planes and intersecting the centers of yellow octahedra, depicted as black crosses. }
    \label{Fig: octoclore geometry}
\end{figure*}

\subsection{Octochlore lattice}

The 3D octochlore lattice is made of corner-sharing octahedra, forming an antiperovskite structure \cite{Benton_Moessner_2021,Szabo2022} which naturally supports a Coulomb phase \cite{Sklan13,Jaubert15c}. As illustrated on Fig.~\ref{Fig: octoclore geometry}(a), Ref.~\cite{Benton_Moessner_2021} introduced a model with a constrainer connecting all spins of a central octahedron and its six neighboring octahedra,
\begin{equation}
    \bm{\mathcal{C}}_o = \sum_{i \in o} \mathbf{S}_i + \alpha \sum_{i \in \langle o \rangle} \mathbf{S}_i + \beta \sum_{i \in \langle \langle o \rangle \rangle} \mathbf{S}_i,
    \label{Eq: constrainer octochlore}
\end{equation}
which is known to host a pinch line for $\alpha = -1/2$ and $\beta = 1$ \cite{Niggemann_2023}. As opposed to all previous lattices presented in this paper, the octochlore lattice has cubic symmetry and its pinch lines propagate along the four equivalent (111) directions in reciprocal space. It means they cross at $\mathbf{q}=0$ and we cannot define a plane orthogonal to all pinch lines at the same time. The real-space theoretical framework to build an effective electric field can nonetheless be applied here.\\

Let us first compute the coarse-grained electric field using the generic formula (\ref{Eq : Electric vector q^*}):
\begin{equation}
        \mathbf{E}(q) = \cos^2 \left(q d\right)  \sin \left(q d\right) \mathbf{E}_\perp^0
        \label{Eq: E octochlore}
\end{equation}
which corresponds to a modulated electric field 
\begin{equation}
    \mathbf{E}_\perp^0 =
    \frac{3}{\sqrt{2}} \begin{pmatrix}
             -\mathbf{S}_1 + \mathbf{S}_2 \\
            (\mathbf{S}_1 + \mathbf{S}_2 - 2 \mathbf{S}_3)/\sqrt{3} \\
            0
        \end{pmatrix}. \\
\end{equation}
$q \equiv \mathbf{q}\cdot \mathbf{e}_{111}$ is the wave vector component along the pinch line, while $d$ denotes the distance between two consecutive kagome planes along $111$ direction. This electric field being a modulation of a unique field $\mathbf{E}_\perp^0$ independant from $q$ tells us there must exist a single 2D flux construction where all sites are interfering the same way, yet the complex periodicity of this modulation as a function of $dq$ suggests this construction to be non trivial.

\begin{figure}
    \centering  \includegraphics[width= \linewidth]{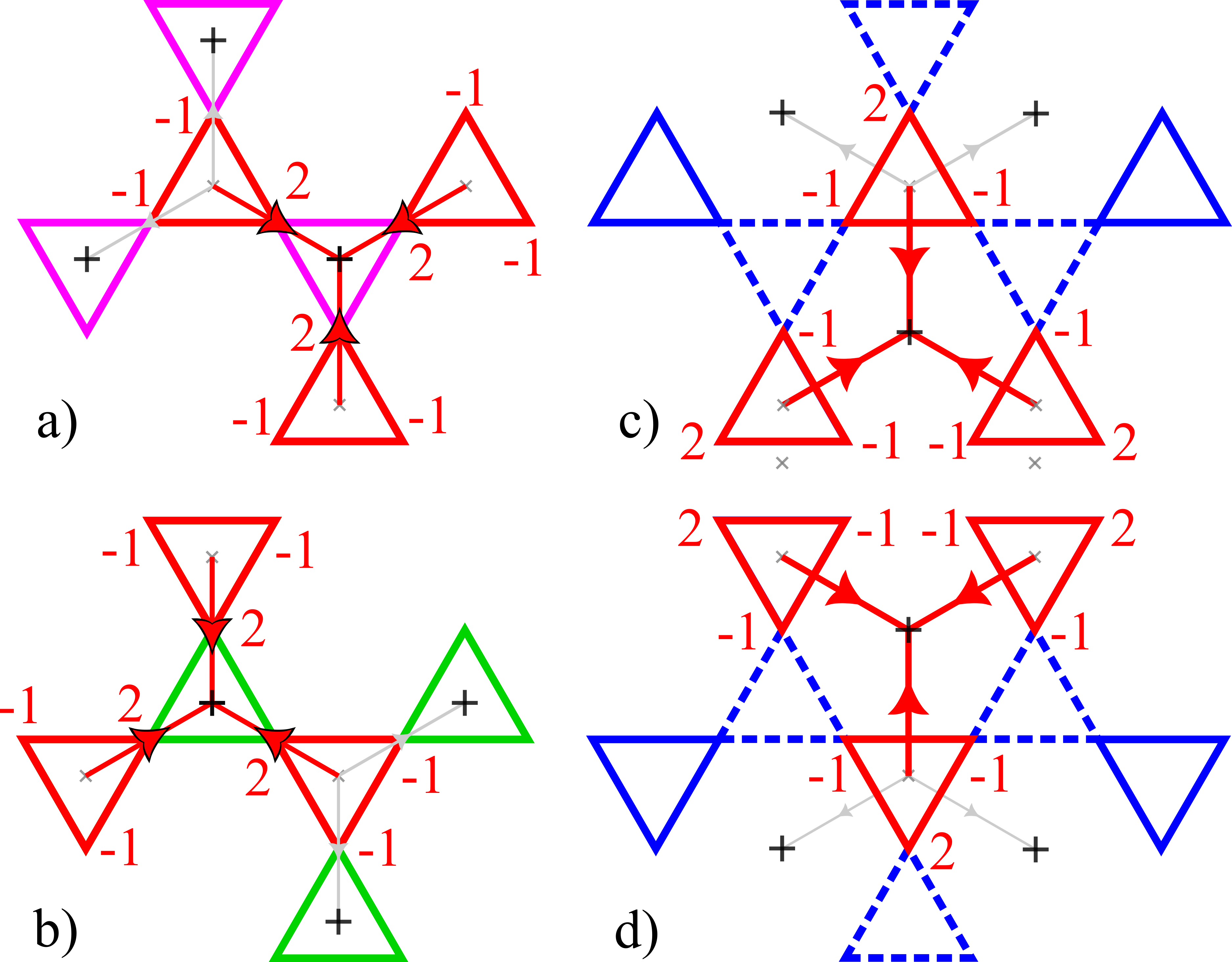}
    \caption{a) Type $A$ kagome plane projected in the plane $P_c$ intersecting the centers of yellow octahedra, depicted as black crosses $+$, see Fig.~\ref{Fig: octoclore geometry}(b,c). Each central point $+$ is linked to the three neighboring triangles centers by a bong oriented inward, to which is attached a flux composed of the three spins composing this triangle, weighted with coefficients $2, -1, -1$. These specific weights allow the three fluxes exiting a neighboring triangle (depicted here in red) to sum to zero. b) Similar construction for $B$ planes, that are related to A planes by inversion symmetry. c) Analog construction for $C$ planes located in front of the yellow octahedra, where centers of these octahedra project at the middle of hexagons. The fluxes can be built in a similar fashion than for $A$ and $B$ planes, bringing the spins from three neighboring triangles, once again counting these spins among these fluxes using weights chosen for the fluxes exiting a triangle to sum up to zero.  d) Similar construction but for the $C$ plane located behind the yellow octahedra, where this time fluxes coming from down triangles only must enter into the hexagon.}
    \label{fig: 2D octoclore fluxes}
\end{figure}

This construction is based on the kagome planes that appear when the octochlore lattice is intersected by planes perpendicular to the (111) direction, see Fig.~\ref{Fig: octoclore geometry}(b, c). Our goal is to construct a field such that its divergence is zero in any point of space. As the field construction associated with a pinch line is intrinsically two dimensional, it must be realized in a 2D reference plane. As this plane must encapsulate the centers of the octahedra, the natural choice is the plane $P_c$ depicted on Fig.~\ref{Fig: octoclore geometry}(b) that passes across the centers of the yellow octahedra. Focusing on this reference plane, we look at the projection of the neighboring $CABC$ kagome planes on it, resulting on the intricate lattice depicted on Fig.~\ref{Fig: octoclore geometry}(c) where the black crosses denote the octahedra centers. In each of these kagome layers, we search for an in-plane flux construction that ensures zero divergence not only at the centers of the octahedra, but also at all other lattice points supporting the fluxes. 
 For octahedra centers, the logic to apply is the one already used for previous examples: we require the sum of the incoming fluxes to equate the full constrainer (\ref{Eq: constrainer octochlore}). For this, in each kagome slice $k$ we require that the fluxes converging toward the cluster center account for all spins located at the intersection between the 3D octahedral cluster $o$ and the kagome plane $k$. This requirement can be expressed as
\begin{equation}
    \sum_{i \to o} \Pi_{i,k}^\alpha = 2\mathcal{C}_{o \cap k} ^\alpha .
    \label{Eq : Fluxes sum in central cluster}
\end{equation}
In A and B layers, the cluster intersects the kagome plane to form a central triangle surrounded by three neighboring triangles. The constraint above then implies that the fluxes entering the central triangle must carry the contributions of the nine spins belonging to the three surrounding triangles, see Figs. \ref{fig: 2D octoclore fluxes}(a,b). 
The C layers intersect the cluster to form three triangles surrounding an hexagon, the constraint (\ref{Eq : Fluxes sum in central cluster}) thus imposes the fluxes entering in this hexagon must bring in the nine spins from the three neighboring triangles, see Figs. \ref{fig: 2D octoclore fluxes}(c,d). The two C planes intersecting the cluster are related by inversion symmetry, so the constructions based on them are related by the same symmetry, as depicted on Figs. \ref{fig: 2D octoclore fluxes}(c,d). The simplest way to construct such fluxes is as follows: in A and B planes, we draw oriented bonds from the centers of neighboring triangles toward the cluster center. We then assign to these bonds fluxes that encapsulate the three spins of the originating triangle, as illustrated on Figs.~\ref{fig: 2D octoclore fluxes}(a,b). The construction is similar for $C$ planes as depicted on Figs. \ref{fig: 2D octoclore fluxes}(c,d).
However, while this construction can be tuned for any values of $\alpha$ and $\beta$ to ensure that the sum of incoming fluxes at the cluster center is zero, thus guaranteeing divergence-free behavior at that point, it does not a priori ensure that the outgoing fluxes from the centers of the neighboring triangles also sum to zero. For the entire field to be divergence-free at every lattice point, we must design the in-plane fluxes such that, at each triangle center, the three outgoing fluxes always sum to zero, regardless of the cluster's geometry. This is achieved by weighting the spin contributions using the coefficients shown in Figs.~\ref{fig: 2D octoclore fluxes}(a–d). This way for $A$ and $B$ planes triangles, each vertex site is outgoing one time with a weight of 2, but entering also two times with a unit weight, such that the sum of the three fluxes outgoing from a triangle is always zero. A similar scheme applies to the $C$ layers. With these weight ratios fixed, we can enforce the condition (\ref{Eq : Fluxes sum in central cluster}) for each kagome layer. This imposes that the flux weights satisfy $\alpha = -1/2$ when applied on $A$ and $B$ layers, and $\beta = - 2\alpha = 1$ on $C$ layers. These specific values explain why this layered field construction is only feasible at one point in the phase diagram—and why pinch lines appear along the (111) directions. Finally, summing the electric fields from the four kagome layers gives
\begin{equation}
    \mathbf{E} = \mathbf{E}_{C, \vartriangle} +  \mathbf{E}_{A} + \mathbf{E}_{B} + \mathbf{E}_{C, \triangledown}
    \label{Eq: E octo as a superposition of CABC fields}
\end{equation}
which satisfies the divergence-free condition
\begin{equation}
    \bm{\nabla} \cdot \mathbf{E} \sim \sum_k \sum_{i \to o} \Pi_{i,k}^\alpha = 2\mathcal{C}_o^\alpha = 0
\end{equation}
at each lattice point for ground state configurations. The global electric field can be finally built as a superposition of these fields constructed for each (111) reference plane, summing them as in (\ref{Eq: E interference} ). Because for this model the layer fields are themselves are made of four components with distinct in-plane geometries, see Eq. (\ref{Eq: E octo as a superposition of CABC fields}), this explains the complex intensity modulation of the electric field along the pinch line observed in the direct calculation of the electric field (\ref{Eq: E octochlore}). Since the construction works equally well for any equivalent (111) direction, this accounts for the observed pinch lines in all such directions in reciprocal space. 

\begin{figure}[h]
    \centering
    \includegraphics[width=0.9\linewidth]{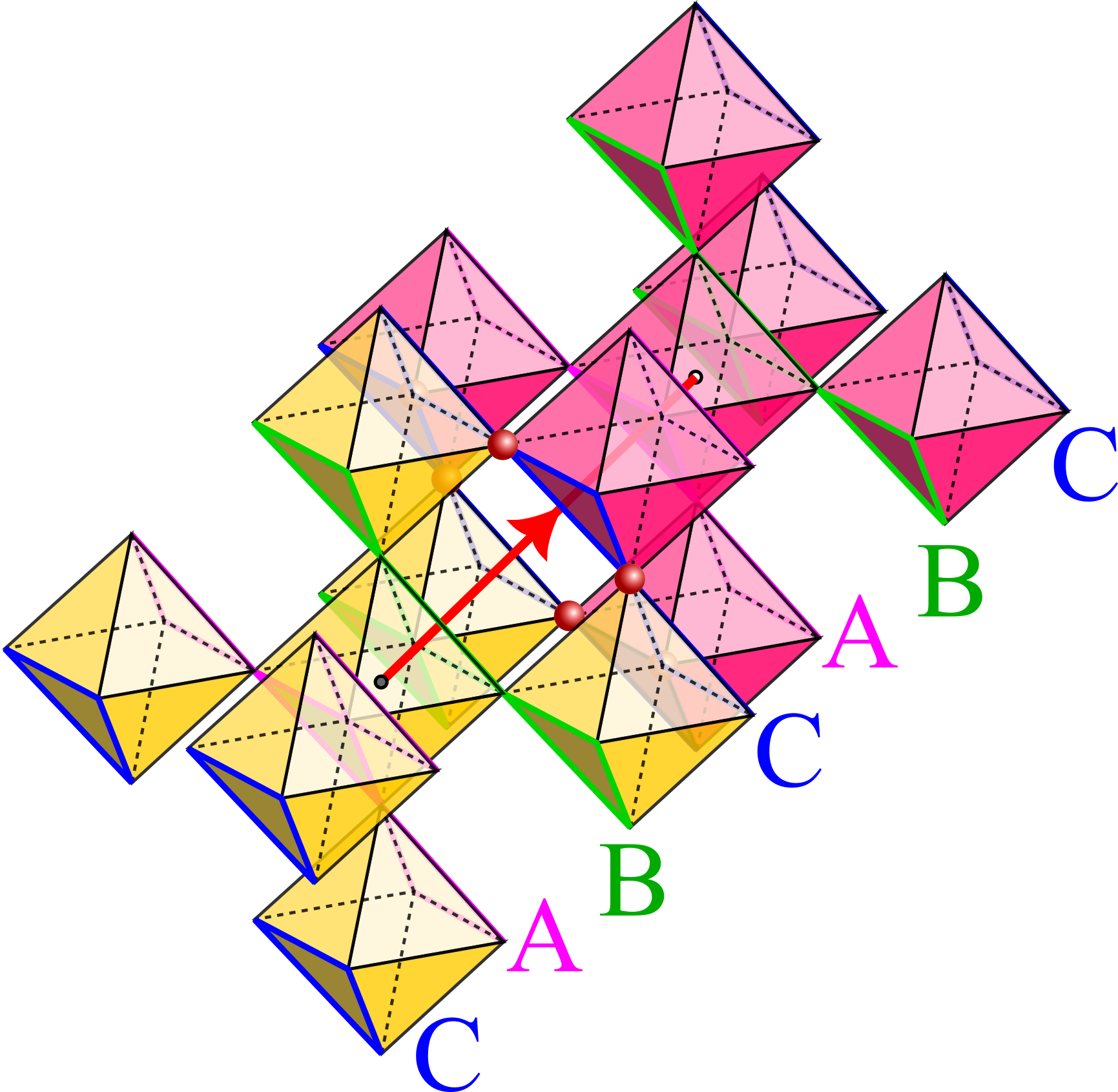}
    \caption{Possible vector flux linking two neighboring clusters along 111 direction. The two clusters are depicted in yellow and magenta. The centers of these clusters can be linked by a bond oriented along 111 direction, as depicted by the red link. This link can only carry a flux composed of the spins shared by the two clusters, in order for the construction to be identical for both clusters. The six spins shared by the two clusters, depicted as red dots, are entering in the constrainer of each cluster with a coefficient $\alpha$ see Fig.~\ref{Fig: octoclore geometry}. }
    \label{fig: octochlore 111 flux}
\end{figure}

In the octochlore case, the vanishing of the longitudinal component of the electric field does not result from the cancellation of fluxes, as it did in previously discussed situations. Instead, it arises from a structural constraint. A flux connecting two neighboring clusters can only involve spins that lie in the intersection of these two clusters. For a link oriented along the (111) direction these shared spins correspond to the six spins forming the hexagon located in the intermediate C plane, equidistant between the two cluster centers, see Fig.~\ref{fig: octochlore 111 flux}. Each of these six spins is included in the constrainer of both clusters with a coefficient $\alpha$. However, there are also twelve additional spins located between the two clusters (see Fig.~\ref{Fig: octoclore geometry}) that are not shared by these two neighboring clusters, yet still contribute to the constrainer of each cluster with the same coefficient $\alpha$. Because of this imbalance, incorporating longitudinal fluxes along (111) into the 2D flux construction would break the requirement that the total flux entering a cluster must match the constrainer. The only possibility to conserve the divergence less condition for the field obtained with such a flux construction is then to consider only trivial fluxes along 111, that is to build an electric field that has no component along the pinch line direction. The only way to preserve the divergence-free condition of the resulting electric field is therefore to restrict ourselves to trivial fluxes along the (111) direction—that is, to construct an electric field with no component in the direction of the pinch line.

\section{Discussion} 

Pinch-line spin liquids form a family of exotic magnetic textures with well-defined one-dimensional singularities in reciprocal space. Very few models have been proposed so far \cite{Benton2016,Niggemann_2023, Yan_2024_long, Fang_2024}. In this work, we expose a simple and generic framework to transform most 2D algebraic spin liquids into 3D pinch-line spin liquids. Our theory authorizes a high degree of design and is confirmed numerically on three concrete examples [Fig.~\ref{fig: 3D kagome latticeSQ}] and [Fig.~\ref{fig:Sq}] which do not order down to the lowest temperatures in Monte Carlo simulations, see Fig. \ref{fig:specific_heat} and [Appendix \ref{Appendix C: Monte Carlo simulations}]. In addition to a generic framework, our results open a platform for exotic forms of higher-rank gauge fields, that can interfere along the direction of the pinch line, producing pinch points superpositions that continuously evolve along the line in reciprocal space. This results into a coexistence of distinct emergent electric fields, presenting different tensorial structures while emerging from a unique system. Our recipe is sufficient to generate pinch lines but not necessary. The real-space discussion derived from this recipe is however extended beyond this layered framework, allowing to understand the underlying physics associated with any pinch line, as illustrated with a pinch-line model presenting cubic symmetry and non-parallel pinch lines \cite{Niggemann_2023}. 

As a next step, the recipe presented here could probably be adapted using the topological-quantum-chemistry route recently proposed in Ref.~[\onlinecite{Fang_2024}] to allow for more tunability. 

A promising aspect of our framework is that for each 2D algebraic parent spin liquid, multiple 3D structures are conceivable. Moreover, the choice to consider intermediate layers with only one sublattice was only out of pedagogical convenience; multiple sublattices are perfectly possible, as long as all $L_{m>n}^c$ components respect condition (\ref{condition for pinch lines}). Hence, while the cluster form of Hamiltonian (\ref{eq:hamcl}) imposes constraints on the model, it remains reasonably generic for frustrated magnets. The huge diversity of possible pinch-line models that can be derived in this framework make us hopeful to see some of these models realized in materials in the future, especially since the large entropy of classical spin liquids stabilizes them at finite temperature, even away from fine-tuned model parameters. A subtle, but experimentally relevant, aspect of our theory if that the appearance of the pinch line is independent on the ratio between intra- and inter-layers, since only the latter couplings appear in $L^c_{n+1}(\mathbf{q})$. It means that the very low-temperature properties of weakly coupled layers, which is a common perturbation to two-dimensional spin liquids, could be a good place to look for pinch lines. Another advantage is that the framework relies on standard isotropic spin-spin Heisenberg exchange terms. There is no need for three- or four-body interactions that are more difficult to realize in experiments. And for materials with strong spin-orbit couplings, nothing prevents to extend our theory to anisotropic interactions, albeit with a more complex form of emergent electric field \cite{Yan17a, Essafi17b, Yan_2020, Lozano24a, lozanogomez_2024_preprint, chung_2024_preprint}. In the same vein, our theory also applies to Ising and XY spins. Such spin models have a tendency of more easily ordering at low temperature, e.g.~via the order-by-disorder mechanism \cite{Moessner_1998_counting} or confinement-deconfinement transitions, which makes 2D parent models a bit less diverse. But recent developments in Rydberg atoms \cite{shauss2015, scholl2021, ebadi2021, chen2023} offer new avenues to realize such systems experimentally, and motivates to explore the influence of quantum fluctuations on all these pinch-line models.\\

\textit{Acknowledgements -- } N.D. would like to thank Han Yan and Daniel Lozano-G\'omez for useful discussions during the Highly Frustrated Magnetism conference in Toronto. The authors acknowledge support from the CNRS International Research Project COQSYS. N.D. and L.J. are supported by Grant No. ANR-23-CE30-0038-01. F. A. G. A. and H. D. R. are partially supported by CONICET (PIP 2021-112200200101480CO), SECyT UNLP PI+D X947 and Agencia I+D+i (PICT-2020-SERIE A-03205). F. A. G. A. acknowledges support from PIBAA 2872021010 0698CO (CONICET).

\appendix
\section{Lattice definitions}
\label{Appendix A: Lattice definitions}

\textit{3D kagome models --} For the 2D kagome model; see Fig.~\ref{fig: 3D kagome lattice}(a),  the position of the three sites belonging to a down triangle, are given relative to the triangle center by 
\begin{equation*}
    \mathbf{r}^{(p)}_1 = a \begin{pmatrix} 0 \\ -\frac{1}{\sqrt{3}} \\ 0 \end{pmatrix}, \quad 
    \mathbf{r}^{(p)}_2 = \frac{a}{2}\begin{pmatrix} 1 \\ \frac{1}{\sqrt{3}} \\ 0 \end{pmatrix}, \quad 
    \mathbf{r}^{(p)}_3 = \frac{a}{2}\begin{pmatrix} -1 \\ \frac{1}{\sqrt{3}} \\ 0 \end{pmatrix}
\end{equation*}
with $a$ the nearest neighbors distance. 

For the first proposition of 3D generalization, see Fig.~\ref{fig: 3D kagome lattice}(b), the positions of the six fourth sublattice sites that are interacting with the three initial sites among the 3D cluster are given, relative to the cluster center, by 
\begin{equation*}
    \mathbf{r}^{(p)}_{4(7)} = \begin{pmatrix} 0 \\ \frac{2a}{\sqrt{3}} \\ \pm d \end{pmatrix}, \quad 
    \mathbf{r}^{(p)}_{5(8)} = \begin{pmatrix} -a \\ -\frac{a}{\sqrt{3}} \\ \pm d \end{pmatrix}, \quad 
    \mathbf{r}^{(p)}_{6(9)} = \begin{pmatrix} a \\ -\frac{a}{\sqrt{3}} \\ \pm d \end{pmatrix}
\end{equation*}
where $d$ is the inter-layer distance. For an up triangle, all the position can be simply multiplied by $-1$ are the two types of clusters are related by inversion symmetry.

For the second proposition of 3D generalization, see Fig.~\ref{fig: 3D kagome lattice}(c), the positions of the two fourth sublattice sites that are interacting with the three initial sites among the 3D cluster are given, relative to the cluster center, by 
\begin{equation*}
    \mathbf{r}^{(p)}_{4(7)} = \begin{pmatrix} 0 \\ -\frac{a}{\sqrt{3}} \\ \pm d \end{pmatrix}
\end{equation*}
where $d$ is again the inter-layer distance. For an up triangle, all the position can be simply multiplied by $-1$ are the two types of clusters are related by inversion symmetry.


\textit{3D snowflake model --} The positions of the 2D snowflake parent model sites that are considered with a coefficient $1$ in the constrainer definition (\ref{eq:Csnow}), and depicted by blue dots on Fig~\ref{fig:Sq}(a), are given by 
\begin{equation*}
    \begin{split}
        \mathbf{r}^{(p)}_{1(4)} = \pm \begin{pmatrix} a \\ 0 \\ 0 \end{pmatrix},  \; 
    \mathbf{r}^{(p)}_{2(5)} = \pm \frac{a}{2} \begin{pmatrix} \sqrt{3} \\ 1 \\ 0 \end{pmatrix},  \;
    \mathbf{r}^{(p)}_{3(6)} = \mp \frac{a}{2} \begin{pmatrix} \sqrt{3} \\ -1 \\ 0 \end{pmatrix}
    \end{split}
\end{equation*}
while sites taken with a coefficient $\gamma$ in the constrainer definition, depicted by red dots on Fig~\ref{fig:Sq}(a), are sitting at positions
\begin{equation*}
    \begin{split}
        \mathbf{r}^{(p)}_{7(10)} = \pm \begin{pmatrix} 2a \\ 0 \\ 0 \end{pmatrix}, \; 
        \mathbf{r}^{(p)}_{8(11)} = \pm a \begin{pmatrix} \sqrt{3} \\ 1 \\ 0 \end{pmatrix}, \;
        \mathbf{r}^{(p)}_{9(12)} = \mp a \begin{pmatrix} \sqrt{3} \\ -1 \\ 0 \end{pmatrix}
    \end{split}
\end{equation*}
For the 3D generalized system, intermediate layers sites, depicted by squares dots on Fig~\ref{fig:Sq}(a), are located at positions
\begin{equation*}
    \begin{split}
        \mathbf{r}^{(l)}_{0} &= (0,0,d), \hspace{0.24\linewidth} \mathbf{r}^{(l)}_{1(4)} = \left(0, \pm \sqrt{3}a, d \right), \\  \mathbf{r}^{(l)}_{2(5)} &= \left(\mp \frac{3}{2}a, \pm \frac{\sqrt{3}}{2}a, d \right), \quad 
        \mathbf{r}^{(l)}_{3(6)} = \left(\mp \frac{3}{2}a, \mp \frac{\sqrt{3}}{2}a, d \right)
    \end{split}
\end{equation*}
for the layer located above the 2D parent system plane (and with $-d$ as last component for sites located in the intermediate plane located below).

These explicit definitions allow to express the constraint vector components as 
\begin{widetext}
\begin{equation}
    \begin{split}
        L_1(\mathbf{q}) &= \bar{L}_2(\mathbf{q}) = e^{iaq_x} + e^{ia(\sqrt{3}q_x/2 + q_y/2)}  
        + e^{ia(-\sqrt{3}q_x/2 + q_y/2)} 
    + \gamma \left(e^{-2iaq_x} + e^{-ia(\sqrt{3}q_x + q_y)} + e^{-ia(-\sqrt{3}q_x + q_y)} \right), \\
    L_3(\mathbf{q}) &= 2 \delta_1 \cos(d q_z) \left[ 1 + 2\frac{\delta_2}{\delta_1} \left( \cos(\sqrt{3}aq_y) + \cos\left(a\frac{3q_x + \sqrt{3}q_y}{2}\right) + \cos\left(a\frac{3q_x - \sqrt{3}q_y}{2}\right)\right) \right].
    \end{split}
\end{equation}
\end{widetext}

\textit{3D generalized checkerboard model --} The positions of the 2D parent sites that are taken with a coefficient 1 in the constrainer definition (\ref{Cckb}), and depicted as blue dots on Fig~\ref{fig:Sq}(e) are given by the vectors 
\begin{equation*}
        \mathbf{r}^{(p)}_{1(3)} = \pm \frac{a}{2} \left(1,1,0 \right), \qquad \mathbf{r}^{(p)}_{2(4)} = \pm \frac{a}{2} \left(-1,1,0 \right).
\end{equation*}
Sites depicted as red dots on Fig~\ref{fig:Sq}(e), which are counted with a coefficient $\gamma_1$ in the constrainer definition, are located at positions
\begin{equation*}
    \begin{split}
        \mathbf{r}^{(p)}_{5(9)} &= \pm \frac{a}{2} \left(3,1,0 \right), \hspace{0.113 \linewidth} \mathbf{r}^{(p)}_{6(10)} = \pm \frac{a}{2} \left(1,3,0 \right), \\
        \mathbf{r}^{(p)}_{7(11)} &= \pm \frac{a}{2} \left(-1,3,0 \right), \qquad \mathbf{r}^{(p)}_{8(12)} = \pm \frac{a}{2} \left(-3,1,0 \right).
    \end{split}
\end{equation*}
Finally, the sites encapsulated in the constrainer definition (\ref{Cckb}), depicted by yellow dots on Fig~\ref{fig:Sq}(e), are sitting at positions 
\begin{equation*}
        \mathbf{r}^{(p)}_{13(15)} = \pm \frac{a}{2} \left(3,3,0 \right), \qquad \mathbf{r}^{(p)}_{14(16)} = \pm \frac{a}{2} \left(-3,3,0 \right).
\end{equation*}
The four sites located in the intermediate plane located above the parent system, depicted as green squares on Fig~\ref{fig:Sq}(e) are located at positions 
\begin{equation*}
        \mathbf{r}^{(l)}_{1(3)} =  \left(\pm a,0,d \right), \qquad \mathbf{r}^{(l)}_{2(4)} = \left(0,\pm a,d \right).
\end{equation*}
and with $-d$ as last component for sites located in the intermediate plane located below.
These explicit positions allow to build the constraint vector that expresses as 
\begin{widetext}
\begin{equation}
    \begin{split}
        L_1(\mathbf{q}) & = \cos\left(a\frac{q_x+q_y}{2}\right) + \gamma_1 \left[ \cos\left(a\frac{3q_x-q_y}{2}\right) + \cos \left(a\frac{q_x-3q_y}{2}\right) \right] + \gamma_2 \cos\left(3a\frac{q_x+q_y}{2}\right) ,\\
        L_2(\mathbf{q}) & = \cos\left(a\frac{q_x-q_y}{2}\right) + \gamma_1 \left[ \cos\left(a\frac{3q_x+q_y}{2}\right) + \cos \left(a\frac{q_x+3q_y}{2}\right) \right] + \gamma_2 \cos\left(3a\frac{q_x-q_y}{2}\right) ,\\ 
        L_3(\mathbf{q}) &= 4\delta \cos(d q_z) \left[ \cos(a q_x)+\cos(a q_y) \right].
    \end{split}
\end{equation}
\end{widetext}


\section{Definition of the structure factor}
\label{Appendix B: Definition of the structure factor} 

The structure factors depicted in the right panels of Fig.~\ref{fig:Sq} are computed analytically using Henley's projective method \cite{Henley2005}; in the zero-temperature limit, the spin-spin correlation functions are expected to be proportional to the projector $\Pi$ into the space orthogonal to the constraint vectors defined for each cluster following  Eq.~(\ref{eq:Lu}). 
It can be built from the matrix $M$ composed of the constraint vectors as columns, through the expression\cite{Henley2005, Davier_2023} 
\begin{equation}
    \Pi = I - M \left( M^\dagger M \right)^{-1} M^\dagger.
\end{equation}
The static structure factor can then be expressed, up to a multiplicative constant, as 
\begin{equation}
    \mathcal{S}(\mathbf{q}) \propto \sum_{m,p} \Pi_{mp}.
\end{equation}
For the last two systems studied in the present work, there is a unique constraint vector and the projector is then simply equal to 
\begin{equation}
    \Pi = I - \mathbf{L} \frac{1}{\mathbf{L}^\dagger  \mathbf{L}} \mathbf{L}^\dagger
\end{equation}
where $I$ denotes the identity matrix.
The structure factor can then be simplified as a sum over projector elements
\begin{equation}
    \mathcal{S}(\mathbf{q}) \propto \sum_{m,p} \Pi_{mp} = n - 1 - \sum_{m \neq p}\frac{L_m(\mathbf{q}) \bar{L}_p(\mathbf{q})}{\|\mathbf{L}(\mathbf{q})\|^2}
\end{equation}
where $n$ is the number of $\mathbf{L}$ components, that is, the number of sublattices, equal to 3 for the last two systems studied in this work. $\bar{L}_p(\mathbf{q})$ denotes the complex conjugate of $L_p(\mathbf{q})$. 


\section{Monte Carlo simulations}
\label{Appendix C: Monte Carlo simulations}
To investigate the temperature-dependent behavior of the systems, we performed Monte Carlo simulations employing the Metropolis algorithm combined with overrelaxation techniques. The simulations were conducted within an annealing framework, where the temperature $T$ was systematically reduced. All temperatures are expressed in units of the largest absolute value of the coupling constant, $J_{max}$, for each specific parameter set in the MC simulations. For the 3D Checkerboard model, $J_{max}$ in units of $J$ (where $J$ is the plaquette coupling constant) is defined as $J_{max}=4\gamma_1=4$ for $\gamma_1=1$. This corresponds to the second-neighbor coupling introduced by $\gamma_1$.  For the 2D Snowflake model, where $\gamma=1/2$, $J_{max}=2+2\gamma=3$. The systems were thermalized for a minimum of $10^6$ MC steps, followed by $2 \times 10^6$ MC steps for calculating mean values. The system size $N$ was defined as $N = n_s \times L_x \times L_x \times L_z$, where $n_s$ represents the number of sites within the unit cell ($n_s = 3$ for both generalized snowflake honeycomb and checkerboard lattices). $L_x$ and $L_z$ denote the linear dimensions of the system, with $ 15\leq L_x \leq 30$ and $10 \leq L_z \leq 20$.  

To confirm the spin-liquid behavior of both 3D models, we computed two key quantities: the specific heat per spin $C$, given by $C =(\langle E^2\rangle-\langle E\rangle^2)/NT^2$ where $E$ is the energy of the system, and the static structure factor $\mathcal{S}(\mathbf{q})$, defined as $\mathcal{S}(\mathbf{q})=\frac{1}{N}\sqrt{\langle|\sum_j{\bf S}_j\,e^{i{\bf q}\cdot{\bf r}_j}|^2\rangle}$, whose corresponding plots are presented in the main text [Fig.~\ref{fig:Sq}].

Figure \ref{fig:specific_heat} shows the temperature dependence of the specific heat $C$ for the four studied models. In all cases, the specific heats monotonously increase upon cooling; there is no peak and thus no indication of any phase transition in both models. Furthermore, $C$ reaches a plateau at a value of $C \simeq 1/2$ which is noticeably lower than the value of 1 traditionally expected for classical Heisenberg models. 
To further confirm the absence of magnetic order, we found no Bragg peaks in the structure factor of both models down to $T=0.0002 J_{max}$, as shown in the left panels of Figs.~\ref{fig:Sq}(b,c,d,f,g,h).

\section{Rank-2 flux construction for the checkerboard lattice}
\label{Appendix D : ckb rank 2 flux construction}

 \begin{figure}[h]
     \centering
     \includegraphics[width=\linewidth]{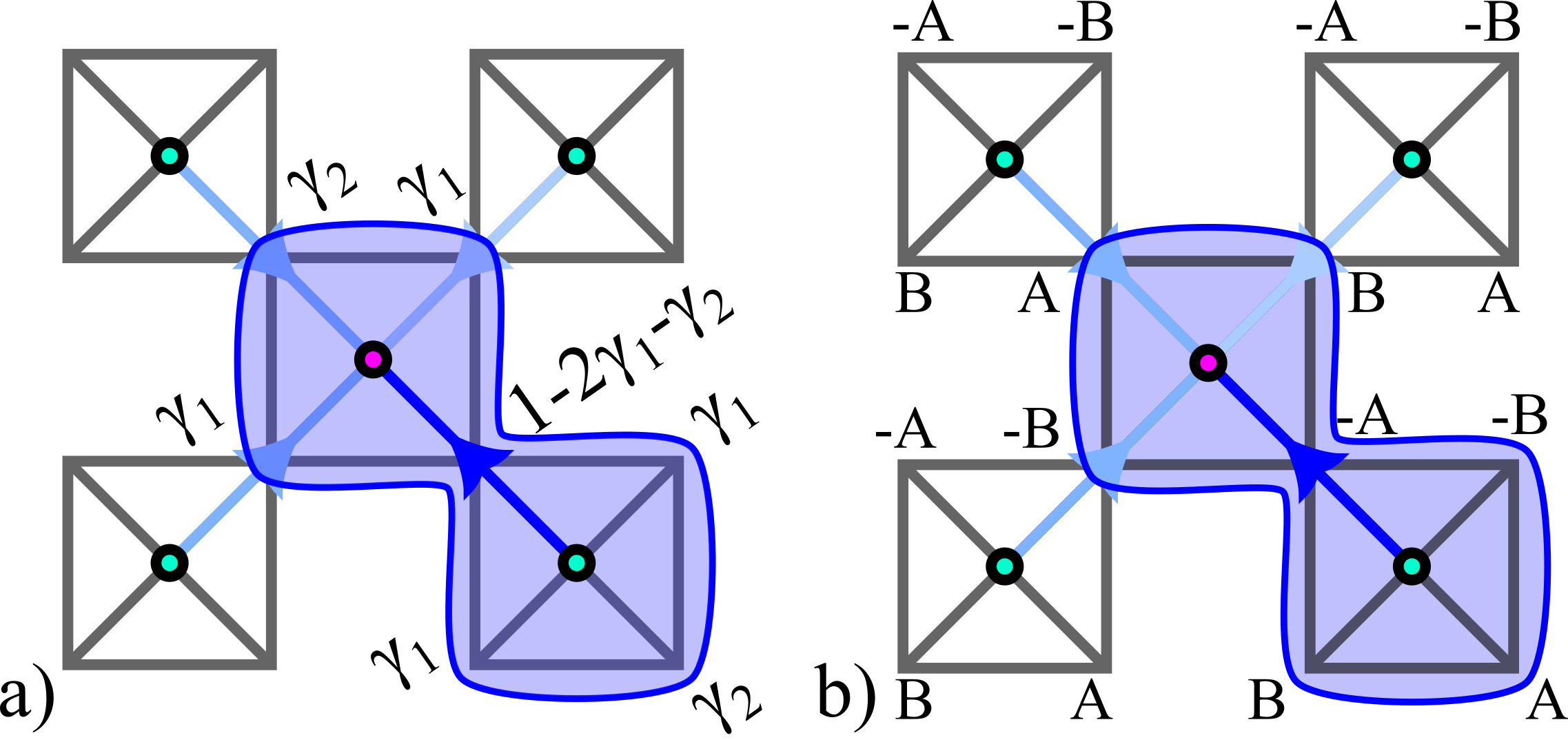}
     \caption{a) 2D generalized model vector flux construction. The checkerboard premedial lattice is a bipartite square lattice, which links can be oriented as depicted by blue arrows. These links can next be used to support fluxes constituted from the seven neighboring sites spin components, counted with coefficients $1-2\gamma_1-\gamma_2$, $\gamma_1$ and $\gamma_2$ depending on their disposition relative to the bond. b) Letters A and B indicate here two flavors of spins associated with the two sublattices of the checkerboard lattice. Each motif is alternating from one site of a sublattice to the next, corresponding to the $M$ modes associated with wave vector $\mathbf{q}_M = (\pi/a, 0)$ that are allowed when there is a contact point in the band structure located in BZ corner point $M$.}
     \label{fig: ckb vector fluxes}
 \end{figure}

We start by the rank-1 vector flux construction on the 2D generalized checkerboard lattice for pedagogical reasons. This construction relies on the premedial lattice of the checkerboard lattice that is simply a bipartite square lattice, which links can then be oriented similarly to the case of honeycomb lattice, see Fig.~\ref{fig: ckb vector fluxes}(a). As the 2D generalized checkerboard model is built from clusters of 16 sites, the vector fluxes to be attached with these links must at least\cite{Davier_2023} contain spin components from the seven neighboring sites, as depicted on Fig.~\ref{fig: ckb vector fluxes}(a). These seven spins must enter in these fluxes with coefficients indicated on Fig.~\ref{fig: ckb vector fluxes}(a) to guaranty that the sum of the fluxes in one cluster center equates the 2D constrainer\cite{Davier_2023}. This construction is always possible for this parent system, regardless of the position and rank of the pinch points. What occurs for the precise sub-parameter space region $\gamma_2 = (1 - 2 \gamma_1)/3$ that have been used to apply our pinch line recipe is that in this case for spins configurations associated with a contact point located at BZ corner point $M$, with coordinates $\mathbf{q}_M = (\pi/a, 0)$ and depicted on Fig.~\ref{fig: ckb vector fluxes}(b), is that the flux norm associated with one link becomes trivial 
 \begin{equation}
     \begin{split}
         \Pi &= 2\gamma_1(B-B) + 2 \gamma_2 A + (1 -2 \gamma_1 - \gamma_2)(-A)\\ 
         &= A(-1+2\gamma_1 +3 \gamma_2) = 0.
     \end{split}
 \end{equation}
This means that the electric field built from these fluxes will be trivial if $\gamma_2 = (1 - 2 \gamma_1)/3$, which explains why there are no two-fold pinch points located in $M$ for this specific line in the phase diagram of the 2D generalized checkerboard model\cite{Davier_2023}. Making a similar construction for interlayers sites requires to build vector fluxes encapsulating a pair of neighboring interlayers sites. If these two sites carry spins in a $\mathbf{q}_M$ configuration, meaning that they are identical but with opposed signs, the associated flux will also be trivial, explaining why there is no vector field emerging from interlayers spins neither. 

\begin{figure}[h]
     \centering
     \includegraphics[width=\linewidth]{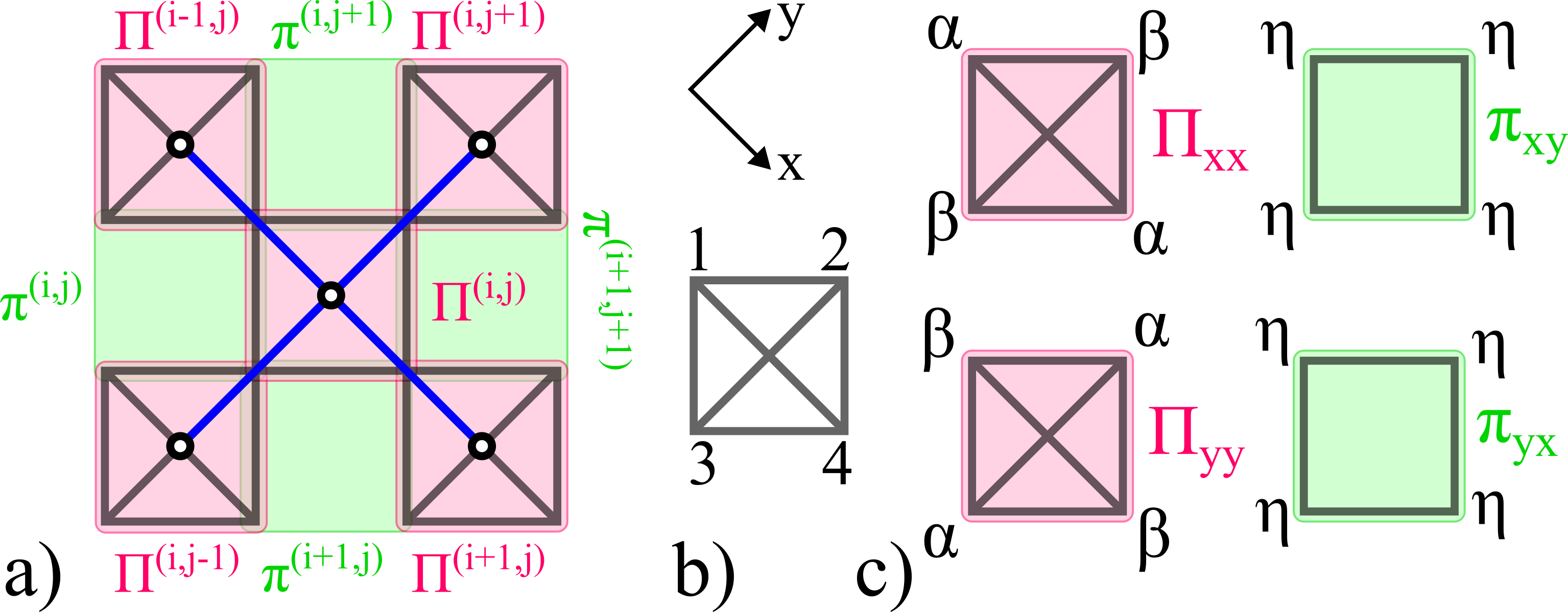}
     \caption{a) 2D generalized checkerboard model rank-2 flux construction. The checkerboard premedial lattice is a bipartite square lattice, which vertices and links as depicted as white dots and blue blue lines. This lattice can be used to support rank-2 fluxes sitting either on vertices or plaquettes, each encapsulating four spins. b) Axis and spin numeration conventions considered for the rank-2 flux construction. Note that these axis are rotated from 45° compared to the convention used in Subsec.~\ref{subsec: 3D ckb} and Appendix.~\ref{Appendix A: Lattice definitions}.  d) Internal weights structure of the different rank-2 fluxes introduced in (a). Note that these weights must be symmetric with respect with the bond vector associated with considered components (e.g $\Pi_{xx}$ component weights must be symmetric relative with bonds oriented along $x$ direction). }
     \label{Fig: ckb rank-2 fluxes}
 \end{figure}

In such situation it is natural to look for a rank-2 flux construction. The latter relies on the same premedial lattice, but the links do not need to be oriented, and asking for a second order derivative to be well defined in any lattice vertex requires no more to place fluxes on lattice bonds. The natural way to built a rank-2 electric field necessitates instead to place rank-2 fluxes on lattice vertices and plaquettes, as depicted on Fig.\ref{Fig: ckb rank-2 fluxes}(a). These fluxes are matrices defined as 
\begin{equation}
    \begin{split}
            \bm{\Pi}^{(i,j)} &= \begin{pmatrix}
            \Pi_{xx}^{(i,j)} & 0 \\ 0 & \Pi_{yy}^{(i,j)}
        \end{pmatrix} 
        ,\\ 
        \bm{\pi}^{(i+a,j+b)} &= \begin{pmatrix}
            0 & \pi_{xy}^{(i,j)} \\ \pi_{yx}^{(i,j)} & 0
        \end{pmatrix},
    \end{split}
\end{equation}
with $\pi_{xy} = \pi_{yx}$. The fluxes components $\Pi_{\mu \mu}^{(i,j)}$ and $\pi_{\mu \nu}^{(i,j)}$ can be defined as a weighted sum over the four spins surrounding a lattice site or a plaquette, with weights depicted on Fig.\ref{Fig: ckb rank-2 fluxes}(c). For example 
\begin{equation}
    \Pi^{(i,j)}_{xx} = \alpha\left( \mathbf{S}_1^{(i,j)} +  \mathbf{S}_3^{(i,j)}\right) + \beta\left( \mathbf{S}_2^{(i,j)} +  \mathbf{S}_4^{(i,j)}\right)
\end{equation}
with spins numerated following Fig.\ref{Fig: ckb rank-2 fluxes}(b). Note that the weights must be symmetric relative to the underlying vertex/plaquette, in order to have the same definition for every equivalent vertex/plaquette among the lattice. An electric field can be next built at cluster center $\mathbf{r}_{i,j}$ as the sum of the 9 neighboring fluxes
\begin{equation}
    \mathbf{E}(\mathbf{r}_{i,j}) = \sum_{k, l \in \langle (i,j) \rangle } \left( \bm{\Pi}^{(k,l)} + \bm{\pi}^{(k,l)} \right).
\end{equation}
Computing its rank-2 divergence from
\begin{equation}
    \begin{split}
        \partial_x\partial_x E_{x x}^{(i,j)} &\simeq \Pi_{xx}^{(i-1,j)} - 2 \Pi_{xx}^{(i,j)} + \Pi_{xx}^{(i+1,j)}, \\
        \partial_y\partial_y E_{yy}^{(i,j)} &\simeq \Pi_{yy}^{(i,j-1)} - 2 \Pi_{yy}^{(i,j)} + \Pi_{yy}^{(i,j+1)}, \\
        \partial_x\partial_yE_{xy}^{(i,j)} &\simeq \pi_{xy}^{(i+1,j+1)} - \pi_{xy}^{(i+1,j)} - \left( \pi_{xy}^{(i,j+1)} - \pi_{xy}^{(i,j)} \right), \\
        \partial_y\partial_x E_{yx}^{(i,j)} &\simeq \pi_{yx}^{(i+1,j+1)} - \pi_{yx}^{(i,j+1)} - \left( \pi_{yx}^{(i+1,j)} - \pi_{yx}^{(i,j)} \right) \\
        &= \partial_x\partial_y E_{xy}^{(i,j)},
    \end{split}
\end{equation}
and asking it to equate the constrainer $\bm{\mathcal{C}}_{(i,j)}$ finally allows to settle the internal structure of the fluxes. It imposes the relations
\begin{equation}
    \begin{split}
        1 &= -2\beta - \alpha \\
        \gamma_1 &= \beta + 2\eta \quad \& \quad \gamma_1 = \beta - 2\eta \\
        \gamma_2 &= \alpha
    \end{split}
\end{equation}
for the three types of sites entering in the constrainer (\ref{Cckb}). These relations impose to fix the fluxes weights as 
\begin{equation}
    \begin{split}
        \eta &= 0 \quad \Rightarrow \quad \bm{\pi}^{(i,j)} = 0,\\
        \beta &= \gamma_1 ,\\
        \alpha &= \gamma_2 ,\\
        1 &= -2\gamma_1 - \gamma_2.
    \end{split}
\end{equation}
This rank-2 flux construction is then only possible for $\gamma_2 = -1-2\gamma_1$, which appears to be the only parameter space region for which fourfold pinch points are observed\cite{Davier_2023}. Note that these fourfold pinch points are located in BZ center point $\Gamma$, in good agreement with the fact the rank-2 flux construction makes no difference between premedial sublattice sites (and is therefore a $\mathbf{q}_\Gamma =0$ construction), contrary to the rank-1 construction relying on an oriented lattice where the bond orientation alternates from one cluster to the next $\mathbf{q}_M$ construction). It is in fact natural that no fourfold pinch points could be observed at BZ corner points $M$, as the associated $\mathbf{q}_M = (\pi/a, 0)$ configurations, depicted on Fig.~\ref{fig: ckb vector fluxes}(b), impose that symmetric four-spins square fluxes, are necessarily containing two pairs of opposed fluxes, counted with the same coefficients, and are thus trivial. 

\begin{figure}[h]
     \centering
     \includegraphics[width=0.5\linewidth]{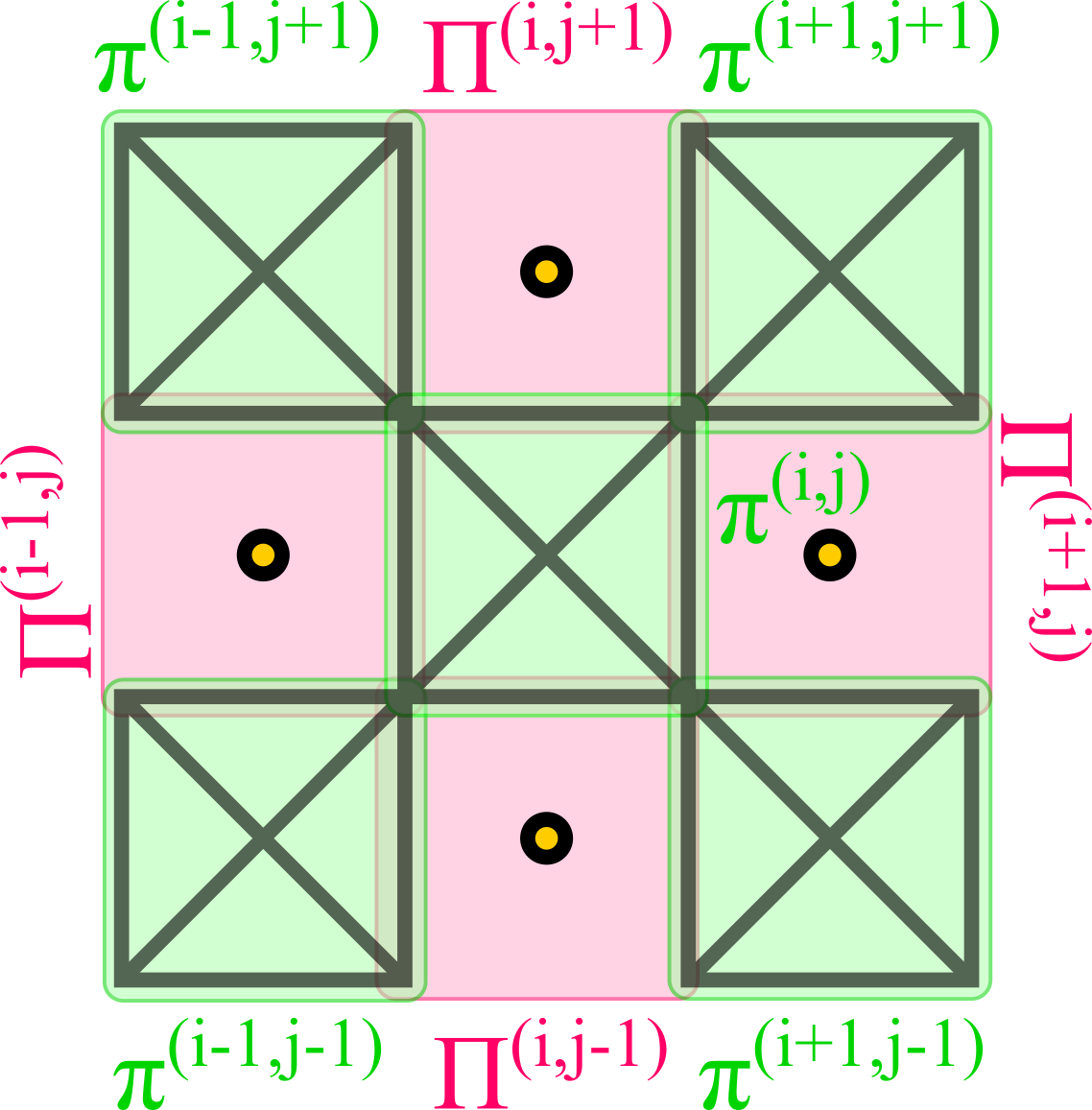}
     \caption{2D rank-2 flux construction for interlayer sites of the 3D checkerboard model. Note that the axis have been rotated from 45° compared to the parent layer construction. The intermediate layer spins, which locations are depicted with yellow dots, are encapsulated in the rank-2 fluxes $\bm{\Pi}$, while fluxes $\bm{\pi}$ contain no spins and are therefore trivial. }
     \label{Fig: ckb rank-2 fluxes interlayers}
 \end{figure}

Now, this 2D rank-2 flux construction can be adapted to intermediate layers, rotating axis from 45° compared with the parent layers construction while building similar square fluxes, see Fig.~\ref{Fig: ckb rank-2 fluxes interlayers}. This way the four cluster interlayer sites will naturally enter in diagonal fluxes $\bm{\Pi}^{(i\pm1, j)}$ and $\bm{\Pi}^{(i, j\pm1)}$ as 
\begin{equation}
    \bm{\Pi}^{(i+1, j)} = \begin{pmatrix}
        \delta \mathbf{S}_{(i+1, j)} & 0 \\ 0 & \delta \mathbf{S}_{(i+1, j)}
    \end{pmatrix}
\end{equation}
while diagonal fluxes will be trivial $\bm{\pi}^{(i,j)} = 0$. This allows to have as required
\begin{equation}
    \begin{split}
        \sum_{\mu', \nu'} \partial_{\mu'} \partial_{\nu'} E_{\mu' \nu'}^{(i,j)} 
        &= \sum_{\mu' =x',y'} \partial_{\mu'} \partial_{\mu'} E_{\mu' \mu'}^{(i,j)} \\
    &\simeq \Pi_{x'x'}^{(i+1,j)} + \Pi_{x'x'}^{(i-1,j)} - 2 \pi_{x'x'}^{(i,j)} 
    + (y'y')\\
    &= \delta \sum_{s \in \langle (i,j) \rangle} \mathbf{S}_s 
    = \bm{\mathcal{C}}_{(i,j)}^{il}
    \end{split}
\end{equation}
for each intermediate layer. As there is a unique spin encapsulated in each flux, there is no reason for the flux to be trivial. The rank-2 electric field emerging from intermediate spins can therefore never become trivial, explaining why the expansion of $L_3(\mathbf{q^\star_\perp})$ cannot be more than second order regardless of the chosen pinch point position $q^\star_\perp$ among the BZ. Note that the effective field produced by this construction has only diagonal terms in the $x',y'$ basis, in good agreement with Eq.~(\ref{Eq: E il checkerboard}) obtained in Subsec.~\ref{subsec: 3D ckb} discussing the 3D checkerboard model, where the $x',y'$ convention where also used.

There must finally exist a rank-3 flux construction, where parent fluxes are no more trivial, that will carry both parent and interlayers spins, producing, as for rank-$n$ pinch lines, a layer field $E_l$ for each parent layer $l$. These fields will interfere together to produce two distinct rank-3 electric fields, one relying on parent sites only, and the second encapsulating all lattice site as discussed for a rank-$n$ pinch line.


\bibliographystyle{apsrev4-2}
\bibliography{BibliotheseP}

\begin{thebibliography}{47}%
\makeatletter
\providecommand \@ifxundefined [1]{%
 \@ifx{#1\undefined}
}%
\providecommand \@ifnum [1]{%
 \ifnum #1\expandafter \@firstoftwo
 \else \expandafter \@secondoftwo
 \fi
}%
\providecommand \@ifx [1]{%
 \ifx #1\expandafter \@firstoftwo
 \else \expandafter \@secondoftwo
 \fi
}%
\providecommand \natexlab [1]{#1}%
\providecommand \enquote  [1]{``#1''}%
\providecommand \bibnamefont  [1]{#1}%
\providecommand \bibfnamefont [1]{#1}%
\providecommand \citenamefont [1]{#1}%
\providecommand \href@noop [0]{\@secondoftwo}%
\providecommand \href [0]{\begingroup \@sanitize@url \@href}%
\providecommand \@href[1]{\@@startlink{#1}\@@href}%
\providecommand \@@href[1]{\endgroup#1\@@endlink}%
\providecommand \@sanitize@url [0]{\catcode `\\12\catcode `\$12\catcode `\&12\catcode `\#12\catcode `\^12\catcode `\_12\catcode `\%12\relax}%
\providecommand \@@startlink[1]{}%
\providecommand \@@endlink[0]{}%
\providecommand \url  [0]{\begingroup\@sanitize@url \@url }%
\providecommand \@url [1]{\endgroup\@href {#1}{\urlprefix }}%
\providecommand \urlprefix  [0]{URL }%
\providecommand \Eprint [0]{\href }%
\providecommand \doibase [0]{https://doi.org/}%
\providecommand \selectlanguage [0]{\@gobble}%
\providecommand \bibinfo  [0]{\@secondoftwo}%
\providecommand \bibfield  [0]{\@secondoftwo}%
\providecommand \translation [1]{[#1]}%
\providecommand \BibitemOpen [0]{}%
\providecommand \bibitemStop [0]{}%
\providecommand \bibitemNoStop [0]{.\EOS\space}%
\providecommand \EOS [0]{\spacefactor3000\relax}%
\providecommand \BibitemShut  [1]{\csname bibitem#1\endcsname}%
\let\auto@bib@innerbib\@empty
\bibitem [{\citenamefont {Wen}(2002)}]{Wen_2002}%
  \BibitemOpen
  \bibfield  {author} {\bibinfo {author} {\bibfnamefont {X.-G.}\ \bibnamefont {Wen}},\ }\href {https://doi.org/10.1103/PhysRevB.65.165113} {\bibfield  {journal} {\bibinfo  {journal} {Phys. Rev. B}\ }\textbf {\bibinfo {volume} {65}},\ \bibinfo {pages} {165113} (\bibinfo {year} {2002})}\BibitemShut {NoStop}%
\bibitem [{\citenamefont {Castelnovo}\ \emph {et~al.}(2008)\citenamefont {Castelnovo}, \citenamefont {Moessner},\ and\ \citenamefont {Sondhi}}]{castelnovo08a}%
  \BibitemOpen
  \bibfield  {author} {\bibinfo {author} {\bibfnamefont {C.}~\bibnamefont {Castelnovo}}, \bibinfo {author} {\bibfnamefont {R.}~\bibnamefont {Moessner}},\ and\ \bibinfo {author} {\bibfnamefont {S.~L.}\ \bibnamefont {Sondhi}},\ }\href {https://doi.org/DOI 10.1038/nature06433} {\bibfield  {journal} {\bibinfo  {journal} {Nature}\ }\textbf {\bibinfo {volume} {451}},\ \bibinfo {pages} {42} (\bibinfo {year} {2008})}\BibitemShut {NoStop}%
\bibitem [{\citenamefont {Kitaev}(2006)}]{Kitaev06a}%
  \BibitemOpen
  \bibfield  {author} {\bibinfo {author} {\bibfnamefont {A.}~\bibnamefont {Kitaev}},\ }\href {https://doi.org/http://dx.doi.org/10.1016/j.aop.2005.10.005} {\bibfield  {journal} {\bibinfo  {journal} {Annals of Physics}\ }\textbf {\bibinfo {volume} {321}},\ \bibinfo {pages} {2 } (\bibinfo {year} {2006})}\BibitemShut {NoStop}%
\bibitem [{\citenamefont {Knolle}\ and\ \citenamefont {Moessner}(2019)}]{Knolle2019}%
  \BibitemOpen
  \bibfield  {author} {\bibinfo {author} {\bibfnamefont {J.}~\bibnamefont {Knolle}}\ and\ \bibinfo {author} {\bibfnamefont {R.}~\bibnamefont {Moessner}},\ }\href {https://doi.org/10.1146/annurev-conmatphys-031218-013401} {\bibfield  {journal} {\bibinfo  {journal} {Annual Review of Condensed Matter Physics}\ }\textbf {\bibinfo {volume} {10}},\ \bibinfo {pages} {451} (\bibinfo {year} {2019})},\ \Eprint {https://arxiv.org/abs/https://doi.org/10.1146/annurev-conmatphys-031218-013401} {https://doi.org/10.1146/annurev-conmatphys-031218-013401} \BibitemShut {NoStop}%
\bibitem [{\citenamefont {Henley}(2010)}]{henley2010coulomb}%
  \BibitemOpen
  \bibfield  {author} {\bibinfo {author} {\bibfnamefont {C.~L.}\ \bibnamefont {Henley}},\ }\href@noop {} {\bibfield  {journal} {\bibinfo  {journal} {Annu. Rev. Condens. Matter Phys.}\ }\textbf {\bibinfo {volume} {1}},\ \bibinfo {pages} {179} (\bibinfo {year} {2010})}\BibitemShut {NoStop}%
\bibitem [{\citenamefont {Pretko}(2017)}]{Pretko_2_2017}%
  \BibitemOpen
  \bibfield  {author} {\bibinfo {author} {\bibfnamefont {M.}~\bibnamefont {Pretko}},\ }\href {https://doi.org/10.1103/PhysRevB.96.035119} {\bibfield  {journal} {\bibinfo  {journal} {Phys. Rev. B}\ }\textbf {\bibinfo {volume} {96}},\ \bibinfo {pages} {035119} (\bibinfo {year} {2017})}\BibitemShut {NoStop}%
\bibitem [{\citenamefont {Pretko}\ \emph {et~al.}(2020)\citenamefont {Pretko}, \citenamefont {Chen},\ and\ \citenamefont {You}}]{pretko20a}%
  \BibitemOpen
  \bibfield  {author} {\bibinfo {author} {\bibfnamefont {M.}~\bibnamefont {Pretko}}, \bibinfo {author} {\bibfnamefont {X.}~\bibnamefont {Chen}},\ and\ \bibinfo {author} {\bibfnamefont {Y.}~\bibnamefont {You}},\ }\href {https://doi.org/10.1142/S0217751X20300033} {\bibfield  {journal} {\bibinfo  {journal} {International Journal of Modern Physics A}\ }\textbf {\bibinfo {volume} {35}},\ \bibinfo {pages} {2030003} (\bibinfo {year} {2020})}\BibitemShut {NoStop}%
\bibitem [{\citenamefont {Nandkishore}\ and\ \citenamefont {Hermele}(2019)}]{Nandkishore19a}%
  \BibitemOpen
  \bibfield  {author} {\bibinfo {author} {\bibfnamefont {R.~M.}\ \bibnamefont {Nandkishore}}\ and\ \bibinfo {author} {\bibfnamefont {M.}~\bibnamefont {Hermele}},\ }\href {https://doi.org/10.1146/annurev-conmatphys-031218-013604} {\bibfield  {journal} {\bibinfo  {journal} {Annual Review of Condensed Matter Physics}\ }\textbf {\bibinfo {volume} {10}},\ \bibinfo {pages} {295} (\bibinfo {year} {2019})}\BibitemShut {NoStop}%
\bibitem [{\citenamefont {Chamon}(2005)}]{Chamon05a}%
  \BibitemOpen
  \bibfield  {author} {\bibinfo {author} {\bibfnamefont {C.}~\bibnamefont {Chamon}},\ }\href {https://doi.org/10.1103/PhysRevLett.94.040402} {\bibfield  {journal} {\bibinfo  {journal} {Phys. Rev. Lett.}\ }\textbf {\bibinfo {volume} {94}},\ \bibinfo {pages} {040402} (\bibinfo {year} {2005})}\BibitemShut {NoStop}%
\bibitem [{\citenamefont {Xu}(2006)}]{Xu06a}%
  \BibitemOpen
  \bibfield  {author} {\bibinfo {author} {\bibfnamefont {C.}~\bibnamefont {Xu}},\ }\href {https://doi.org/10.1103/PhysRevB.74.224433} {\bibfield  {journal} {\bibinfo  {journal} {Phys. Rev. B}\ }\textbf {\bibinfo {volume} {74}},\ \bibinfo {pages} {224433} (\bibinfo {year} {2006})}\BibitemShut {NoStop}%
\bibitem [{\citenamefont {Haah}(2011)}]{Haah11a}%
  \BibitemOpen
  \bibfield  {author} {\bibinfo {author} {\bibfnamefont {J.}~\bibnamefont {Haah}},\ }\href {https://doi.org/10.1103/PhysRevA.83.042330} {\bibfield  {journal} {\bibinfo  {journal} {Phys. Rev. A}\ }\textbf {\bibinfo {volume} {83}},\ \bibinfo {pages} {042330} (\bibinfo {year} {2011})}\BibitemShut {NoStop}%
\bibitem [{\citenamefont {Benton}\ \emph {et~al.}(2016)\citenamefont {Benton}, \citenamefont {Jaubert}, \citenamefont {Yan},\ and\ \citenamefont {Shannon}}]{Benton2016}%
  \BibitemOpen
  \bibfield  {author} {\bibinfo {author} {\bibfnamefont {O.}~\bibnamefont {Benton}}, \bibinfo {author} {\bibfnamefont {L.~D.~C.}\ \bibnamefont {Jaubert}}, \bibinfo {author} {\bibfnamefont {H.}~\bibnamefont {Yan}},\ and\ \bibinfo {author} {\bibfnamefont {N.}~\bibnamefont {Shannon}},\ }\href {https://doi.org/10.1038/ncomms11572} {\bibfield  {journal} {\bibinfo  {journal} {Nature Communications}\ }\textbf {\bibinfo {volume} {7}},\ \bibinfo {pages} {11572} (\bibinfo {year} {2016})}\BibitemShut {NoStop}%
\bibitem [{\citenamefont {Slagle}\ and\ \citenamefont {Kim}(2017)}]{Slagle17a}%
  \BibitemOpen
  \bibfield  {author} {\bibinfo {author} {\bibfnamefont {K.}~\bibnamefont {Slagle}}\ and\ \bibinfo {author} {\bibfnamefont {Y.~B.}\ \bibnamefont {Kim}},\ }\href {https://doi.org/10.1103/PhysRevB.96.165106} {\bibfield  {journal} {\bibinfo  {journal} {Phys. Rev. B}\ }\textbf {\bibinfo {volume} {96}},\ \bibinfo {pages} {165106} (\bibinfo {year} {2017})}\BibitemShut {NoStop}%
\bibitem [{\citenamefont {Yan}\ \emph {et~al.}(2020)\citenamefont {Yan}, \citenamefont {Benton}, \citenamefont {Jaubert},\ and\ \citenamefont {Shannon}}]{Yan_2020}%
  \BibitemOpen
  \bibfield  {author} {\bibinfo {author} {\bibfnamefont {H.}~\bibnamefont {Yan}}, \bibinfo {author} {\bibfnamefont {O.}~\bibnamefont {Benton}}, \bibinfo {author} {\bibfnamefont {L.~D.~C.}\ \bibnamefont {Jaubert}},\ and\ \bibinfo {author} {\bibfnamefont {N.}~\bibnamefont {Shannon}},\ }\href {https://doi.org/10.1103/PhysRevLett.124.127203} {\bibfield  {journal} {\bibinfo  {journal} {Phys. Rev. Lett.}\ }\textbf {\bibinfo {volume} {124}},\ \bibinfo {pages} {127203} (\bibinfo {year} {2020})}\BibitemShut {NoStop}%
\bibitem [{\citenamefont {Benton}\ and\ \citenamefont {Moessner}(2021)}]{Benton_Moessner_2021}%
  \BibitemOpen
  \bibfield  {author} {\bibinfo {author} {\bibfnamefont {O.}~\bibnamefont {Benton}}\ and\ \bibinfo {author} {\bibfnamefont {R.}~\bibnamefont {Moessner}},\ }\href {https://doi.org/10.1103/PhysRevLett.127.107202} {\bibfield  {journal} {\bibinfo  {journal} {Phys. Rev. Lett.}\ }\textbf {\bibinfo {volume} {127}},\ \bibinfo {pages} {107202} (\bibinfo {year} {2021})}\BibitemShut {NoStop}%
\bibitem [{\citenamefont {Niggemann}\ \emph {et~al.}(2023)\citenamefont {Niggemann}, \citenamefont {Iqbal},\ and\ \citenamefont {Reuther}}]{Niggemann_2023}%
  \BibitemOpen
  \bibfield  {author} {\bibinfo {author} {\bibfnamefont {N.}~\bibnamefont {Niggemann}}, \bibinfo {author} {\bibfnamefont {Y.}~\bibnamefont {Iqbal}},\ and\ \bibinfo {author} {\bibfnamefont {J.}~\bibnamefont {Reuther}},\ }\href {https://doi.org/10.1103/PhysRevLett.130.196601} {\bibfield  {journal} {\bibinfo  {journal} {Phys. Rev. Lett.}\ }\textbf {\bibinfo {volume} {130}},\ \bibinfo {pages} {196601} (\bibinfo {year} {2023})}\BibitemShut {NoStop}%
\bibitem [{\citenamefont {Desrochers}\ \emph {et~al.}(2023)\citenamefont {Desrochers}, \citenamefont {Chern},\ and\ \citenamefont {Kim}}]{Desrochers23a}%
  \BibitemOpen
  \bibfield  {author} {\bibinfo {author} {\bibfnamefont {F.}~\bibnamefont {Desrochers}}, \bibinfo {author} {\bibfnamefont {L.~E.}\ \bibnamefont {Chern}},\ and\ \bibinfo {author} {\bibfnamefont {Y.~B.}\ \bibnamefont {Kim}},\ }\href {https://doi.org/10.1103/PhysRevB.107.064404} {\bibfield  {journal} {\bibinfo  {journal} {Phys. Rev. B}\ }\textbf {\bibinfo {volume} {107}},\ \bibinfo {pages} {064404} (\bibinfo {year} {2023})}\BibitemShut {NoStop}%
\bibitem [{\citenamefont {Davier}\ \emph {et~al.}(2023)\citenamefont {Davier}, \citenamefont {G\'omez~Albarrac\'{\i}n}, \citenamefont {Rosales},\ and\ \citenamefont {Pujol}}]{Davier_2023}%
  \BibitemOpen
  \bibfield  {author} {\bibinfo {author} {\bibfnamefont {N.}~\bibnamefont {Davier}}, \bibinfo {author} {\bibfnamefont {F.~A.}\ \bibnamefont {G\'omez~Albarrac\'{\i}n}}, \bibinfo {author} {\bibfnamefont {H.~D.}\ \bibnamefont {Rosales}},\ and\ \bibinfo {author} {\bibfnamefont {P.}~\bibnamefont {Pujol}},\ }\href {https://doi.org/10.1103/PhysRevB.108.054408} {\bibfield  {journal} {\bibinfo  {journal} {Phys. Rev. B}\ }\textbf {\bibinfo {volume} {108}},\ \bibinfo {pages} {054408} (\bibinfo {year} {2023})}\BibitemShut {NoStop}%
\bibitem [{\citenamefont {Yan}\ \emph {et~al.}(2024{\natexlab{a}})\citenamefont {Yan}, \citenamefont {Benton}, \citenamefont {Moessner},\ and\ \citenamefont {Nevidomskyy}}]{Yan_2024_short}%
  \BibitemOpen
  \bibfield  {author} {\bibinfo {author} {\bibfnamefont {H.}~\bibnamefont {Yan}}, \bibinfo {author} {\bibfnamefont {O.}~\bibnamefont {Benton}}, \bibinfo {author} {\bibfnamefont {R.}~\bibnamefont {Moessner}},\ and\ \bibinfo {author} {\bibfnamefont {A.~H.}\ \bibnamefont {Nevidomskyy}},\ }\href {https://doi.org/10.1103/PhysRevB.110.L020402} {\bibfield  {journal} {\bibinfo  {journal} {Phys. Rev. B}\ }\textbf {\bibinfo {volume} {110}},\ \bibinfo {pages} {L020402} (\bibinfo {year} {2024}{\natexlab{a}})}\BibitemShut {NoStop}%
\bibitem [{\citenamefont {Yan}\ \emph {et~al.}(2024{\natexlab{b}})\citenamefont {Yan}, \citenamefont {Benton}, \citenamefont {Nevidomskyy},\ and\ \citenamefont {Moessner}}]{Yan_2024_long}%
  \BibitemOpen
  \bibfield  {author} {\bibinfo {author} {\bibfnamefont {H.}~\bibnamefont {Yan}}, \bibinfo {author} {\bibfnamefont {O.}~\bibnamefont {Benton}}, \bibinfo {author} {\bibfnamefont {A.~H.}\ \bibnamefont {Nevidomskyy}},\ and\ \bibinfo {author} {\bibfnamefont {R.}~\bibnamefont {Moessner}},\ }\href {https://doi.org/10.1103/PhysRevB.109.174421} {\bibfield  {journal} {\bibinfo  {journal} {Phys. Rev. B}\ }\textbf {\bibinfo {volume} {109}},\ \bibinfo {pages} {174421} (\bibinfo {year} {2024}{\natexlab{b}})}\BibitemShut {NoStop}%
\bibitem [{\citenamefont {Fang}\ \emph {et~al.}(2024)\citenamefont {Fang}, \citenamefont {Cano}, \citenamefont {Nevidomskyy},\ and\ \citenamefont {Yan}}]{Fang_2024}%
  \BibitemOpen
  \bibfield  {author} {\bibinfo {author} {\bibfnamefont {Y.}~\bibnamefont {Fang}}, \bibinfo {author} {\bibfnamefont {J.}~\bibnamefont {Cano}}, \bibinfo {author} {\bibfnamefont {A.~H.}\ \bibnamefont {Nevidomskyy}},\ and\ \bibinfo {author} {\bibfnamefont {H.}~\bibnamefont {Yan}},\ }\href {https://doi.org/10.1103/PhysRevB.110.054421} {\bibfield  {journal} {\bibinfo  {journal} {Phys. Rev. B}\ }\textbf {\bibinfo {volume} {110}},\ \bibinfo {pages} {054421} (\bibinfo {year} {2024})}\BibitemShut {NoStop}%
\bibitem [{\citenamefont {Lozano-Gómez}\ \emph {et~al.}(2024{\natexlab{a}})\citenamefont {Lozano-Gómez}, \citenamefont {Noculak}, \citenamefont {Oitmaa}, \citenamefont {Singh}, \citenamefont {Iqbal}, \citenamefont {Reuther},\ and\ \citenamefont {Gingras}}]{Lozano24a}%
  \BibitemOpen
  \bibfield  {author} {\bibinfo {author} {\bibfnamefont {D.}~\bibnamefont {Lozano-Gómez}}, \bibinfo {author} {\bibfnamefont {V.}~\bibnamefont {Noculak}}, \bibinfo {author} {\bibfnamefont {J.}~\bibnamefont {Oitmaa}}, \bibinfo {author} {\bibfnamefont {R.~R.~P.}\ \bibnamefont {Singh}}, \bibinfo {author} {\bibfnamefont {Y.}~\bibnamefont {Iqbal}}, \bibinfo {author} {\bibfnamefont {J.}~\bibnamefont {Reuther}},\ and\ \bibinfo {author} {\bibfnamefont {M.~J.~P.}\ \bibnamefont {Gingras}},\ }\href {https://doi.org/10.1073/pnas.2403487121} {\bibfield  {journal} {\bibinfo  {journal} {Proceedings of the National Academy of Sciences}\ }\textbf {\bibinfo {volume} {121}},\ \bibinfo {pages} {e2403487121} (\bibinfo {year} {2024}{\natexlab{a}})},\ \Eprint {https://arxiv.org/abs/https://www.pnas.org/doi/pdf/10.1073/pnas.2403487121} {https://www.pnas.org/doi/pdf/10.1073/pnas.2403487121} \BibitemShut {NoStop}%
\bibitem [{\citenamefont {Chung}(2024)}]{chung_2024_preprint}%
  \BibitemOpen
  \bibfield  {author} {\bibinfo {author} {\bibfnamefont {K.~T.~K.}\ \bibnamefont {Chung}},\ }\href {https://arxiv.org/abs/2411.03429} {\bibinfo {title} {Mapping the phase diagram of a frustrated magnet: Degeneracies, flat bands, and canting cycles on the pyrochlore lattice}} (\bibinfo {year} {2024}),\ \Eprint {https://arxiv.org/abs/2411.03429} {arXiv:2411.03429 [cond-mat.str-el]} \BibitemShut {NoStop}%
\bibitem [{\citenamefont {Lozano-Gómez}\ \emph {et~al.}(2024{\natexlab{b}})\citenamefont {Lozano-Gómez}, \citenamefont {Benton}, \citenamefont {Gingras},\ and\ \citenamefont {Yan}}]{lozanogomez_2024_preprint}%
  \BibitemOpen
  \bibfield  {author} {\bibinfo {author} {\bibfnamefont {D.}~\bibnamefont {Lozano-Gómez}}, \bibinfo {author} {\bibfnamefont {O.}~\bibnamefont {Benton}}, \bibinfo {author} {\bibfnamefont {M.~J.~P.}\ \bibnamefont {Gingras}},\ and\ \bibinfo {author} {\bibfnamefont {H.}~\bibnamefont {Yan}},\ }\href {https://arxiv.org/abs/2411.03547} {\bibinfo {title} {An atlas of classical pyrochlore spin liquids}} (\bibinfo {year} {2024}{\natexlab{b}}),\ \Eprint {https://arxiv.org/abs/2411.03547} {arXiv:2411.03547 [cond-mat.str-el]} \BibitemShut {NoStop}%
\bibitem [{\citenamefont {Prem}\ \emph {et~al.}(2018)\citenamefont {Prem}, \citenamefont {Vijay}, \citenamefont {Chou}, \citenamefont {Pretko},\ and\ \citenamefont {Nandkishore}}]{Prem_2018}%
  \BibitemOpen
  \bibfield  {author} {\bibinfo {author} {\bibfnamefont {A.}~\bibnamefont {Prem}}, \bibinfo {author} {\bibfnamefont {S.}~\bibnamefont {Vijay}}, \bibinfo {author} {\bibfnamefont {Y.-Z.}\ \bibnamefont {Chou}}, \bibinfo {author} {\bibfnamefont {M.}~\bibnamefont {Pretko}},\ and\ \bibinfo {author} {\bibfnamefont {R.~M.}\ \bibnamefont {Nandkishore}},\ }\href {https://doi.org/10.1103/PhysRevB.98.165140} {\bibfield  {journal} {\bibinfo  {journal} {Phys. Rev. B}\ }\textbf {\bibinfo {volume} {98}},\ \bibinfo {pages} {165140} (\bibinfo {year} {2018})}\BibitemShut {NoStop}%
\bibitem [{Note1()}]{Note1}%
  \BibitemOpen
  \bibinfo {note} {Note that there exists a copy of this electric field for each one of the three spin components.}\BibitemShut {Stop}%
\bibitem [{Note2()}]{Note2}%
  \BibitemOpen
  \bibinfo {note} {From a general point of view, the correct way to obtain the critical vector $\protect \mathbf {l}^c$ if one wants to compute the associated Gauss law is to look for the eigenvector\cite {Yan_2024_long} of of the dispersive band that admits a contact point in $\protect \mathbf {q}_\perp ^*$. In the case of the kagome lattice, this leads to consider $\protect \mathbf {l}^c(\protect \mathbf {q}_\perp ) = \protect \mathbf {l}^\triangledown (\protect \mathbf {q}_\perp ) - e^{i \phi } \protect \mathbf {l}^\vartriangle (\protect \mathbf {q}_\perp )$ with $\phi (\protect \mathbf {q}_\perp ) = \protect \text {Arg} \left ( \protect \mathbf {l}^\triangledown (\protect \mathbf {q}_\perp ) \cdot \protect \mathbf {l}^\triangledown (\protect \mathbf {q}_\perp ) \right )$.}\BibitemShut {Stop}%
\bibitem [{\citenamefont {Chalker}\ \emph {et~al.}(1992)\citenamefont {Chalker}, \citenamefont {Holdsworth},\ and\ \citenamefont {Shender}}]{Chalker1992}%
  \BibitemOpen
  \bibfield  {author} {\bibinfo {author} {\bibfnamefont {J.~T.}\ \bibnamefont {Chalker}}, \bibinfo {author} {\bibfnamefont {P.~C.~W.}\ \bibnamefont {Holdsworth}},\ and\ \bibinfo {author} {\bibfnamefont {E.~F.}\ \bibnamefont {Shender}},\ }\href {https://doi.org/10.1103/PhysRevLett.68.855} {\bibfield  {journal} {\bibinfo  {journal} {Phys. Rev. Lett.}\ }\textbf {\bibinfo {volume} {68}},\ \bibinfo {pages} {855} (\bibinfo {year} {1992})}\BibitemShut {NoStop}%
\bibitem [{\citenamefont {Zhitomirsky}(2002)}]{Zhitomirsky2002}%
  \BibitemOpen
  \bibfield  {author} {\bibinfo {author} {\bibfnamefont {M.~E.}\ \bibnamefont {Zhitomirsky}},\ }\href {https://doi.org/10.1103/PhysRevLett.88.057204} {\bibfield  {journal} {\bibinfo  {journal} {Phys. Rev. Lett.}\ }\textbf {\bibinfo {volume} {88}},\ \bibinfo {pages} {057204} (\bibinfo {year} {2002})}\BibitemShut {NoStop}%
\bibitem [{\citenamefont {Zhitomirsky}(2008)}]{Zhitomirsky2008}%
  \BibitemOpen
  \bibfield  {author} {\bibinfo {author} {\bibfnamefont {M.~E.}\ \bibnamefont {Zhitomirsky}},\ }\href {https://doi.org/10.1103/PhysRevB.78.094423} {\bibfield  {journal} {\bibinfo  {journal} {Phys. Rev. B}\ }\textbf {\bibinfo {volume} {78}},\ \bibinfo {pages} {094423} (\bibinfo {year} {2008})}\BibitemShut {NoStop}%
\bibitem [{\citenamefont {Henley}(2005)}]{Henley2005}%
  \BibitemOpen
  \bibfield  {author} {\bibinfo {author} {\bibfnamefont {C.~L.}\ \bibnamefont {Henley}},\ }\href {https://doi.org/10.1103/PhysRevB.71.014424} {\bibfield  {journal} {\bibinfo  {journal} {Phys. Rev. B}\ }\textbf {\bibinfo {volume} {71}},\ \bibinfo {pages} {014424} (\bibinfo {year} {2005})}\BibitemShut {NoStop}%
\bibitem [{\citenamefont {Moessner}\ and\ \citenamefont {Chalker}(1998)}]{Moessner_1998_counting}%
  \BibitemOpen
  \bibfield  {author} {\bibinfo {author} {\bibfnamefont {R.}~\bibnamefont {Moessner}}\ and\ \bibinfo {author} {\bibfnamefont {J.~T.}\ \bibnamefont {Chalker}},\ }\href {https://doi.org/10.1103/PhysRevB.58.12049} {\bibfield  {journal} {\bibinfo  {journal} {Phys. Rev. B}\ }\textbf {\bibinfo {volume} {58}},\ \bibinfo {pages} {12049} (\bibinfo {year} {1998})}\BibitemShut {NoStop}%
\bibitem [{\citenamefont {Rehn}\ \emph {et~al.}(2016)\citenamefont {Rehn}, \citenamefont {Sen}, \citenamefont {Damle},\ and\ \citenamefont {Moessner}}]{Rehn_Moessner_2016}%
  \BibitemOpen
  \bibfield  {author} {\bibinfo {author} {\bibfnamefont {J.}~\bibnamefont {Rehn}}, \bibinfo {author} {\bibfnamefont {A.}~\bibnamefont {Sen}}, \bibinfo {author} {\bibfnamefont {K.}~\bibnamefont {Damle}},\ and\ \bibinfo {author} {\bibfnamefont {R.}~\bibnamefont {Moessner}},\ }\href {https://doi.org/10.1103/PhysRevLett.117.167201} {\bibfield  {journal} {\bibinfo  {journal} {Phys. Rev. Lett.}\ }\textbf {\bibinfo {volume} {117}},\ \bibinfo {pages} {167201} (\bibinfo {year} {2016})}\BibitemShut {NoStop}%
\bibitem [{\citenamefont {G\'omez~Albarrac\'{\i}n}\ and\ \citenamefont {Pujol}(2018)}]{AlbaPujol2018}%
  \BibitemOpen
  \bibfield  {author} {\bibinfo {author} {\bibfnamefont {F.~A.}\ \bibnamefont {G\'omez~Albarrac\'{\i}n}}\ and\ \bibinfo {author} {\bibfnamefont {P.}~\bibnamefont {Pujol}},\ }\href {https://doi.org/10.1103/PhysRevB.97.104419} {\bibfield  {journal} {\bibinfo  {journal} {Phys. Rev. B}\ }\textbf {\bibinfo {volume} {97}},\ \bibinfo {pages} {104419} (\bibinfo {year} {2018})}\BibitemShut {NoStop}%
\bibitem [{\citenamefont {G\'omez~Albarrac\'{\i}n}\ and\ \citenamefont {Rosales}(2021)}]{AlbaRosales2021}%
  \BibitemOpen
  \bibfield  {author} {\bibinfo {author} {\bibfnamefont {F.~A.}\ \bibnamefont {G\'omez~Albarrac\'{\i}n}}\ and\ \bibinfo {author} {\bibfnamefont {H.~D.}\ \bibnamefont {Rosales}},\ }\href {https://doi.org/10.1088/1361-648X/abf19c} {\bibfield  {journal} {\bibinfo  {journal} {Journal of Physics: Condensed Matter}\ }\textbf {\bibinfo {volume} {33}},\ \bibinfo {pages} {185801} (\bibinfo {year} {2021})}\BibitemShut {NoStop}%
\bibitem [{Note3()}]{Note3}%
  \BibitemOpen
  \bibinfo {note} {This calculation does not account for weather-wane modes on kagome responsible for quartic modes; as $T\rightarrow 0^+$, $C_h\rightarrow 11/12$ \cite {Chalker1992}.}\BibitemShut {Stop}%
\bibitem [{Note4()}]{Note4}%
  \BibitemOpen
  \bibinfo {note} {If the premedial lattice is not bipartite, such a construction is still possible yet more complicated, see for example Fig. 26 from \cite {Davier_2023} for the construction associated with $K$ pinch points in snowflake model.}\BibitemShut {Stop}%
\bibitem [{Note5()}]{Note5}%
  \BibitemOpen
  \bibinfo {note} {Examples of order $n$ divergences : \begin {equation*} \begin {split} &n = 1 : \hskip 2em\relax \protect \bm {\nabla }\cdot \protect \mathbf {E} \\ &n = 2 : \hskip 2em\relax \partial _i \partial _j E_{ij} \\ &n = 3 : \hskip 2em\relax \partial _i \partial _j \partial _k E_{ijk} \\ \end {split} \end {equation*} with Einstein's summation rule over repeated indices.}\BibitemShut {Stop}%
\bibitem [{\citenamefont {Szab\'o}\ \emph {et~al.}(2022)\citenamefont {Szab\'o}, \citenamefont {Orlandi},\ and\ \citenamefont {Manuel}}]{Szabo2022}%
  \BibitemOpen
  \bibfield  {author} {\bibinfo {author} {\bibfnamefont {A.}~\bibnamefont {Szab\'o}}, \bibinfo {author} {\bibfnamefont {F.}~\bibnamefont {Orlandi}},\ and\ \bibinfo {author} {\bibfnamefont {P.}~\bibnamefont {Manuel}},\ }\href {https://doi.org/10.1103/PhysRevLett.129.247201} {\bibfield  {journal} {\bibinfo  {journal} {Phys. Rev. Lett.}\ }\textbf {\bibinfo {volume} {129}},\ \bibinfo {pages} {247201} (\bibinfo {year} {2022})}\BibitemShut {NoStop}%
\bibitem [{\citenamefont {Sklan}\ and\ \citenamefont {Henley}(2013)}]{Sklan13}%
  \BibitemOpen
  \bibfield  {author} {\bibinfo {author} {\bibfnamefont {S.~R.}\ \bibnamefont {Sklan}}\ and\ \bibinfo {author} {\bibfnamefont {C.~L.}\ \bibnamefont {Henley}},\ }\href {https://doi.org/10.1103/PhysRevB.88.024407} {\bibfield  {journal} {\bibinfo  {journal} {Phys. Rev. B}\ }\textbf {\bibinfo {volume} {88}},\ \bibinfo {pages} {024407} (\bibinfo {year} {2013})}\BibitemShut {NoStop}%
\bibitem [{\citenamefont {Jaubert}(2015)}]{Jaubert15c}%
  \BibitemOpen
  \bibfield  {author} {\bibinfo {author} {\bibfnamefont {L.~D.~C.}\ \bibnamefont {Jaubert}},\ }\href {https://doi.org/10.1142/S2010324715400056} {\bibfield  {journal} {\bibinfo  {journal} {Spin}\ }\textbf {\bibinfo {volume} {05}},\ \bibinfo {pages} {1540005} (\bibinfo {year} {2015})}\BibitemShut {NoStop}%
\bibitem [{\citenamefont {Yan}\ \emph {et~al.}(2017)\citenamefont {Yan}, \citenamefont {Benton}, \citenamefont {Jaubert},\ and\ \citenamefont {Shannon}}]{Yan17a}%
  \BibitemOpen
  \bibfield  {author} {\bibinfo {author} {\bibfnamefont {H.}~\bibnamefont {Yan}}, \bibinfo {author} {\bibfnamefont {O.}~\bibnamefont {Benton}}, \bibinfo {author} {\bibfnamefont {L.}~\bibnamefont {Jaubert}},\ and\ \bibinfo {author} {\bibfnamefont {N.}~\bibnamefont {Shannon}},\ }\href {https://doi.org/10.1103/PhysRevB.95.094422} {\bibfield  {journal} {\bibinfo  {journal} {Phys. Rev. B}\ }\textbf {\bibinfo {volume} {95}},\ \bibinfo {pages} {094422} (\bibinfo {year} {2017})}\BibitemShut {NoStop}%
\bibitem [{\citenamefont {Essafi}\ \emph {et~al.}(2017)\citenamefont {Essafi}, \citenamefont {Benton},\ and\ \citenamefont {Jaubert}}]{Essafi17b}%
  \BibitemOpen
  \bibfield  {author} {\bibinfo {author} {\bibfnamefont {K.}~\bibnamefont {Essafi}}, \bibinfo {author} {\bibfnamefont {O.}~\bibnamefont {Benton}},\ and\ \bibinfo {author} {\bibfnamefont {L.~D.~C.}\ \bibnamefont {Jaubert}},\ }\href {https://doi.org/10.1103/PhysRevB.96.205126} {\bibfield  {journal} {\bibinfo  {journal} {Phys. Rev. B}\ }\textbf {\bibinfo {volume} {96}},\ \bibinfo {pages} {205126} (\bibinfo {year} {2017})}\BibitemShut {NoStop}%
\bibitem [{\citenamefont {Schauß}\ \emph {et~al.}(2015)\citenamefont {Schauß}, \citenamefont {Zeiher}, \citenamefont {Fukuhara}, \citenamefont {Hild}, \citenamefont {Cheneau}, \citenamefont {Macrì}, \citenamefont {Pohl}, \citenamefont {Bloch},\ and\ \citenamefont {Gross}}]{shauss2015}%
  \BibitemOpen
  \bibfield  {author} {\bibinfo {author} {\bibfnamefont {P.}~\bibnamefont {Schauß}}, \bibinfo {author} {\bibfnamefont {J.}~\bibnamefont {Zeiher}}, \bibinfo {author} {\bibfnamefont {T.}~\bibnamefont {Fukuhara}}, \bibinfo {author} {\bibfnamefont {S.}~\bibnamefont {Hild}}, \bibinfo {author} {\bibfnamefont {M.}~\bibnamefont {Cheneau}}, \bibinfo {author} {\bibfnamefont {T.}~\bibnamefont {Macrì}}, \bibinfo {author} {\bibfnamefont {T.}~\bibnamefont {Pohl}}, \bibinfo {author} {\bibfnamefont {I.}~\bibnamefont {Bloch}},\ and\ \bibinfo {author} {\bibfnamefont {C.}~\bibnamefont {Gross}},\ }\href {https://doi.org/10.1126/science.1258351} {\bibfield  {journal} {\bibinfo  {journal} {Science}\ }\textbf {\bibinfo {volume} {347}},\ \bibinfo {pages} {1455} (\bibinfo {year} {2015})},\ \Eprint {https://arxiv.org/abs/https://www.science.org/doi/pdf/10.1126/science.1258351} {https://www.science.org/doi/pdf/10.1126/science.1258351} \BibitemShut {NoStop}%
\bibitem [{\citenamefont {Scholl}\ \emph {et~al.}(2021)\citenamefont {Scholl}, \citenamefont {Schuler}, \citenamefont {Williams}, \citenamefont {Eberharter}, \citenamefont {Barredo}, \citenamefont {Schymik}, \citenamefont {Lienhard}, \citenamefont {Henry}, \citenamefont {Lang}, \citenamefont {Lahaye}, \citenamefont {Läuchli},\ and\ \citenamefont {Browaeys}}]{scholl2021}%
  \BibitemOpen
  \bibfield  {author} {\bibinfo {author} {\bibfnamefont {P.}~\bibnamefont {Scholl}}, \bibinfo {author} {\bibfnamefont {M.}~\bibnamefont {Schuler}}, \bibinfo {author} {\bibfnamefont {H.~J.}\ \bibnamefont {Williams}}, \bibinfo {author} {\bibfnamefont {A.~A.}\ \bibnamefont {Eberharter}}, \bibinfo {author} {\bibfnamefont {D.}~\bibnamefont {Barredo}}, \bibinfo {author} {\bibfnamefont {K.-N.}\ \bibnamefont {Schymik}}, \bibinfo {author} {\bibfnamefont {V.}~\bibnamefont {Lienhard}}, \bibinfo {author} {\bibfnamefont {L.-P.}\ \bibnamefont {Henry}}, \bibinfo {author} {\bibfnamefont {T.~C.}\ \bibnamefont {Lang}}, \bibinfo {author} {\bibfnamefont {T.}~\bibnamefont {Lahaye}}, \bibinfo {author} {\bibfnamefont {A.~M.}\ \bibnamefont {Läuchli}},\ and\ \bibinfo {author} {\bibfnamefont {A.}~\bibnamefont {Browaeys}},\ }\href {https://doi.org/10.1038/s41586-021-03585-1} {\bibfield  {journal} {\bibinfo  {journal} {Nature}\ }\textbf {\bibinfo {volume} {595}},\ \bibinfo {pages} {233} (\bibinfo {year} {2021})}\BibitemShut {NoStop}%
\bibitem [{\citenamefont {Ebadi}\ \emph {et~al.}(2021)\citenamefont {Ebadi}, \citenamefont {Wang}, \citenamefont {Levine}, \citenamefont {Keesling}, \citenamefont {Semeghini}, \citenamefont {Omran}, \citenamefont {Bluvstein}, \citenamefont {Samajdar}, \citenamefont {Pichler}, \citenamefont {Ho}, \citenamefont {Choi}, \citenamefont {Sachdev}, \citenamefont {Greiner}, \citenamefont {Vuletić},\ and\ \citenamefont {Lukin}}]{ebadi2021}%
  \BibitemOpen
  \bibfield  {author} {\bibinfo {author} {\bibfnamefont {S.}~\bibnamefont {Ebadi}}, \bibinfo {author} {\bibfnamefont {T.~T.}\ \bibnamefont {Wang}}, \bibinfo {author} {\bibfnamefont {H.}~\bibnamefont {Levine}}, \bibinfo {author} {\bibfnamefont {A.}~\bibnamefont {Keesling}}, \bibinfo {author} {\bibfnamefont {G.}~\bibnamefont {Semeghini}}, \bibinfo {author} {\bibfnamefont {A.}~\bibnamefont {Omran}}, \bibinfo {author} {\bibfnamefont {D.}~\bibnamefont {Bluvstein}}, \bibinfo {author} {\bibfnamefont {R.}~\bibnamefont {Samajdar}}, \bibinfo {author} {\bibfnamefont {H.}~\bibnamefont {Pichler}}, \bibinfo {author} {\bibfnamefont {W.~W.}\ \bibnamefont {Ho}}, \bibinfo {author} {\bibfnamefont {S.}~\bibnamefont {Choi}}, \bibinfo {author} {\bibfnamefont {S.}~\bibnamefont {Sachdev}}, \bibinfo {author} {\bibfnamefont {M.}~\bibnamefont {Greiner}}, \bibinfo {author} {\bibfnamefont {V.}~\bibnamefont {Vuletić}},\ and\ \bibinfo {author} {\bibfnamefont {M.~D.}\ \bibnamefont {Lukin}},\ }\href
  {https://doi.org/10.1038/s41586-021-03582-4} {\bibfield  {journal} {\bibinfo  {journal} {Nature}\ }\textbf {\bibinfo {volume} {595}},\ \bibinfo {pages} {227} (\bibinfo {year} {2021})}\BibitemShut {NoStop}%
\bibitem [{\citenamefont {Chen}\ \emph {et~al.}(2023)\citenamefont {Chen}, \citenamefont {Bornet}, \citenamefont {Bintz}, \citenamefont {Emperauger}, \citenamefont {Leclerc}, \citenamefont {Liu}, \citenamefont {Scholl}, \citenamefont {Barredo}, \citenamefont {Hauschild}, \citenamefont {Chatterjee}, \citenamefont {Schuler}, \citenamefont {Läuchli}, \citenamefont {Zaletel}, \citenamefont {Lahaye}, \citenamefont {Yao},\ and\ \citenamefont {Browaeys}}]{chen2023}%
  \BibitemOpen
  \bibfield  {author} {\bibinfo {author} {\bibfnamefont {C.}~\bibnamefont {Chen}}, \bibinfo {author} {\bibfnamefont {G.}~\bibnamefont {Bornet}}, \bibinfo {author} {\bibfnamefont {M.}~\bibnamefont {Bintz}}, \bibinfo {author} {\bibfnamefont {G.}~\bibnamefont {Emperauger}}, \bibinfo {author} {\bibfnamefont {L.}~\bibnamefont {Leclerc}}, \bibinfo {author} {\bibfnamefont {V.~S.}\ \bibnamefont {Liu}}, \bibinfo {author} {\bibfnamefont {P.}~\bibnamefont {Scholl}}, \bibinfo {author} {\bibfnamefont {D.}~\bibnamefont {Barredo}}, \bibinfo {author} {\bibfnamefont {J.}~\bibnamefont {Hauschild}}, \bibinfo {author} {\bibfnamefont {S.}~\bibnamefont {Chatterjee}}, \bibinfo {author} {\bibfnamefont {M.}~\bibnamefont {Schuler}}, \bibinfo {author} {\bibfnamefont {A.~M.}\ \bibnamefont {Läuchli}}, \bibinfo {author} {\bibfnamefont {M.~P.}\ \bibnamefont {Zaletel}}, \bibinfo {author} {\bibfnamefont {T.}~\bibnamefont {Lahaye}}, \bibinfo {author} {\bibfnamefont {N.~Y.}\ \bibnamefont {Yao}},\ and\ \bibinfo {author} {\bibfnamefont
  {A.}~\bibnamefont {Browaeys}},\ }\href {https://doi.org/10.1038/s41586-023-05859-2} {\bibfield  {journal} {\bibinfo  {journal} {Nature}\ }\textbf {\bibinfo {volume} {616}},\ \bibinfo {pages} {691} (\bibinfo {year} {2023})}\BibitemShut {NoStop}%
\end{thebibliography}%


\end{document}